%% file: ms.tex
\documentclass[tighten,twocolumn,appendixfloats,apjs]{likeapj}
\pdfoutput=1 %ArXiv please use pdflatex
\usepackage{etoolbox}
\makeatletter

\patchcmd{\NAT@citex}
  {\@citea\NAT@hyper@{%
     \NAT@nmfmt{\NAT@nm}%
     \hyper@natlinkbreak{\NAT@aysep\NAT@spacechar}{\@citeb\@extra@b@citeb}%
     \NAT@date}}
  {\@citea\NAT@nmfmt{\NAT@nm}%
   \NAT@aysep\NAT@spacechar\NAT@hyper@{\NAT@date}}{}{}

\patchcmd{\NAT@citex}
  {\@citea\NAT@hyper@{%
     \NAT@nmfmt{\NAT@nm}%
     \hyper@natlinkbreak{\NAT@spacechar\NAT@@open\if*#1*\else#1\NAT@spacechar\fi}%
       {\@citeb\@extra@b@citeb}%
     \NAT@date}}
  {\@citea\NAT@nmfmt{\NAT@nm}%
   \NAT@spacechar\NAT@@open\if*#1*\else#1\NAT@spacechar\fi\NAT@hyper@{\NAT@date}}
  {}{}

  \makeatother
\usepackage{graphicx,bm,mathptmx,amsmath,float}

\hypersetup{linkcolor=blue,citecolor=blue,filecolor=cyan,urlcolor=blue,bookmarksnumbered=true,bookmarksopen=true}
\usepackage[all]{hypcap}
\usepackage{natbib}
\usepackage{graphicx}
\usepackage{epsfig}
\usepackage{array}
\newcolumntype{C}[1]{>{\centering}m{#1}}

\include{newcomm}

\newcommand{\bne}{{\sc bn}{\fontsize{7}{12}\selectfont 11}}
\newcommand{\bnet}{{\sc bn}{\fontsize{6}{12}\selectfont 11}}

\defcitealias{torrejon2010}{T10} 
\newcommand{\ctt}{\citetalias{torrejon2010}}

\received{2017 October 12}
\revised{2018 January 15}
\accepted{2018 January 23}
\published{2018}
\shorttitle{New Constraints from HETG Spectroscopy of \x\ Binaries}
\shortauthors{Tzanavaris \& Yaqoob}

\begin{document}
\title{New Constraints on the Geometry and Kinematics of Matter
  Surrounding the Accretion Flow in X-ray Binaries from
  \textit{Chandra} HETG X-ray Spectroscopy}

\author{P.~Tzanavaris}%\altaffiliation{Not-A Fellow}
\affil{University of Maryland, Baltimore County, 1000
  Hilltop Circle, Baltimore, MD 21250, USA}
\affil{Laboratory for X-ray Astrophysics, NASA/Goddard
  Spaceflight Center, Mail Code 662, Greenbelt, Maryland, 20771, USA;
  \email{panayiotis.tzanavaris-1@nasa.gov}}

\author{T.~Yaqoob}
\affil{University of Maryland, Baltimore County, 1000
  Hilltop Circle, Baltimore, MD 21250, USA}
\affil{Laboratory for X-ray Astrophysics, NASA/Goddard
  Spaceflight Center, Mail Code 662, Greenbelt, Maryland, 20771, USA}
\affil{Department of Physics and Astronomy, The Johns
  Hopkins University, Baltimore, MD 21218, USA}

\begin{abstract}
The narrow, neutral \feka\ fluorescence emission line in X-ray
binaries (XRBs) is a powerful probe of the geometry, kinematics and Fe
abundance of matter around the accretion flow. In a recent study it
has been claimed, using \chandra\ High-Energy Transmission Grating
(HETG) spectra for a sample of XRBs, that the circumnuclear material
is consistent with a solar-abundance, uniform, spherical
distribution. It was also claimed that the \feka\ line was unresolved
in all cases by the HETG.  However, these conclusions were based on ad
hoc models that did not attempt to relate the global column density to
the \feka\ line emission. We revisit the sample and test a
self-consistent model of a uniform, spherical X-ray reprocessor
against HETG spectra from 56 observations of 14 Galactic XRBs. We find
that the model is ruled out in 13/14 sources because a variable Fe
abundance is required. In 2 sources a spherical distribution is viable
but with non-solar Fe abundance.  We also applied a solar-abundance
Compton-thick reflection model, which can account for the spectra that
are inconsistent with a spherical model, but spectra with a broader
bandpass are required to better constrain model parameters. We also
robustly measured the velocity width of the \feka\ line and found full
width half maximum values up to $\sim 5000$ \kmps. Only in
some spectra was the \feka\ line unresolved by the HETG.
\end{abstract}


\section{Introduction}
Galactic\footnote{Although extragalactic XRBs have also been detected
  with \chandra, their distance precludes any high-resolution spectral
  analysis.}\fspace\ X-ray binaries (XRBs) that exhibit absorption and
fluorescent line emission in their X-ray spectra present a powerful
means of studying the matter around the accretion flow.  XRBs consist
of a compact object, either neutron star or black hole, and a donor
star. The type of donor specifies the general XRB classification. XRBs
with massive (e.g. OB-type) donors are known as high-mass XRBs
(HMXBs); low-mass donors are found in low-mass XRBs (LMXBs). HMXBs
usually show prominent stellar wind accretion, giving rise to the
absorbing and line-emitting matter. In contrast, in LMXBs accretion is
usually via Roche Lobe overflow.

In this paper, we revisit 14 Galactic XRBs that have been observed
with \chandra, namely 4U1700$-$37, 4U1822$-$371, 4U1908$+$075,
Cen~X$-$3, Cir~X$-$1, Cyg~X$-$1, Cyg~X$-$3, $\gamma$~Cas, GX~301$-$2,
GX~1$+$4, Her~X$-$1, LMC~X$-$4, OAO~1657$-$415 and Vela~X$-$1.

Regardless of the details of the accretion process, both XRB types are
known to often exhibit fluorescent emission due to neutral and low
ionization iron \citep{gottwald1995}. Fe atoms can exhibit X-ray
fluorescence via absorption of energetic \x\ photons that remove
K-shell electrons. The latter are replaced by upper level electrons,
that in the process emit fluorescent lines at $\sim 6.4$~keV (\feka,
$\sim 1.94$~\AA, L~$\rightarrow$~K electronic transition) and $\sim
7.058$~keV (\fekb,$\sim 1.75$~\AA, M~$\rightarrow$~K transition).

The \feka\ line may have two components, namely a broad, relativistic
component with full-width at half-maximum (FWHM) that can exceed a keV
\citep[e.g., see the reviews by][and references
  therein]{miller2007,reynolds2014,reynolds2016} and a narrow one with
FWHM of up to several tens of eV ($< 2000$~\kmps). The broad
\feka\ line profile probes the innermost regions of the putative
accretion disk in the vicinity of the compact object due to the
imprints of characteristic gravitational and Doppler energy shifts. On
the other hand, the narrow \feka\ line probes the geometry, dynamics,
and element abundances of material surrounding the accretion flow that
is much further from the compact object than the material shaping the
broad line.
In addition to providing access to this circumnuclear
matter, it is necessary to model the narrow \feka\ line with the best
available spectral resolution in order to correctly model the broad
\feka\ line since the features are convolved together when both lines
are present. However, the narrow \feka\ line is too narrow to study
with X-ray CCDs (which typically have a spectral resolution of FWHM
$\sim 7000$~\kmps\ at the \feka\ line energy), and is currently
best-studied with the \chandra\ high-energy transmission grating
spectrometer \citep[HETGS,][no narrow \feka\ line in an X-ray binary
  was observed with the SXS calorimeter aboard {\it
    Hitomi}]{canizares2005}. The \chandra\ HETGS provides a spectral
resolution at the \feka\ line energy of 0.012 and 0.023\AA, FWHM, for
the high- and medium-energy grating, HEG and MEG, respectively. This
is equivalent to $\sim 1800$ and $\sim 3600$ \kmps\ for the HEG and
MEG, respectively. In the present paper we are concerned with the
narrow \feka\ line. Hereafter, the term ``\feka\ line'' will refer to
the narrow line, unless otherwise stated. This emission line
originates in neutral matter, indicated by its centroid energy being
consistent with 6.4 keV. Although in some observations of some XRBs,
narrow emission lines appear at higher energies originating in ionized
Fe, the focus of the present paper is on the \feka\ line from the
neutral matter distribution, which must be distinct from the region
producing ionized lines.

The small effective area of the \chandra\ HETG in the Fe K band
compared to CCD detectors makes it unsuitable to study the
relativistically broadened \feka\ lines reported in the literature for
some of the sources in our sample (details can be found in
Appendix~\ref{app-indi}). The low contrast of such broad lines against
the continuum, combined with limited signal-to-noise ratio render HETG
spectral fits insensitive to modeling the broad lines. On the other
hand, the high spectral resolution of the HETG data means that narrow
\feka\ lines present a higher contrast against the continuum. Thus,
narrow lines that are comparable to, or narrower than, the spectral
resolution are well-suited for studying with the HETG.  Moreover,
model-fitting results for the narrow lines are typically not sensitive
to the presence or absence of a relativistically broadened line.

The only comprehensive study of the narrow \feka\ emission in XRBs
using the \chandra\ HETGS is by \citet[][hereafter, T10]{torrejon2010},
who presented the analysis of 41 XRB \chandra\ HETG spectra (10 HMXBs
and 31 LMXBs), detecting \feka\ emission in all HMXBs and 4
LMXBs. Some sources in the study had multiple observations. The
authors concluded that for all the observations of all of the sources,
the \feka\ emission line (when detected) is produced in a uniform
spherical distribution of matter, with solar abundance. Moreover,
\citetalias{torrejon2010}
concluded that in all sources the \feka\ emission line (when detected)
was unresolved by the \chandra\ HEG.  If true, such robust and sweeping
conclusions about XRBs in general would have far-reaching implications
in terms of the physical conditions, geometry, and dynamics of the
circumnuclear matter if alternative scenarios are strongly ruled
out. However, the conclusions concerning the spherical geometry and
the solar element abundances were based on {\it ad hoc} modeling of
the X-ray spectra using a simple Gaussian model for the \feka\
emission line. No attempt was made to model the Compton-scattered
(reflection) continuum that would be associated with the line-emitting
material, and therefore the flux of the \feka\ emission line was a
free parameter, yet in reality it should be determined by the physical
properties of the material, as is the shape and magnitude of the
reflection continuum. Based on the {\it ad hoc} modeling, \ctt\ simply
showed that the equivalent width (EW) of the \feka\ emission line and
the fitted line-of-sight (l.o.s.) column density (the modeling of which also
ignored Compton scattering), were consistent with a simple analytical
model of a uniform spherical matter distribution. However, the
analytical calculation was strictly based on the assumption that the
radial column density was optically-thin at the \feka\ line energy and
did not account for radiative transfer effects in material out of the
line-of-sight.  Departures from the assumptions in the analytic
formulation can lead to significant errors in the predicted
\feka\ line flux and EW, especially since the l.o.s. column density 
may be different to the global column density in some geometries
\citep[e.g.][]{yaqoob2010}.

The goal of the present paper is to take the sources in the \ctt\ study
in which a \feka\ line was detected and to test whether the
\chandra\ HETGS observations are truly consistent with a uniform,
solar-abundance, spherical distribution of matter by applying a
physically self-consistent model of the \feka\ line emission,
continuum absorption, and Compton scattering. 
Whereas \ctt\ stated that the
\feka\ lines were unresolved by HETGS, without providing case-by-case 
upper limits on the line widths, we provide explicit measurements
or upper limits on the \feka\ line FWHM for spectra that have a
sufficiently high signal-to-noise ratio to constrain the line
width. We find that the spherical X-ray reprocessor model with solar
abundance does not in fact provide an acceptable description of the
spectra for many of the sources, and when it does, it is not a unique
interpretation of the data. Therefore, for each data set we also give
the results of applying a self-consistent toroidal model of the X-ray
reprocessor,
although, as will be explained later, the reprocessor geometry
is not necessarily toroidal.

The use of such self-consistent physical models is completely lacking
in the literature of XRBs, with the notable exception of
\citet{motta2017}, who apply the \myt\ model to the \x\ spectra of
V404 Cygni.  As these authors state, the use of such a model is not
restricted to any specific size scale, and can thus be applied to any
axisymmetric distribution of matter centrally illuminated by \x s.

\renewcommand{\arraystretch}{0.85}
\input{tab_sample}

\renewcommand{\arraystretch}{1}

The structure of the paper is as follows. Section \ref{sec-sample}
presents the sample of XRBs used in our study. Section \ref{sec-red}
sets out the details of the data reduction. Section \ref{sec-anal} presents
details of the self-consistent X-ray reprocessor models used in this
work, and our spectral fitting methodology. Section \ref{sec-res} presents 
and discusses our results. 
A summary and conclusions are given in Section \ref{sec-summ}. Details
of the results for spectral fits to each observation, as well as
pertinent discussion of historical observations, are given in
Appendix~\ref{app-indi}.

\section{Sample}\label{sec-sample}
Since the primary purpose of the present study is to investigate
whether the results and conclusions of the study of HETG observations
of 14 Galactic XRBs by \ctt\ withstand improved modeling, our sample is
based on the same sources. However, we do not use all of the
observations in the \ctt\ sample but we include additional observations
that were not in the \ctt\ sample. The neutral \feka\ fluorescent line
is the main feature that will constrain the self-consistent models
that we apply and any additional features due to ionized Fe can
introduce uncertainty if blending is a factor. Our sample avoids
sources that have spectra dominated by features from ionized Fe.
These sources were
identified in a preliminary analysis of 60 HETG observations of the 14
XRBs. For example, we exclude some observations of \cirxI\ that show
significant P-Cygni absorption profiles, and of \cygxIII\ that show
spectra dominated by features from non-neutral material.

We further exclude observations where the determination of the
position of the zeroth-order image leads to an offset between the $-1$
and $+1$ orders that is greater than the instrumental resolution.
However, we do not in general exclude observations in which there
appears to be no neutral \feka\ emission line (see
\scr{sec-detectability}) because the use of self-consistent models
means that the absence of a line detection must have implications for
some of the other spectral parameters in the model.

Our final sample consists of 56 individual observations. Since \ctt\ did
not use our criteria for excluding observations, and did exclude
spectra with no \feka\ line detections, our sample largely overlaps
theirs, but not completely.  \tr{tab-sample} presents the full list of
observations, XRB types, exposure times, fitting bands, count rates
and line-of-sight 
(non-intrinsic)
neutral hydrogen column densities.  In most cases,
the latter is simply the Galactic column density, \nhgal, obtained
with the FTOOL {\tt nh}, 
based on the LAB survey described in \citet{kalberla2005}. 
However, 
for two sources (Cyg X-1 and Her X-1), it was found that in most
observations the spectrum demanded a smaller column density than this
value.  Since the Galactic column density in the model is degenerate
with any other line-of-sight column density in the model, in these
cases we had to fix it at an arbitrary, low value of
$10^{18}$~\cunits\ (see Sections \ref{sec-cygxi} and \ref{sec-herxi}).

\input{fig-NHZFe}

\section{Data reduction}\label{sec-red}
The data reduction and analysis was carried out using standard X-ray
analysis tools.  First, level 2 event files were produced using the
\chandra\ X-ray Center's CIAO 4.6 data reduction and analysis suite and
our own modified version of the pipeline script {\tt chandra\_repro}.
Specifically, the default, fixed width of the HEG mask in the
cross-dispersion direction (parameter {\tt width\_factor\_hetg}) often
leads to a premature termination at the high-energy end of the
spectrum because the HEG region strip intersects the MEG region strip
too soon.  This leads to compromising the HEG data for the \feka\ line
and Fe K edge region of the spectrum.  We addressed this issue by
modifying the script to accept a variable width.

For a few observations, this procedure failed to correctly identify
the position of the zeroth-order image, leading to poor wavelength
calibration and mismatch between the $-1$ and $+1$ orders.  The
original detection method for the zeroth-order image is a sliding
square ``detect'' cell whose size is matched to the instrument PSF
({\tt celldetect} in CIAO). As this may fail for bright, piled-up, or
blocked zeroth-order images, an alternative method is to find the
intersection of one of the grating arms with the detector readout
streak\footnote{\href{http://space.mit.edu/cxc/analysis/findzo/Data/memo_fzero_1.4.pdf}{http://space.mit.edu/cxc/analysis/findzo/Data/memo\_fzero\_1.4.pdf}}.
{\tt chandra\_repro} uses the script {\tt tgdetect2} to decide which
detection method is appropriate for a given observation. The choice is
based on an empirical relation between the zeroth-order image count
rate and the dispersed spectrum rate. Although tests suggest that the
correct method is chosen in all but 2\%\ of the cases, the tool
clearly fails for two observations of \gxIII\ (obs. IDs 103 and 2733)
and one of \fuXIXbt.  By interactively fitting a two-dimensional
Gaussian surface brightness profile to the zeroth-order image, we
managed to eliminate the problem for all observations, except \gxIII,
obs. ID 103. Therefore this observation was not included in our
sample.

As we are interested in the highest spectral resolution possible, we
only use the HEG data in this paper, rebinned to 0.0025\AA, well below
the theoretical and observationally established instrumental
resolution. The $\pm1$ spectral orders were then individually
extracted and combined for each observation individually in order to
maximize the signal-to-noise.

The zeroth-order image of some of these sources is clearly piled up
but we do not use the zeroth-order data for any scientific
analysis. For higher orders of grating spectra, pile-up will be more
of a concern where the effective area is higher, i.e. at $\sim 2$ keV,
and for MEG data, whereas all of our analysis is above 2 keV using HEG
data. Even if pile-up can be identified, data reconstruction has high
systematic uncertainties \citep{schulz2016}. We thus only flag
observations that may suffer from pile-up by using the validation and
verification pile-up warning provided on the online
\chandra\ Transmission Grating Data Catalog and Archive
\citep[TGCat,][]{huenemoerder2011}. According to this, only 5 out of
56 observations might be affected by serious pile-up. These are one
observation of \cirxI\ (1905) and four observations of \cygxI, namely
13219, 2741, 2742 and 2743. However, this does not affect our overall
results, as the \feka\ line is not detected in any of these
observations. 

In principle, one can further combine spectra for individual
observations to produce a single co-added spectrum for each target.
However, these sources often show significant variability between
observations, and may even have been selected to be in different
spectral states and/or orbital configurations.  As a result, spectral
slopes, column densities and \feka\ emission may not be consistent for
the same target, but instead depend on observation date. This
precludes a physical interpretation of a single, averaged spectrum. We
thus choose to analyze each observation independently, leading to 56
individually analyzed observations.

\section{Analysis Strategy and Spectral-Fitting Models}\label{sec-anal}

We fit spectral models to the HEG spectrum from each observation using
\xspec, version 12.8.1g \citep{arnaud1996}. We use the $C$-statistic
for minimization and statistical error analysis since some of the
spectra have regimes in which the counts per spectral bin are too
low for the $\chi^{2}$ statistic. For models that involve absorption
or scattering we use the \citet{verner1996} photoionization absorption
cross-sections and \citet{anders1989} abundances.  The upper energy of
the spectral-fitting band is 8.0~keV in all cases because the detector
sensitivity and effective area falls off sharply above this energy.
Although it would be desirable to extend the spectral-fitting down to
$\sim 0.9$~keV (the effective lower end of the usable HEG bandpass), 
preliminary examination of the spectra showed many cases
of soft X-ray emission-line complexes. The ionized material responsible
for this soft X-ray emission is distinct from the material
producing the neutral \feka\ line. Modeling these complex soft X-ray
spectra can add considerable burden to the running time and stability of
the spectral-fitting analysis, yet it may not affect modeling of the
neutral \feka\ line. We choose to use 2.4~keV as the lower limit of
the spectral bandpass, and this restriction will always be borne
in mind in our interpretations of the spectral-fitting results and
discussed on a case-by-case basis.

We use two particular models of the X-ray reprocessing of the primary
X-ray continuum, which is responsible for producing the neutral
\feka\ fluorescent emission line, as well as the X-ray absorption and
reflection that are associated with the line-emitting material.  One
of these models is the uniform spherical matter distribution model by
\citet[][hereafter, \bne]{brightman2011}. The other model is a
toroidal reflector, as implemented by the \myt\ model \citep[as
  described in][]{murphy2009}.  The two models will be described
further below. Both models treat the \feka\ line emission and the
Compton-scattered continuum self-consistently. They differ in one key
aspect: that is, the \myt\ model can provide a reflection-dominated
X-ray spectrum since the X-ray source is external to the X-ray
reprocessor, whereas the uniform spherical model cannot give a
reflection-dominated X-ray spectrum because the X-ray source is
embedded at the center of the X-ray reprocessor. Our use of the
\myt\ model is not intended to imply that the geometry is necessarily
actually toroidal.  It is simply one manifestation of possible
geometries and physical scenarios that can give rise to a
reflection-dominated X-ray spectrum.
The particular setup we are using can well be interpreted as
mimicking
a clumpy, patchy, not necessarily toroidal reprocessor \citep[][see also \scr{sec-mytorus}]{yaqoob2012}.
However, the very limited
bandpass of the HETG data means that there is considerable degeneracy
for different specific geometries, so exploration of different
reflection geometries is not warranted and requires simultaneous
higher energy coverage to constrain the continuum shape. In the
present study we use the \myt\ model in preference to available
disk-geometry models \citep[such as {\sc pexmon}, see][]{nandra2007},
because the \myt\ model allows exploration of X-ray reprocessing in a
finite-column density medium, whereas the available disk models force
an infinite column density. This can be particularly important for
high-resolution spectroscopy because the flux of the Compton-shoulder
relative to the core of the \feka\ line is dependent on column
density.  In addition, the \myt\ model properly treats the \feka\ line
as a doublet, with the K$\alpha_{1}$ and K$\alpha_{2}$ components each
forming their own Compton shoulder (see \scr{sec-fekalineenergy} for
details). This allows a more accurate determination of velocity
broadening since a simplistic, single-line treatment would incorrectly
add artificial apparent velocity broadening.

In some of the XRB spectra in our sample, one or both of the X-ray
reprocessor models alone may be insufficient to account for the HEG
spectrum in the fitted bandpass, and in such cases one or more of two
additional types of model components are also used.  One is an
additional (power-law) continuum that is subject to a different
line-of-sight absorption (which could be zero) to that applied to the
primary continuum. This second continuum is needed for spectra that
rise towards low energies, often despite the spectrum below a few keV
being flat due to absorption. The other type of spectral component
uses a simple Gaussian emission-line model, and is used to empirically
account for emission lines that are not included in the X-ray
reprocessor model but may appear in a given spectrum.  One or more
Gaussian model components are included on an as-needed basis. In many
cases these additional emission lines are not needed but when they
are, they are most often associated with the \fetwofive\ He-like
triplet around $\sim 6.7$~keV, or with the \fetwosix\ Ly$\alpha$ line,
at $\sim 6.97$~keV. In general the centroid energy and flux of the
Gaussian components are free parameters but the width is fixed at 100~\kmps\ FWHM if the line is unresolved. The best-fitting parameters of
these additional emission lines from ionized material are not dicussed
in detail and full statistical errors are not derived because they are
only incidental to the primary goal of the present analysis (which is
to determine the constraints on the neutral matter distribution from
the neutral \feka\ line).  The ionized lines are often sufficiently
isolated that they do not affect modeling of the \feka\ line but in
some instances they do have a potential impact and these will be
discussed on a case-by-case basis. In some observations narrow,
isolated ionized lines were detected below 4~keV and we simply excised
the data in narrow energy bands containing such lines.  Again, such
cases will be explicitly pointed out. Other potential emission lines
that can be expected to be detected in some of the spectra are the
fluorescent emission lines from elements other than Fe that originate
in the same neutral material that produces the \feka\ line.  The \bne\
model already includes the neutral fluorescent lines of the common
elements but the \myt\ model does not, so for this model any
fluorescent lines detected from elements other than Fe have to be
explicitly modeled by additional Gaussian components.

For consistency, identical sets of model components are used for all
observations of a given target, although in practice the fitted
parameters for any of these components may effectively remove the
component.  Statistical errors will be given for 90\%\ confidence, one
interesting parameter (corresponding to a \delc\ criterion of
2.706). In some cases this may result in only a lower or upper limit
on a parameter.  Results of the \xspec\ {\tt goodness} command will
also be given, based on 2000 random realizations of the data. The
value gives the percentage of these simulations with the fit statistic
less than that for the original data. In other words, high values of
the goodness, near 100\%, indicate a poor fit, while low values of the
goodness that tend to zero indicate an over-parametrized fit.

\subsection{Uniform Spherical Model}
\label{sec-spherical}
We now briefly summarize the uniform (fully covering), spherical X-ray
reprocessor model of \bne. In this model an isotropic X-ray source is
located at the center of a spherical distribution of neutral
matter. The model is implemented using an \xspec\ emission table
(``atable'' {\tt sphere0708.fits}, see \bne\ for details). This table
is calculated for the central X-ray source having an intrinsic
power-law continuum with a termination energy of 500 keV.  The model
is characterized by the power-law continuum photon index, \gammasph,
and its normalization, \normgammasph, a global radial equivalent
hydrogen column density, \nhsph, a redshift parameter, \zsph, and two
elemental abundances, \zfe\ and \zm, relative to solar. The former is
the Fe abundance relative to the adopted solar value, and the latter
is a single abundance multiplier for C, O, Ne, Mg, Si, S, Ar, Ca, Cr
and Ni relative to their respective solar values. Consistent with
other model components that we use that involve absorption, the
\bne\ model uses the abundances of \citet{anders1989} and
photoelectric absorption cross-sections of \citet{verner1996}.  For
all modeling in this paper, we assume \zm~$=$~1, and only allow \zfe\ to
be a free parameter during spectral fitting. While it is certainly
possible that elements other than Fe may have nonsolar abundances, the
fluorescent lines from the other elements are generally too weak to
constrain \zm. We note that the table model is only valid for the
pre-calculated range of Fe relative abundance, namely 0.1~--~10.0.
Also, the valid ranges of \gammasph\ and \nhsph\ are 1.0~--~3.0 and
$10^{20}$ to $10^{26}$ \cunits\ respectively.  None of the
parameters can vary beyond the valid ranges during spectral fitting.

The \bne\ model self-consistently calculates the fluorescent emission
lines of all of the above-mentioned elements, most importantly the
\feka\ and \fekb\ lines. Compton scattering of continuum and line
photons is explicitly included in the model. However, an important
restriction of the \bne\ table model is that the fluorescent lines,
Compton-scattered continuum, and direct line-of-sight (or
zeroth-order) continuum cannot be separated from each other. One
consequence is that time delays between variations in the zeroth-order
continuum and response of the reflection continuum and fluorescent
lines cannot be accommodated by the model.

\subsection{{\fontsize{7}{12}\selectfont MYTORUS} model}
\label{sec-mytorus}

In the \myt\ model, an isotropic X-ray source is situated at the
center of a toroidal \x\ reprocessor.  The baseline geometry consists
of a torus with circular cross section.  The torus diameter is
characterized by the equatorial column density, \nhs. The global
covering factor of the reprocessor is 0.5, corresponding to a solid
angle subtended at the central \x\ source of $2\pi$ and an opening
half-angle of $60^{\circ}$. The model self-consistently calculates the
\feka\ and K$\beta$ fluorescent emission-line spectrum, as well as the
effects of absorption and Compton scattering on the \x\ continuum and
line emission. As in the \bne\ model, element abundances are from
\citet{anders1989}, and photoelectric absorption cross-sections from
\citet{verner1996}. A limitation of the model is that, in contrast to
the \bne\ model, currently none of the element abundances can be
varied.  For further details of the model we refer the reader to
\citet{murphy2009}, \citet{yaqoob2010} and \citet{yaqoob2012}. More
examples of applications of the model can be found in
\citet{lamassa2014} and \citet{yaqoob2015}.  In the present
application we test the HETG data against the model in a restricted
part of parameter space since the lack of data above 8~keV precludes
allowing too many parameters to be free because the high-energy part
of the Compton-scattered continuum is out of the observed bandpass.
Specifically, we fix the orientation of the torus to be face-on.  The
Compton-scattered (or reflection) continuum is provided by the
\myt\ model component \myts, while the \feka\ and \fekb\ line emission
is provided by the component \mytl\ \citep[see][for
  details]{yaqoob2012}.  For a given column density of the reflecting
medium, the shape of the reflection spectrum viewed along lines of
sight that do not intercept the medium is not very sensitive to the
exact geometry or orientation
\citep[e.g. see][]{liu2014,yaqoob2015,yaqoob2016}. Thus, the face-on
torus reflection spectrum could be representative of other physical
scenarios, such as same-side reflection in a distribution of clouds.
An additional component of the \myt\ suite, \mytz, provides
l.o.s. extinction of the direct continuum due to absorption and
Compton scattering in additional material that is not part of the
torus, characterized by a column density \nhz. This component is the
zeroth-order continuum, which does not depend on geometry (because it
is line-of-sight only), and has no fluorescent line emission (since the
solid angle subtended at the source is by definition
negligible). Moreover, the \mytz\ component is simply an
energy-dependent multiplicative factor so it is also independent of
the shape of the incident intrinsic continuum. The set-up of the
\myt\ model just described is a version of the so-called decoupled
mode of applying the model \citep[see][for
  details]{yaqoob2012,yaqoob2015}.  The zeroth-order continuum (\mytz)
is implemented as an \xspec\ table (``etable'' {\tt
  mytorus\_Ezero\_v00.fits}).

A major advantage of the \myt\ model relative to \bne\ is that it
allows free relative normalizations between different model
components. This can be used to accommodate for differences in the
actual geometry compared to the specific model assumptions used in the
original calculations, as well as for time delays between direct
continuum, Compton-scattered continuum, and fluorescent line photons.
The Compton-scattered continuum (\myts) is implemented as an additive
\xspec\ table, (``atable'' {\tt
  mytorus\_scatteredH500\_v00.fits}), which corresponds to a power-law
incident continuum with a termination energy of 500 keV, and a photon
index, \gammas, in the range $1.4-2.6$. Finally, the \feka\ and
K$\beta$ emission lines (\mytl) are implemented with another additive
\xspec\ table ({\tt mytl\_V000010nEp000H500\_v00.fits}).  Both of
these last two tables are calculated from the same self-consistent
Monte Carlo simulations and have the same five model parameters,
namely, the incident power-law normalization \normgammas, the
power-law photon index \gammas, the toroidal column density \nhs, the
inclination angle of the torus axis relative to the l.o.s. \thobss,
and a redshift parameter \zs. Each of the corresponding parameters in
the \myts\ and \mytl\ components are tied together (and \thobss\ fixed at
$0^{\circ}$), but \myts\ and \mytl\ are multiplied by the relative
normalization factors $A_{S}$ and $A_{L}$, respectively, before being
added to the direct, zeroth-order continuum.  In our modeling we
enforce $A_{L} \equiv A_{S}$, otherwise the self-consistency of the
reflection continuum and \feka\ line emission is broken.  We allow
$A_{S}$ to be free, but this parameter does not have a simple physical
meaning, except for the value of 1.0, which corresponds to a torus
with a covering factor of 0.5, either illuminated by a time-steady
X-ray source, or by a variable source with light-crossing times across
the X-ray reprocessor that are much less than the integration time of
the fitted spectrum. Departures from $A_{S}=1.0$ could be due to a
number of factors, including a different covering factor, nonsolar
abundances, a time lag between the direct and reflected continua, or a
different geometry entirely. For example, the \myts\ and \mytl\ components
could correspond to emission observed through ``holes'' in a
patchy, possibly amorphous, distribution, corresponding to reflection
and fluorescence respectively, from the back-side of material on the
far side of the X-ray source.  Material inbetween the X-ray source and
the observer of such a patchy reprocessor could then correspond to the
\mytz\ component if it lies in the l.o.s. We note that, in any geometry,
if the direct continuum is suppressed and tends to zero, the value of
$A_{S}$ would tend to infinity as the observed spectrum becomes more
and more reflection-dominated.

\input{fig-compA}

The table models for the \myts\ and \mytl\ components do not allow the
photon index, \gammas, to go outside of the pre-calculated range of
1.4 to 2.6. There is no such restriction for the \mytz\ component since
it does not depend on the intrinsic continuum parameters. During
preliminary exploratory spectral fitting we found that many of the
HETG spectra in our sample are very flat, and often require a
power-law photon index less than 1.4. In some cases a value steeper
than 2.6 was required. Therefore we allow the \mytz\ component to be
associated with a photon index, \gammaz, that is independent of the
photon index associated with \myts\ and \mytl\ (\gammas). The outcome of
this approximation is that it is always possible to fit the observed
continuum well, at the expense of accuracy in the shape of the
Compton-scattered continuum. The \feka\ and \fekb\ line fluxes
predicted for a given \gammas\ and \nhs\ are also different if \gammas\
and \gammaz\ are different but the line flux anomaly is assimilated
into the parameter $A_{S}$, to which we are already not assigning a
physical interpretation.  In summary, the caveats described above
prevent a robust interpretation of continuum parameters and column
densities that are obtained from fitting the \myt\ model to the HETG
data in our sample. However, the real value of applying the
\myt\ model is to obtain robust constraints on the velocity broadening
of the \feka\ line because of the superior treatment of the separate
line components and their Compton shoulders, compared to other
available models. For spectra which cannot be explained by a uniform,
fully covering spherical X-ray reprocessor the \myt\ model is thus a
favorable choice.

\subsection{\feka\ Line Detectability}\label{sec-detectability}

In out HETG sample the signal-to-noise ratio in the \feka\ line
emission shows a large range, and in a few of the spectra the
\feka\ line is not detected.  Although even the absence of an
\feka\ emission line in itself can yield important constraints when
applying self-consistent X-ray reprocessing models, clearly, a weak or
absent \feka\ line in the data affects the stability of a spectral fit
since one or more of the model parameters may become
unconstrained. For example, if the \feka\ line is not detected in a
spectrum, it does not make sense for the line width to be a free
parameter. Thus, for each spectrum we derive a quantitative measure of
the statistical quality of the \feka\ line in order to serve as a
guide on the most robust approach to spectral
fitting, and to facilitate the interpretation of the model fitting for
each spectrum. To this end, we fit the restricted energy range
$5.0-6.6$ keV with a simple power-law continuum, and then fit again
with an additional Gaussian model component added to the power-law
continuum. Apart from any \feka\ emission, the above energy range is
usually free from other prominent emission features.  For the purpose
of this exercise, we fixed the peak energy and FWHM of the Gaussian
component at 6.4~keV and 100~\kmps, respectively.  While the actual
data for a given observation may demand a different peak energy and/or
FWHM, this approach provides a very simple assessment with a uniform
treatment of the data across all observations. We use the $C$-statistic
for finding the best-fits, and for each pair of fits for a given
spectrum we then calculate the difference in the $C$-statistic between
the two fits:

\exi 
\Delta C \equiv C_{\rm without \ line} - C_{\rm with \ line} \ .
\exo 

The results are shown in \tr{tab-sample}. We consider \delc~$\ge
6.63$, corresponding to a 99\%\ level confidence for one parameter, to
formally indicate a significant detection of the \feka\ line. Even for
larger values of $\Delta C$ the \feka\ line FWHM may have to be fixed,
and this will be discussed on a case-by-case basis.  In
\tr{tab-sample}, 12 observations have \delc~$< 6.63$.

\subsection{The \feka\ Line Energy}
\label{sec-fekalineenergy}

In the \bne\ and \myt\ models, the centroid energy of the \feka\ line
emission is not a free parameter since it is explicitly modeled as
originating in neutral matter.  The same is true for the \fekb\ line.
In fact, in the \myt\ model, the \feka\ line is explicitly modeled as
the doublet K$\alpha_{1}$ at 6.404 keV and K$\alpha_{2}$ at 6.391 keV,
with a branching ratio of 2:1, respectively, which results in a
weighted mean centroid energy of $6.400$~keV \citep[see][for
  details]{murphy2009}.  However, in practice the peaks of the
\feka\ and \fekb\ emission lines in the actual data may be offset
relative to the baseline model because of instrumental systematics in
the energy scale and/or mild ionization. In addition, the practical
implementation of the \feka\ line in the \bne\ table model is rather
inadequate compared to the precision of the HETG data.  Not only does
the \bne\ model approximate the Fe K$\alpha_{1}$ and Fe K$\alpha_{2}$
doublet as a single line, but the entire line core is covered only by
three energy grid points, each separated by $\sim 10$~eV. Moreover,
the peak energy bin center is not at 6.400 keV, it is at 6.397 keV.
Considering these factors, we make use of the redshift parameter
associated with the fluorescent line emission in the \bne\ and
\myt\ models, allowing it to vary independently as a free parameter
(and allowing it to be positive or negative). The energy shift is
applied to the Compton-scattered continuum as well as to the
fluorescent lines.  In practice, after finding the best-fitting
redshift during spectral fitting, the redshift is frozen at that value
before deriving statistical errors on the other parameters of the
model.  In the tables of spectral-fitting results that we will
present, the redshift offset will be given as the effective
\feka\ line energy offset. A positive shift
means that the \feka\ line centroid energy is higher than the expected
6.400~keV.  For each fitted redshift value, $z$, we calculate the
corresponding energy and velocity shifts ($\Delta E$ and $\Delta v$
respectively) using $\Delta E = E_0 (\frac{-z}{1+z})$ and $\Delta v =
cz$, where $E_{0}$ is the unshifted \feka\ line centroid energy.

\subsection{The \feka\ Line Velocity Width}
\label{sec-fekalinewidth}

For both models, we implement line broadening by using a Gaussian
convolution kernel ({\tt gsmooth} in \xspec) which has an energy width
$\sigma_E = \sigma_L (\frac{E}{6 {\rm keV}})^{\alpha}$.  Here
$\sigma_L$ and $\alpha$ are the two free parameters of the Gaussian
model. We assume a velocity width that is independent of energy,
attained by fixing $\alpha=1$.  We use $\sigma_L$ to estimate a
full-width at half-maximum (FWHM) in velocity units using ${\rm FWHM}
= 2.354 \, c \, (\frac{\sigma_L}{6})$, where $c$ is the speed of light. This
is equivalent to ${\rm FWHM} = 117,700 \, \sigma_{L}{\rm (keV)}$~\kmps.
In cases where either the line is not detected or
the line width cannot be constrained (leading to unstable spectral
fits), we formally fix $\sigma_L$ at $8.5\times 10^{-4}$~keV,
equivalent to $\rm FWHM = 100 $ \kmps, well below the HEG spectral resolution
in the Fe~K energy band.

The \bne\ model does not allow the fluorescent lines to be separated
from the continuum, so in that case the broadening ({\tt gsmooth}) had
to be applied to the \bne\ model continuum as well as the
lines. However, this is not strictly correct because the continuum and
fluorescent lines would not necessarily be expected to have the same
broadening (e.g. the l.o.s. material may have different kinematics to
material out of the l.o.s.).  If we were to include a second, free,
broadening component in the \myt\ model for the continuum, the extra
free parameter in an already complex model would not be
well-constrained. However, since the reflection continuum is broad and
the feature around the Fe~K edge in the reflection continuum is weak
compared to the total continuum, the continuum broadening would not
impact the key model parameters. Moreover, the signal-to-noise ratio
in the HETG spectra drops sharply above $\sim 7$~keV, making it even
less likely for the additional broadening to have an impact.
Therefore, the broadening convolution model was not applied to the
reflection continuum.

\subsection{The \feka\ Line Flux and Equivalent Width}
\label{sec-fekalineflux}

In both the \bne\ and \myt\ models, the \feka\ emission-line flux is
not an explicit model parameter because it is determined
self-consistently by other parameters in the model. Rather, the
\feka\ line flux must be calculated indirectly from the best-fitting
spectra. Moreover, in the \bne\ model table, the fluorescent lines and
continuum cannot be separated so it is not possible to determine
statistical errors on the \feka\ line flux (and EW). For the \myt\ model
we adopt the fractional error on the parameter $A_L$ and assume that
the same fractional error applies to the line flux, as an estimate of
its statistical uncertainty. In cases where $A_L$ is frozen, we untie
it from $A_S$ and allow it to be free temporarily, and then use the
standard procedure in \xspec\ to get the statistical error on $A_{L}$.

For the \bne\ model we estimate the \feka\ line photon flux
by calculating the total photon flux in a narrow energy
band that contains the line (including the Compton shoulder) and
subtracting from this the estimated continuum flux in the same energy
band. The actual energy range depends on the specific spectrum because
it should not be so large that the line flux is a small difference of
two large numbers (in which case it would be subject to unnecessary
uncertainty), and the energy range should not be so narrow that it
does not include all of the line flux. Typically the energy range is
$\sim 6.0-6.5$~keV. The narrow range also excludes contributions of
any line emission from highly ionized Fe. The total flux in the narrow
energy band is obtained using the {\tt flux} commmand in \xspec, and the
continuum flux in the same band is estimated by extracting from the
best-fitting model the average value of the monochromatic fluxes at the
extreme ends of the bandpass and multiplying by the energy width of
the bandpass.

For the \myt\ model we estimate the \feka\ line photon flux by turning off
all continuum components (leaving only the velocity-broadened line
table, \mytl, in place), and using the \xspec\ {\tt flux} command over the
energy range $6.0-6.5$~keV.  The exact range is not critical since the
continua are turned off, but the upper end excludes the \fekb\ line,
which is part of the same table model.  Again, the \myt\ \feka\ line
flux includes any Compton shoulder emission.

The \xspec\ {\tt flux} command also gives energy fluxes in addition to
photon fluxes and the former are used to calculate an estimate for the
energy flux of the \feka\ line, \ffeka.

For both the \bne\ and \myt\ models, the EW of the \feka\ line is
calculated by dividing the \feka\ flux by the monochromatic continuum
flux at the line peak energy. Values of the latter are estimated from
inspection of the best-fitting models. No statistical errors on the EW
can be calculated for the \bne\ model (for the same reason as for the
line flux), and for the \myt\ model the statistical errors on the EW
are calculated using the fractional statistical errors on $A_{L}$.

\subsection{Observed and Intrinsic Continuum Fluxes}
\label{sec-contfluxes}
For each observation, we calculate two different fluxes for each of
the applied models (\bne\ and \myt).
\begin{enumerate}
\item \fcobs\ (\funits), 
the total observed continuum flux.
\item \fcintr\ (\funits), the continuum flux
  originating in the source that would be measured without any
  interaction with the X-ray reprocessor (i.e. the observed flux
  corrected for any absorption and scattering).
\end{enumerate}

In the \bne\ table model the continuum and line emission cannot be
separated, so we first estimate the total observed flux, \ftot, of
each best-fitting model in the $2-10$~keV band, by using the {\tt
  flux} command in \xspec. Although we have not fitted the data beyond
8.0~keV, we perform this extrapolation for the purposes of comparison
with other work. The observed continuum flux between $2-10$~keV is
then estimated by simply subtracting the energy flux of the
\feka\ line (\ffeka) from the total flux, or
\fcobs~$\sim$~\ftot~$-$~\ffeka. This procedure somewhat over-estimates
the continuum flux because it still includes the fluxes of \fekb\ line
and other weak fluorescent lines. However, any additional Gaussian
emission-line components that were included in the model are turned
off. We estimate the intrinsic continuum flux (i.e. with no absorption
or scattering), \fcintr, by setting up a power law with a photon index
equal to the best-fitting values of \gammasph\ and normalization,
\normgammasph, and then calculating the associated flux with the {\tt
  flux} command.

For the \myt\ model, \fcobs\ is calculated by turning off all of the
emission lines in the best-fitting model and using the {\tt flux}
command in \xspec.  The intrinsic continuum flux, \fcintr, is
calculated by using the {\tt flux} command on the direct power-law
component with a photon index (\gammaz) and normalization
(\normgammaz) from the best-fitting \myt\ model.  We do not give
statistical errors on any of the continuum fluxes for any model
because absolute continuum fluxes are dominated by systematic
uncertainties which are not well quantified but are typically of the
order of $\sim 10-20\%$, considering multimission calibration studies
\citep[e.g.][see also the HETG team webpage\footnote{\href{http://space.mit.edu/CXC/calib/hetgcal.html}{http://space.mit.edu/CXC/calib/hetgcal.html}}]{madsen2017,tsujimoto2011}.

\section{Results}\label{sec-res}

We now present overviews and salient aspects of the spectral-fitting
results.  A more detailed description of fitting for individual
systems can be found in Appendix~\ref{app-indi}. A list of plots with
data and fitted models can be found in Appendix~\ref{app-ufda}. Plots
of confidence contours showing constraints on the \feka\ line width
are given in Appendix~\ref{app-con}.  Below we give the key results
separately for the spherical (\bne) and \myt\ models, and
subsequently compare the results and discuss some implications for the
\feka\ line from both models.

\renewcommand{\arraystretch}{0.875}
\input{taball_trans}
\renewcommand{\arraystretch}{1.}

\subsection{Overview of Spherical Model Results\label{sec-sview}}
\tr{tab-trans} shows the best-fitting results for the spherical
\bne\ model.

The overwhelming result is that a uniform spherical distribution of
matter with solar abundances is ruled out in 8 out of 14 sources
(corresponding to a total of 16 observations).  The remaining 6
sources have at least one observation for which the Fe abundance is
within 20\%\ of the solar value, within the statistical errors. The
sources with one such observation are \fuXIX, \cirxI, \cygxI, \velxI,
while \gxIII\ and \herxI\ have two such observations each. This
leaves 31 observations of these 6 sources for which the spherical,
solar abundance model is rejected.  Thus, we can say that 47 out of 56
spectra are not consistent with the \feka\ emission line originating
in a uniform spherical distribution of matter with an Fe abundance
that is solar within 20\%.  These observations fit into two
categories: either the derived Fe abundance is non-solar (by more than
20\%), with robust lower and upper bounds, or the Fe abundance is
unconstrained. The latter category is itself comprised of two
sub-categories: one in which the Fe abundance only has a lower or
upper bound, and one in which the Fe abundance has reached the maximum
model table value of 10.0~\zfesun.  In the latter sub-category, the fitted
column density is so small that the Fe abundance is driven to the
highest value in the model table in order to attempt to account for
the \feka\ line flux, but this abundance is still insufficient to fit
the line. In such cases (20 observations of 9 sources) we do not
derive statistical errors for the other model parameters because the
fit has failed in its objective to account for the \feka\ line. There
are 6 observations with only a lower or upper bound on the Fe
abundance, leaving 20 observations of 11 sources that have a non-solar
Fe abundance with both lower and upper bounds.  These Fe abundance
values range 
from a factor of $\sim 0.3$  to $\sim 6$ solar, i.e. from sub-solar 
to super-solar.
The corresponding column density measurements, \nhsph,
range from $\sim 10^{21}$ \cunits\ to $\sim 10^{24}$ \cunits,
and \fr{fig-NHZFe} shows \nhsph\ plotted against the Fe
abundance (\zfe) measurements for all the observations that have both
lower and upper bounds on both of the parameters.  These Fe abundance
measurements, especially the non-solar values, may of course be an
artifact of fitting a model that is not appropriate.
If we further
stipulate that an apparently variable Fe abundance among different
observations of the same source is not physical, we are left with only
3 sources out of the sample of 14 (\fuXVII, \fuXVIII, \gxIII) that are
consistent with a uniform, spherical distribution of matter, and of
these only \gxIII\ has an Fe abundance within 20\%\ of the solar value.
In fact, \gxIII\ stands out as the only source that has an
approximately solar Fe abundance in more than one observation.
The derived Fe abundance for the one
\noindent observation of \fuXVII\ is
$0.60^{+0.06}_{-0.05}$ \zfesun, and the Fe abundances for the three
observations of \fuXVIII\ are all consistent with $\sim 3.1$ \zfesun\ (see
\tr{tab-trans}). These non-solar values may be artificial but it is
not possible to determine this definitively with the current
data. However, in \scr{sec-mytoverview} we will show the results of
fitting these and other observations with a solar-abundance Compton
reflector (using the \myt\ model), as opposed to a reprocessor with a
closed spherical geometry.

\tr{tab-trans} (column 6) shows that the primary power-law continuum
is generally very flat, and in about half of the spectra 
(27/56) \gammasph\ reached 1.0, the lowest value in the \bne\ table
model. While this was adequate to describe the continuum, without
higher energy spectral coverage it was not possible to better
constrain the continuum. Hard power-law continua are not uncommon in
X-ray binaries, and in particular for many of the sources in this
paper
\citep[e.g.][]{oosterbroek2001,chakrabarty2002,boroson2003,watanabe2003,ji2011,burderi2000,neilsen2009,smith2012,grinberg2013,paul2005}.

%gawk -F"&" '{if($8!~"ldots") {print $7}}' NAKED_taball_trans.tex | wc -l
\tr{tab-trans} also indicates that about half of the observations
(29/56) also require a second power law.  As already mentioned, this
is often required to fit a continuum rise in the soft part of the
spectrum. However, 3 of these observations have \gammasoft~$< 0$, so
that component might also be compensating for the inability of the
\bne\ model component photon index to access values harder than 1.0.

\subsection{Overview of {\fontsize{7}{12}\selectfont MYTORUS} Model Results}
\label{sec-mytoverview}

Whereas many of the spectra that were fitted with the spherical
\bne\ model did not have sufficient column density to account for the
\feka\ line flux even for an Fe abundance as high as 10~\zfesun, the
\myt\ model can provide unobscured lines of sight to a Compton
reflection continuum and fluorescent line emission.  The model is set
up as described in \scr{sec-mytorus} but it is important to note that
since there are more free parameters than in the \bne\ model, spectra
in which the \feka\ is not detected cannot provide useful
constraints. In the \bne\ model, due to the smaller number of
parameters, the absence of an \feka\ line provided constraints on the
column density and Fe abundance.  Therefore, the \myt\ model is only
applied to those observations listed in \tr{tab-sample} that have a
significant \feka\ line detection, as defined in
\scr{sec-detectability} ($\Delta C\ge6.63$).  We attempted to constrain
the column density of the material out of the line of sight (\nhs) but
found that for most observations it could not be constrained and often
had a spread of over two orders of magnitude. 
This is likely due to geometric degeneracies and the
very limited HETG bandpass
(see discussion at the beginning of \scr{sec-anal}).
For these cases we fixed
\nhs\ at 10\up{25}~\cunits\ (the maximum allowed by the \myt\ model),
specifically testing the data against a Compton-thick reflection
model. The exceptions were \fuXVII, one observation of \gxIII, and
one observation of \velxI.  The best-fitting results for all of the
\myt\ spectral fits are shown in \tr{tab-myun}. Our procedure for
investigating \nhs\ involved obtaining a first, approximate, best-fit,
before inspecting two-parameter, $A_{S}$ versus \nhs, confidence
contours for each observation.  If the 99\%\ confidence contours were
closed, then the best-fitting value of \nhs\ was used as a better
initial guess for finalizing the fit, otherwise \nhs\ was fixed at
10\up{25}~\cunits.

\renewcommand{\arraystretch}{0.875}
\input{taball_myun}

\renewcommand{\arraystretch}{1}

For the same reason that \nhs\ was difficult to constrain, so was the
continuum photon index \gammas, which had to be frozen at a value at
one of the extreme ends of the allowed range (1.4 in 36 cases, and 2.6
in 6 cases). The line-of-sight column densities (\nhz) were easier to
constrain, and \tr{tab-myun} shows that a wide range is found, from 0
to $\sim 1.5 \times 10^{24}$~\cunits.  Acceptable fits
are obtained for all the
observations, showing that an open, reflection-dominated geometry is
generally viable, and in particular it is needed for the many cases
for which a closed geometery, spherical model fails to explain the
data. However, we stress that we are not inferring a literal physical
interpretation of the data in terms of the \myt\ model toroidal
reflector, especially since the systematics resulting from limitations
on the photon index are absorbed into the parameter $A_{S}$. Rather,
the success of the reflection-dominated fits should be seen as paving
the way for more realistic and physically-motivated reflection models
to be investigated in future work.

\
\subsection{Comparison of the Spherical and {\fontsize{7}{12}\selectfont MYTORUS} Fits} 
\label{sec-sphmytcomp}

From the results in \tr{tab-trans} and \tr{tab-myun}, we see that for
a given observation, the \myt\ fit is statistically similar, or better
than the corresponding \bne\ fit, except for 5 observations of
\herxI\ (obs IDs 2703, 2705, 3821, 4375, and 6149) and 1 observation of
\gxIII\ (obs ID 2733). The residuals in the \myt\ fits to the 5
observations of \herxI\ correspond to excesses of data at the highest
energies, relative to the model. This indicates that it is the
restriction on the minimum value of the photon index for the
Compton-scattered continuum (\gammas) that is likely the cause of the
\myt\ fits being somewhat worse than the \bne\ fits.  In the case of
\gxIIIbt, the \myt\ is significantly worse than the \bne\ fit, and it
can be seen from the spectral plots in Appendix~\ref{app-ufda} that
this is because the \myt\ predicted model \feka\ line flux falls
substantially short of the data.  There are some observations in which
the \myt\ fit is only marginally worse than the \bne\ fit, but given
the limitations placed on some of the parameters of the \myt\ model,
we do not interpret these differences as meaningful. We still conclude
that the \bne\ model is ruled out in the sources in which the Fe
abundance is required to vary among observations of the same
source. That leaves \fuXIX\ (1 observation) and \fuXVIII\ (3
observations) that are consistent with \bne\ models with non-solar Fe
abundance, and \gxIII\ (2 observations) with an Fe abundance within
20\%\ of the solar value. The \bne\ and \myt\ fits are statistically
similar for \fuXIX, \fuXVIII, and \gxIIIct. For \gxIIIbt, as mentioned
above, the \bne\ fit is significantly better than the
\myt\ fit. However, for \gxIIIct, the \myt\ fit is better than the
\bne\ fit, with $C$ being lower by 121.95 for the \myt\ fit than for the
\bne\ fit, despite the fact that the \bne\ model fit has only 2 more
free parameters than the \myt\ fit. The goodness values of the
\bne\ and \myt\ fits are 99.6\%\ and 28.3\%\ respectively, further
indicating that the \myt\ fit is better than the \bne\ fit.  Also,
from the spectral plots in Appendix~\ref{app-ufda} it can be seen that
the poorer fit with the spherical model is due largely to a
significant, broad excess in the $\sim 5$--8~keV band compared to the
\myt\ fit.

\subsection{The \feka\ Line}
\label{sec-ironlineresults}

%PT: gawk -F"&" '{print $1,$2,$(NF)}' NAKED_tab_sample.tex | gawk -F\\ '{print $1,$2}' | gawk '{if($3<6.63) {print}}' 
The \feka\ line is not detected in 12 observations according to the
criterion \delc~$<6.63$ (see \tr{tab-sample}). Specifically, the line
is not detected in any of the 5 observations of \cirxI, and it is not
detected in 7 out of 15 of the \cygxI observations.

From the spectral fits described in \scr{sec-sview}
(\bne\ model) and \scr{sec-mytoverview} (\myt\ model), the detailed
results for the parameters of the \feka\ line are shown together for
both models in \tr{tab-feka}. Shown are only results for observations
in which the \feka\ line was detected (based on the $\Delta C>6.63$
criterion), and in which, for the \bne\ fit, the Fe abundance did not
reach its maximum table value of 10~\zfesun\ (since the latter
condition indicates that the model failed to fit the \feka\ line).
Since a single number for the quality of each fit (``goodness''
parameter in \tr{tab-trans} and \tr{tab-myun}) does not necessarily
indicate how well the model profile of the \feka\ line fits the data,
columns 11 and 12 in \tr{tab-feka} show the maximum residual as a
percentage of the model value in the energy range $6.3-6.5$~keV
for the \bne\ and \myt\ models, respectively. In addition, for each
observation and for both of the models, the data overlaid with the
fitted models are shown in Appendix~\ref{app-ufda} zoomed in on the
energy band containing the \feka\ line. The quality of the fits to the
detailed \feka\ line profile are very important because one of the
premises of our study is to determine what the fitted \feka\ line
model implies for the column density of the global matter distribution
in each observation, and whether the resulting predicted model
continuum does not conflict with the observed spectrum.

\subsubsection{\feka\ Line Shift and Peak Energy}
\label{sec-linepeakenergy}

For both the \bne\ and \myt\ models, the fitted shifts of the
\feka\ line peak energy, $\Delta E$, are shown in \tr{tab-feka}, in
columns 3 and 4, respectively.  The equivalent velocity shifts are
shown in columns 5 and 6 of \tr{tab-feka} for the \bne\ and
\myt\ models, respectively. In
\fr{fig-comp}, panel (a),
the energy shifts from the
\myt\ model are plotted against the corresponding shifts obtained from
the \bne\ model. It can be seen that the shifts from the \bne\ model
are systematically higher than those from the \myt\ model by $\sim 3$~eV.  
However, as explained in \scr{sec-spherical}, this is expected
because of the less accurate and less detailed representation of the \feka\ line
in the \bne\ model table. Despite the offset, the relation between the
shifts from one model versus the other follows a linear trend.  The
shifts are spread in the approximate range of $\pm 10$~eV, with one
outlier at $18^{+12.0}_{-11.1}$~eV (\bne\ model). This supports the
origin of the \feka\ line in essentially neutral matter. The shifts
are not biased in the positive or negative directions. They could be
due to systematic errors in the energy scale, mild ionization, or
small velocity shifts.  Mild ionization actually results in a negative
shift of up to $\sim -10$~eV for ionization states up to Fe~{\sc ix},
\renewcommand{\arraystretch}{0.9}
\input{taball_feka}

\renewcommand{\arraystretch}{1}
and overall, the range in shifts of $\pm 10$~eV constrains the highest
ionization state
to be Fe~{\sc xvii} or so 
\citep[e.g. see][]{palmeri2003,kallman2004}.
As for velocity shifts, a range of $\pm 10$~eV corresponds to an
equivalent velocity shift of $\pm 466 \ \rm km \ s^{-1}$.

\subsubsection{\feka\ Line Flux and Equivalent Width}
\label{sec-linefluxew}

The flux of the \feka\ line for each observation and for each of the
two models (\bne\ and \myt) is given in columns 13 and 14 of
\tr{tab-feka}. Also given in \tr{tab-feka} are the EW values of the
\feka\ line for each observation and each model (columns 15 and 16).
The fluxes and EWs were derived using the methods decribed in
\scr{sec-fekalineflux}, and \tr{tab-feka} includes
those observations for which the Fe abundance in the \bne\ spherical
model fits reached the maximum value of 10~\zfesun. The \bne\ model parameters
for these observations are given for reference only and are not
meaningful because the model did not fit the \feka\ line (they can be
identified in \tr{tab-feka} by the fact that their \bne\ parameters
have no statistical errors).  The \feka\ line fluxes and EW values
from the \myt\ model are therefore the most reliable for examining
consistently derived values for the largest number of sources in the
sample.
These \feka\ line fluxes span a range of a factor of $\sim 170$ 
from $\sim 0.5$ to $87 \times 10^{-4}$ \phunits,
and the EW values span a range of a factor of $\sim 120$,
from $\sim 15$~eV to $\sim 1.9$~keV.
In
\fr{fig-comp}, panel (e),
we show the EW
measurements from the \myt\ model versus those from the \bne\ model,
for those observations that had valid measurements of EW from the
\bne\ model. The agreement with the line of equality in the diagram
demonstrates the consistency of the EW values inferred from the two
models for this subset of the observations. In general, large values
of the EW of the order of 1~keV or higher are associated with
reflection-dominated X-ray spectra regardless of geometry, and smaller
values of EW of the order of 10s of eV are associated with spectra
that are dominated by a direct continuum that swamps the reflection
continuum.
  
In \fr{fig-comp}, panel (f), we show the EW measurements from
the \myt\ model versus those from the Gaussian modeling of
\ctt\ (their Table 2) for the 26 observations that have valid
results in both papers. There appears to be consistency of the EW
values inferred from these two models for this subset of the
observations as well, although for ten observations at EW values
$<200$~eV the \ctt\ values are too low with respect to the line of
equality even when quoted errors are taken into account. However, we
do not expect agreement between the \ctt\ and \myt\ EW values in
every case because (a) in many cases \ctt\ fix the line width
to unresolved ($\sigma_L = 0.005$~\AA), and (b) they do not include a Compton
shoulder.

An inspection of column (16) in \tr{tab-feka} reveals that five
observations show extreme values of ${\rm EW_{MYtorus}} \ge 400$
eV\footnote{A sixth observation with an extreme value, \lmcxIVct, has
  an upper 90\%\ EW uncertainty that is too large, and is not shown in
  panels (e) and (f) of \fr{fig-comp}.}.  These extreme EW
values are consistent with variability in the line-of-sight extinction
of the direct continuum for all of these sources, associated with
absorption clumps in the line of sight, located in the stellar wind
from the non-compact companion, on the companion surface, and/or an
accretion disk. Although these large values appear across fitted
models, including the \bne\ model and the Gaussian \feka\ modeling of
\ctt, it is worth noting that the decoupled \myt\ setup used here,
specifically mimics a clumpy reprocessor distribution.  These
observations and the interpretation of the EW values are discussed in
greater detail in \scr{sec-gxIII} for \gxIIIbt, \scr{sec-herxid} for
\herxIdt, \scr{sec-lmcxIV} for \lmcxIVbt, \hspace{2pt}
\scr{sec-oaoXVI} for \oaoXVIat, \scr{sec-velxIa} for \velxIat\ and
\scr{sec-velxIc} for \velxIct.

\subsubsection{\feka\ Line Width}
\label{sec-linewidthresults}
%SUMMARY to distill:
%ALL observations unresolved in: 4U1700, Gamma Cas, GX1+4, OAO1657, Vela X-1. [5]
%ALL observations resolved in: Cyg X-3, GX301.				      [2]
%Resolved in some but not all observations: 
%4U 1822(1), 4U 1908(1), Cen X-3(1), Her X-1(4), LMC X-4(1).		      [5]
%Cyg X-1: resolved in 3, FWHM could not be measured in 4 (sigma was fixed).   [1]
%Line not detected in any observations: Cir X-1.			      [1]

The \feka\ line width parameter $\sigma_{L}$ (see
\scr{sec-fekalinewidth}) is shown in columns 7 and 8 of \tr{tab-feka}
for the \bne\ and \myt\ models, respectively.  The corresponding pairs
of FWHM are shown in columns 9 and 10. In
\fr{fig-comp}, panel (b),
we show the
FWHM obtained from the \myt\ model plotted against the FWHM obtained
from \bne\ model, for those observations for which both a lower and
upper bound could be obtained for both models. In the remaining cases
there is either only an upper limit on the FWHM or else it could not
be constrained and was fixed at $100 \rm \ km \ s^{-1}$. It can be
seen from
\fr{fig-comp}, panel (b),
that the FWHM values obtained from the
\myt\ model are systematically smaller than the corresponding values
obtained from the \bne\ model.  This is expected because, as explained
in \scr{sec-anal}, the implementation of the \feka\ in the \myt\ model
tables is more accurate than that in the \bne\ model table. The latter
represents the \feka\ line as a single component, whereas the
\myt\ model represents the line as the K$\alpha_{1}$ and K$\alpha_{2}$
doublet, each component of the doublet having its own Compton
shoulder. Also, the energy grid points representing the \feka\ line in
the \bne\ model are much coarser than those in the
\myt\ model. Therefore, hereafter, we take the FWHM values from the
\myt\ model as providing the best indicators of the velocity
dispersion of the matter distribution responsible for producing the
\feka\ line.  The approximate range in FWHM among the observations
represented in
\fr{fig-comp}, panel (b),
is $\sim 1000-5000 \ \rm km \ s^{-1}$.
The distributions of FWHM measurements in histogram form are shown in
\fr{fig-comp}, panel (c).
The histograms do not include observations for
which the FWHM has lower and/or upper bounds for only one of the
models.

\renewcommand{\arraystretch}{0.82}
\input{taball_flux}
\renewcommand{\arraystretch}{1}

In order to determine more rigorously whether the \feka\ line is
resolved in a given observation, we must take into consideration the
fact that the Compton shoulder of the \feka\ line has a direct impact
on the inferred velocity width. Since the strentgh of the Compton
shoulder relative to the line core depends on the column density of
the material in which the line is formed, we examine confidence
contours of FWHM (from the \myt\ model) versus the \myt\ column
density, \nhs. The \myt\ model self-consistently calculates the
Compton shoulder for each component of the \feka\ doublet so there is
no arbitrary ad hoc parameter controlling the magnitude of the
shoulders relative to the line cores.  The two-parameter confidence
contours of FWHM versus \nhs\ (for 68\%, 90\%, and 99\%\ confidence)
are shown in Appendix~\ref{app-con} for every observation in which the
\feka\ line was detected.  At 99\%\ confidence, we can infer the
following. The contours show that the \feka\ line is unresolved in all
of the observations of 5 sources, namely \fuXVII, \gamcas, \gxI,
\oaoXVI, and \velxI.
They also show that the \feka\ line is resolved
in all observations of 2 sources, namely \cygxIII\ and \gxIII.  In 5
sources the \feka\ line is resolved in some of the observations and
unresolved in the remaining observations of a given source. These 5
sources are \fuXVIII, \fuXIX, \cenxIII, \herxI\ and \lmcxIV, and the
upper limits on the FWHM for the unresolved line cases are in the
range $\sim2000-5000 \ \rm km \ s^{-1}$. The contours for \cygxI\ show
that the \feka\ line is resolved in 2 observations of the source, but
the FWHM had to be fixed at 100~\kmps\ in 4 of the remaining 6
observations because it could not be constrained. In \cirxI, the
\feka\ line was not detected in any of the observations.

\subsection{Continuum}
\label{sec-continuumresults}

As described in Sections \ref{sec-mytorus} and \ref{sec-sview}, many
of the X-ray spectra in the sample are very hard with a photon index
$<1.5$, and in some cases the photon index reaches the lower limit of
$1.0$ of the \bne\ table model.  Although such flat spectra are not
unusual for the sources in the sample, the fitted models may only be
empirical descriptions, and the photon indices are often very poorly
constrained due to the lack of data beyond the \chandra\ HETG
bandpass.

Thus, continuum fluxes and luminosities based on the
extrapolation of the model spectra beyond $\sim 10$~keV would not be
reliable.  We give the 2--10~keV observed and intrinsic continuum
fluxes and luminosities for both the \bne\ and \myt\ models in
\tr{tab-flux} (see \scr{sec-contfluxes} for details of how they are
derived). In
\fr{fig-comp}, panel (d),
we show the observed and intrinsic
continuum fluxes from the \myt\ model against the corresponding
quantities obtained from the \bne\ model fits.  The fluxes span a
range of $\sim 4$ orders of magnitude, from $\sim 0.03$ to $\sim 300
\times 10^{-10} \ \rm ergs \ cm^{-2} \ s^{-1}$.  As expected there is
good agreement between the observed fluxes from the \bne\ and
\myt\ models because all models that give good fits to the data should
empirically return consistent fluxes since the models have to follow
the same data. However, intrinsic fluxes (i.e. corrected for
absorption and reflection) are highly model-dependent. In the case of
the \bne\ and \myt\ models, the intrinsic luminosties are most
divergent when the two models involve very different magnitudes of
absorption and/or reflection.

\subsection{Line-Emitting Region Size Constraints}
\label{sec-regionsize}

Although the kinematics of the matter distribution producing the
\feka\ line is unknown, we can perform a simple exercise to estimate
the size of the matter distribution under the assumption of Keplerian
motion.  Specifically, we can use \feka\ line FWHM measurements
derived in \scr{sec-linewidthresults}, combined with mass estimates of
the central compact object, assuming that the line-emitting material
is virialized \citep[e.g., as in][]{shu2011}.  If the velocity
dispersion is $\langle v^2\rangle$, the distance, $r$ of this material
to a compact object of mass \mco\ can be estimated via the relation
$GM_{\rm CO}=r\langle v^2\rangle$. Further assuming that the velocity
dispersion is related to FWHM velocity by $\langle
v^2\rangle=\frac{3}{4} v^2_{\rm FWHM}$ \citep[as in][]{netzer1990}, we
obtain $r=\frac{4c^2}{3v^2_{\rm FWHM}} r_g$, where the gravitational
radius $r_g=GM_{\rm CO} /c^2$. The results of the calculations are
shown in \tr{tab-rg}, in which columns 3 and 4 show the radius in
gravitational radii units for the \bne\ and \myt\ models
respectively. The radius of the line-emitting region in units of
gravitational radii does not of course depend on the mass of the
compact object.  The calculated radii range from thousands to tens of
thousands gravitational radii. In order to estimate the radii in
absolute units, we searched the literature and compiled mass estimates
of the compact object for each source in our sample. The masses and
references to the literature are shown in \tr{tab-rg}.  Since, in many
cases, a range in mass estimates exists for a give source, resulting
from different studies, we performed calculations for the lowest
available mass estimate ($M_{\rm CO,1}$, giving a radius $r_{1}$), and
for the highest available mass estimate ($M_{\rm CO,2}$, giving a
radius $r_{2}$). Values of $r_{1}$ for the two models are shown in
columns 6 and 7 of \tr{tab-rg}, and values of $r_{2}$ are shown in
columns 9 and 10 of \tr{tab-rg}. The values of $r_{1}$ and $r_{2}$ are
typically of the order of $10^{5}$~km.

\input{taball_rg}

\section{Summary and Conclusions}\label{sec-summ}

In the present work, we tested two specific conclusions of a study by
\citet{torrejon2010} of the \chandra\ High Energy Grating (HEG)
spectra of 14 X-ray binaries that exhibit \feka\ line emission from
matter surrounding the accretion flow onto the compact object in each
source. The first is their conclusion that all
observations of the 14 sources were consistent with
the \feka\ line originating in a uniform spherical distribution of
matter with solar Fe abundance. However, this conclusion was based on
ad hoc models that did not self-consistently treat the \feka\ line
emission, absoprtion, and Compton scattering.  Their second conclusion
 was that all of the \feka\ emission lines were
unresolved by the HEG (which has a spectral resolution $\sim 1860
\ \rm km \ s^{-1}$ FWHM at 6.4 keV).  In our study we used 56 HEG
observations of the same 14 sources on which the conclusions of
\citet{torrejon2010} were based. However, instead of using
unphysical, ad hoc models we applied the \citet{brightman2011}
self-consistent model of a spherical, uniform distribution of matter
to the spectra. We also tested the data against a solar-abundance
Compton-thick toroidal reflector model \citep[implemented with the
  \myt\ model of][]{murphy2009}. In addition, we rigorously investigated the
confidence contours of the FWHM of the \feka\ line versus the global
column density for both models. Our results can be summarized as
follows.

\begin{enumerate}
\itemsep-4pt 
\item{The \feka\ line was not detected in all 5 observations of
  \cirxI, and 7 out of 15 observations of \cygxI. }
\item{Only 1 source out of the 14, \gxIII, is consistent with a
  uniform spherical distribution of matter with the abundance of Fe
  within 20\%\ of the solar value.  However, 1 of the 2 observations of
  \gxIII\ is better described by a solar-abundance, Compton-thick
  reflection model.}
\item{Two sources, \fuXVII\ and \fuXVIII, are consistent
  with a uniform spherical distribution with a non-solar Fe
  abundance. This abundance relative to solar is $\sim 0.6$ for
  \fuXVII\ and $\sim 3$ for all 3 observations of \fuXVIII.}
\item{A uniform, spherical distribution of matter is ruled out for the
  remaining 11 sources because a variable Fe abundance among
  different observations is required. Of these 11 sources, 5 have at
  least one observation that happens to have an abundance of Fe that
  is within 20\%\ of the solar value. }
%%%%%%%%%%%%%%%%%%%%%%%%%%%%%%%%%%%%%%%%%%%%%%%%%%%%%%%%%%%%%%%%%%%%%%%%%%%%%%%%%
%gawk -F"&" '{if($7~"{10.") {print $1,$2,$7}}' NAKED_taball_trans.tex | wc -l

%gawk -F"&" '{if($7~"{10.") {print $1,$2,$7}}' NAKED_taball_trans.tex  | gawk '{print $1}' | uniq | wc -l
\item{There are 20 observations of 9 sources for which the Fe
  abundance in the uniform spherical model reached 10~\zfesun, the maximum
  value allowed by the model. These 9 sources are among the 11 that
  are already rejected on the basis that the Fe abundance is required
  to vary between observations.}
\item{ The \feka\ line is unresolved in all of the observations of 5
  sources, namely \fuXVII, \gamcas, \gxI, \oaoXVI, and \velxI.}
\item{The \feka\ line is resolved in all observations of 2 sources,
  namely \cygxIII\ and \gxIII. }
\item{In 5 sources the \feka\ line is resolved in some of the
  observations and unresolved in the remaining observations of a given
  source. These 5 sources are \fuXVIII, \fuXIX, \cenxIII, \herxI, and
  \lmcxIV.}
\item{In \cygxI, among the observations in which the \feka\ line was
  detected, it was resolved in 2 observations, but the line width
  could not be constrained and had to be fixed in 4 out of the 6 remaining 
  observations.}
\item{Among the observations in which the \feka\ line was resolved,
  the FWHM covers a wide range, $\sim 1000-5000 \ \rm km \ s
  ^{-1}$.}
\item{If the matter distribution is virialized, the FWHM measurements
  imply a size range of thousands to tens of thousands gravitational
  radii. Using black-hole and neutron-star mass estimates in the
  literature, this typically corresponds to a range of $\sim 3 \times
  10^{4} \rm \ km$ to $\sim 4 \times 10^{5} \rm \ km$.}
\item{The spherical X-ray reprocessor model of \citet{brightman2011}
  introduces artificial energy shifts and broadening of the
  \feka\ line when fitted to HETG data because of the over-simplified
  implementation of the \feka\ line in this model. Specifically, the
  energy grid in the model table covering the \feka\ line is too
  coarse, and the \feka\ line is modeled as a single component instead
  of the doublet, with components K$\alpha_{1}$ and K$\alpha_{2}$ that
  are separated by $\sim 13$~eV.  The \myt\ model implements a much
  more accurate treatment of the \feka\ line and so does not suffer from
  these systematic effects.}
\item{Although the \myt\ model provides a better or statistically
  similar fit for the majority of observations, there are limitations
  in the \myt\ model and the energy bandpass of the HEG data that are
  important to address in future work. Our results should be
  interpreted as a motivation to apply more physically realistic
  models to spectra with a wider bandpass, while still retaining
  sufficient spectral resolution and throughput in the Fe~K band.}
\end{enumerate}

\acknowledgments

We would like to thank the anonymous referee for their constructive
comments and support. We also thank Mihoko Yukita, Keith Arnaud and
David Huenemoerder for
useful discussions.  Support for this work was provided by the
National Aeronautics and Space Administration through \chandra\ Award
Number AR4-15002A issued by the \chandra\ X-ray Observatory Center,
which is operated by the Smithsonian Astrophysical Observatory for and
on behalf of the National Aeronautics and Space Administration under
contract NAS8-03060.  The scientific results reported in this article
are based on data obtained from the \chandra\ Data Archive.

\facilities{\chandra.}

\appendix
\numberwithin{figure}{section} %needs amsmath!

\section{\\Details on Individual Sources and Observations}\label{app-indi}

This Appendix presents specific details on spectral fitting of
individual observations for all targets in this paper, as well as some
historical information and context for each source.  Our results are
based on fitting the uniform spherical model (see \scr{sec-spherical})
and the \myt\ toroidal X-ray reprocessor model (see
\scr{sec-mytorus}).  We do not directly compare the various column
densities we obtain from our fits with previous work on the same data
that applied ad hoc, phenomenological models because such a comparison
is not meaningful. Only column densities obtained with the same model
are comparable but none of the previous works on the sources in our
sample apply the models that we do. We also note that the precision of
\feka\ line EW values in \ctt\ is over-stated compared to the statistical
errors, but we nevertheless quote exact values given by \ctt\ whenever
we compare our EW values with those of \ctt.  We note that existing
X-ray spectroscopy in the literature for some observations of some
of the sources was performed on spectra selected by orbital phase
rather than by observation IDs. Phase-resolved spectroscopy is beyond
the scope of our present study, whose primary goal is to test the
assertion by \ctt\ that the circumstellar matter distribution in all of
the X-ray binaries in our sample is consistent with a uniform,
solar-abundance sphere. If the assertion is true, the orbital phase
should be irrelevant.
However, accounting for observation phase can help understand the
extreme EW values (${\rm EW_{MYtorus}} \ge 400$ eV) registered for
six of these observations (\scr{sec-linefluxew}). Thus, in the
interest of completeness some phase-resolved spectroscopy results
are taken into account in the discussion of these specific
observations below.
  
In the \myt\ spectral fits, unless otherwise stated, the toroidal
column density, \nhs, is fixed at $10^{25}$~\cunits\ because in most
cases the limited bandpass of the data makes this parameter poorly
constrained, leading to unstable fits.

In some of the spectra, in addition to the basic X-ray reprocessor
model and its associated fluorescent emission lines, additional
emission lines are detected due to ionized Fe. These are discussed on
a case-by-case basis whenever the data required the additional model
components. The parameters for these extra emission lines are not
given for the sake of clarity of presentation of the complex
results. Moreover, those paramaters do not affect our inferences of
the properties of the matter distribution producing the
\feka\ emission line, which is the primary focus of our study.

Regarding our analysis results, when we refer to the \feka\ line being
resolved by the HEG we specifically mean that the two-parameter, 99\%\ 
confidence contour of FWHM versus \nhs\ (the equatorial column density
in the \myt\ model) is greater than zero. We showed in \scr{sec-anal}
that the \feka\ line width derived from the \myt\ is more accurate
than that derived from the \bne\ model because the implementation of
the \feka\ line profile in the \bne\ model is compromised by
approximations discussed in \scr{sec-anal}.  The spectral resolution
of the HEG is $\sim 1860$~\kmps\ FWHM at 6.4~keV.  Most of our
\feka\ line width measurements cannot be compared with previous work
because the HEG line width for the sources in our sample has not been
rigorously measured before for most of the sources in the sample.

In the last few years, a number of papers have presented analyses
for some of these objects using broadband \x\ data from missions
such as \nustar\ or \suzaku. Thus \citet{jaisawal2015} analyze
\suzaku\ data for \fuXVII; \citet{sasano2014} analyze \suzaku\ data
for \fuXVIII; \citet{naik2011} analyze \suzaku, and
\citet{farinelli2016} \suzaku\ and \nustar\ data for \cenxIII;
\citet{tomsick2014} and \citet{parker2015} analyze \suzaku\ and
\nustar\ data for \cygxI, while \citet{walton2016} \nustar-only
data for the same object; \citet{suchy2012} analyze \suzaku\ data
for \gxIII; \citet{yoshida2017} analyze \suzaku\ data for \gxI;
\citet{fuerst2013} and \citet{wolff2016} analyze \nustar, and
\citet{farinelli2016} \suzaku\ and \nustar\ data for \herxI;
\citet{shtykovsky2017} analyze \nustar, and \citet{hung2010}
\suzaku\ data for \lmcxIV; \citet{jaisawal2014} and
\citet{pradhan2014} analyze \suzaku\ data for \oaoXVI; and
\citet{maitra2013} analyze \suzaku\ data for \velxI.

None of these works use either the \bne\ or the \myt\ model employed
in the present paper. It is also important to stress that the photon
indices in these papers are not directly comparable to ours, which
reflect a limited energy range, and cannot be extrapolated to
calculate fluxes and luminosities over more extended energy ranges.

\subsection{\fuXVII}\label{subsec_fuxvii}
This is a supergiant, eclipsing XRB with a single HETG observation in
our sample, \fuXVIIa.  There is debate in the literature for the type,
as well as the mass, of the compact object. The most recent result is
for a neutron star by \citet{falanga2015}, with \mco~$\sim 2$~\msun,
based on more than ten years of monitoring observations with {\it RXTE} and
{\it INTEGRAL} to refine orbital parameter determination. We also use the
earlier determination of \citet{clark2002} of a higher mass ($\sim
2.4$~\msun), suggesting that this compact object is in a grey area
between high-mass neutron stars and low-mass black holes. This work is
based on a detailed non-LTE analysis of the ultraviolet and optical
spectrum of the O-type high-mass donor of this system.

\citet{boroson2003} analyze the same HETG observation as in our
sample, splitting it into three segments, namely ``flaring'' (F),
``quiescent'' (Q), and ``ending'' (E). They further subdivide F to a
``high'' (H) and ``low'' (L) state, and E to an EL state.  The \feka\ line
centroid energy ranges from $1.936 \pm 0.001$\AA\ to $1.940 \pm
0.002$\AA\ (or $6.391\pm0.007  - 6.404\pm0.003$~keV), and the FWHM
from $900\pm 400$ to $2200\pm 700$~\kmps\ for E and F,
respectively. The line is unresolved in their EL state. The line flux
ranges from $3.50\pm1.30$ to $14.0\pm3.00 \ \times
10^{-4}$~\phunits\ for EL and FH, respectively (\citealt{boroson2003} did not
give EW values).

For this observation \ctt\ obtained a line centroid of
$1.9386\pm0.0006$~\AA\ 
($6.3955\pm0.0020$~keV), a line flux of $8.70\pm 0.81 \ \times
10^{-4}$~\phunits, and an EW of $70.19\pm6.45$~eV.

In our analysis, the \bne\ model fits the data well with an iron
abundance of $\sim 0.6$ solar. Although the \feka\ line is resolved in
the \bne\ fit, in the \myt\ fit the \feka\ line is unresolved, with an
upper limit on the FWHM of $975 \ \rm km \ s^{-1}$. Overall, there are
two distinct solutions for the \myt\ fit, one with a low value of
\nhs\ and one with a high value ($\sim 0.4$ and $\sim 2.6 \times
10^{24}$~\cunits\ respectively). We choose the low-\nhs\ solution for
which we give fit results, as this corresponds to a larger, more
conservative \sigl\ and FWHM upper limit\footnote{Note that, unlike a
  few other cases with two possible \nhs\ solutions, the
  high-\nhs\ solution here does not reach the \myt\ model's upper
  bound, $10^{25}$~\cunits, which we normally chose in such cases.}.
The \feka\ line fluxes from both \bne\ and \myt\ fits are consistent
with those obtained by \citet{boroson2003} and \ctt, both of which used
ad hoc, phenomenological models, and our \feka\ EWs are consistent
with the value given by \ctt.

\citet{vandermeer2005} calculate intrinsic fluxes in the
$0.3-11.5$ keV band with \xmm, ranging from $\sim 10^{-10}$ to
$\sim$\ten{4.9}{-9} \funits, depending on model and observation
phase.  These overlap with our result of \ten{1.35}{-9} \funits\
obtained for our single observation with \myt, although in the $2-10$
keV band.

\subsection{\fuXVIII}
This object belongs to the rare class of eclipsing LMXBs.  It is also
the prototypical Compton-thick accretion disk corona (ADC) source, for
which only X-rays scattered from material above and below the
accretion disk are thought to be able to reach the observer
\citep{white1982}.

There is a range of mass estimates for the neutron star compact object
in the literature. \citet{jonker2003} combine radial velocity
estimates from optical spectroscopic observations with pulse timing
analysis to obtain a lower limit of $0.97\pm0.24$~\msun.
\citet{munoz-darias2005} model the ``K-correction'' for emission lines
to obtain a range $1.61 \le M_{\rm NS}/M_{\odot} \le 2.324$. This
range includes the most recent estimate of $1.69\pm
0.13$~\msun\ \citep{iaria2015} from \x-based estimates of the spin
period and its derivative.

There are 3 HETG observations of \fuXVIII\ in our sample (see
\tr{tab-sample}).  For obs.~ID.~671 the combined Medium- and
High-Energy Transmission Grating analysis of \citet{cottam2001}
obtains an upper limit to the \feka\ line broadening of 1500~\kmps.
These authors also split the observation into different phases, always
obtaining a line centroid consistent with 1.935~\AA\ (6.407~keV).
\citet{ji2011} use all 3 observations and divide them into five phases
and obtain \feka\ line wavelengths ranging from
$1.936$--$1.944$\AA\ (or 6.378--6.404~keV). In one case the
\feka\ line is resolved, with $\sigma=$ ~\aer{5.7}{+2.5}{-2.2} $\times
10^{-3}$\AA, or\footnote{\citet{ji2011} use an undefined symbol
  $\sigma$ throughout, with values in \AA.  For conversion to
  velocities, we assume here that this represents FWHM.} $\rm FWHM \ =
\ 880^{+390}_{-340} \ km \ s^{-1}$. In two cases the line is
unresolved, with only an upper limit on the width, the largest being
$7.2\times 10^{-3}$~\AA, corresponding to a FWHM of 1115~\kmps. In the
other two cases the width is simply given as $10^{-3}$\AA\ (or $\rm
FWHM \ = \ 155 \ km \ s^{-1}$), with no statistical errors.

\ctt\ also only analyze obs.~ID.~671, obtaining a centroid of
$1.9379\pm0.0015$~\AA\ ($6.398\pm0.005$~keV) for the \feka\ line, a
flux of $2.94\pm0.69 \times 10^{-4}$\phunits, and an EW of
$36.61\pm8.52$~eV.

Finally, \citet{iaria2013} use an average HETG spectrum from
obs.~ID.~9858 and obs.~ID.~9076 and obtain a line energy of $6.393 \pm
0.002$~keV, line width $\sigma=18\pm3$~eV
or FWHM~$=1990\pm330 \rm \ km \ s^{-1}$, EW~$=$~$40\pm6$~eV, and an
estimated Keplerian region radius of
\aer{1.9}{+0.8}{-0.5}~$\times10^{10}$~cm.

As discussed by \citet{cottam2001}, the HEG spectra for this object
show discrete emission lines for a range of abundant elements. In
particular, both \feka\ and \fetwosix\ emission lines are detected,
which clearly originate in regions with completely different
ionization states.  The lines are reported to be persistent through
all orbital phases, although their intensities vary \citep{ji2011}.
In our analysis, we utilize all 3 HETG observations (obs.~IDs.~671,
9858, and 9076) and for both the \bne\ and \myt\ models, we include a
Gaussian component to fit the high-ionization He-like line at an
energy of $\sim 6.96$~keV.
We find that the \feka\ line is unresolved in obs.~ID.~671 but
resolved in the other two observations, with the FWHM being $\sim
1400$--$1700 \rm \ km \ s^{-1}$ (see \tr{tab-feka} and
\fr{fig-cont}
in
Appendix~\ref{app-con}). The \feka\ line fluxes and EW values we
obtain for both models are consistent with values in the literature
derived from fitting phenomenological models, as described above.  Our
\bne\ fits yield an Fe abundance relative to solar of $\sim 3$ for
each of the three HETG observations, so a uniform, solar-abundance
spherical matter distribution is ruled out.

\subsection{\fuXIX}
This XRB belongs to the rare class of HMXBs with an OB-supergiant
companion \citep{levine2004,morel2005}.  There are no mass estimates
for the primary of this source in the literature. Following standard
practice, we assume a canonical neutron-star mass of
1.4~\msun\ \citep[e.g.][]{levine2004,morel2005,martinez-nunez2015}.
The \x\ emission is hard and highly absorbed with a column density
varying by factors of a few between phases.

Among the 3 HETG observations of this source in the present study,
the \bne\ model requires an Fe abundance that varies among
observations, from $\sim 0.24$--3 solar, so we reject the uniform,
spherical, model (see \tr{tab-trans}). In the \myt\ fits the
line-of-sight column density varies between observations (as found
from previous studies) and the \feka\ line flux varies by an order of
magnitude (the latter is consistent with the phenomenological
modeling of \ctt). The \myt\ fits, corresponding to a Compton-thick
reflector with an Fe abundance that is solar, work because the
line-of-sight absorption is decoupled from the global column
density. In contrast, in the spherical model, the line-of-sight
absorption, which affects the low-energy continuum shape, is the same
as the global column density, which affects the flux of the
\feka\ line. In order to remain self-consistent, the spherical model
can only vary the Fe abundance in order to try to fit the \feka\ line
flux with a column density that must also explain the shape of the
contnuum.  We note that obs.~IDs.~5477 and 6336 also have Compton-thin
reflection solutions that cannot be statistically distinguished from
the Compton-thick solutions with \nhs\ fixed at $10^{25}$~\cunits. We
choose to present the Compton-thick solutions for all 3 observations,
consistent with most of the 56 observations in the present sample.

The \feka\ line is resolved in 1 observation, with FWHM $\sim 2500$~\kmps\ 
(\fuXIXa), but unresolved in the remaining 2 observations.

\subsection{\cenxIII}
This prototypical ``standard'' HMXB is one of the brightest
accreting \x\ pulsars. It is believed to have a hard, phase-dependent
\x\ spectrum, \citep[e.g.][]{burderi2000,iaria2005}.  The observation
of eclipses has permitted the determination of all orbital and stellar
parameters for this system. %\citep{thompson2009} 
In particular the
compact object mass is estimated to be between
\mns~$=1.21\pm0.21$~\msun\ \citep{ash1999} and
$1.57\pm0.16$~\msun\ \citep{falanga2015}.

There are 3 HETG observations of \cenxIII\ in our sample. Using
\cenxIIIa\ \citet{iaria2005} estimate a line centroid of
\aer{6.3975}{+0.0033}{-0.0033}~keV,
\sigl=~\aer{0.0115}{+0.0045}{-0.0042}~keV and
EW~$=$~\aer{13.2}{+2.1}{-1.9} eV.  \citet{wojdowski2003} divide
observation 705 into ``dim'', ``bright'' and ``eclipse''. They obtain
individual estimates for \feka\ line velocity shifts relative to 6.40~keV
ranging from $\sim 80$ to $\sim 2000$ \kmps, and an average shift of
$350\pm200$ \kmps. The line FWHM ranges from $\sim20$ to $\sim 3200$
\kmps, with an average of $900\pm300$ \kmps.

Consistent with earlier work \citep{iaria2005,wojdowski2003},
we detect prominent emission from the \fetwofive\ He-like complex of
forbidden, intercombination and resonance emission lines between
$\sim 6.61$ and $\sim 6.72$ keV in all three observations presented here.
Additional Gaussian components were included in each model to accommodate
these features. In addition, in \cenxIIIc\ we fitted emission
from \fetwosix\ Ly$\alpha$ around $\sim 6.9$~keV with an additional
Gaussian component.

In the \bne\ fits, for 2 of the observations (\cenxIIIa\ and
\cenxIIIc), the Fe abundance reached the maximum table value of 10,
meaning that the column density required to model the continuum
curvature is insufficient to provide enough \feka\ line flux to fit
the data. The other observation (\cenxIIIb) gave an Fe abundance of
$0.74^{+0.15}_{-0.11}$ solar. The fact that the Fe abundance is so
high in 2 of the observations, and that it is required to vary between
observations rules out the uniform, spherical model for the
\feka\ line. On the other hand, the Compton-thick, solar abundance
\myt\ model fits can account for the data.
\ctt\ presented results for
\cenxIIIa\ and \cenxIIIb\ but obtained \feka\ line fluxes (using
phenomenological models) that are lower than ours by $\sim 50\%$ and
$\sim 40\%$ respectively.

The \feka\ line is unresolved for \cenxIIIa\ and \cenxIIIb\ but
resolved for \cenxIIIc\ (see Appendix~\ref{app-con},
\fr{fig-cont})
with a FWHM of $1752^{+301}_{-287}$~\kmps\ (see \tr{tab-feka}).

\subsection{\cirxI}
\cirxI\ is formally a HMXB since it has a massive secondary
\citep{johnston1999,jonker2007}. However, in the literature it is
classified as a LMXB because its \x\ phenomenology corresponds to that
of an accreting low magnetic field compact object, more commonly found
in LMXBs \citep{asai2014,boutloukos2006,fridriksson2015}.  The system
is known for its peculiar \x\ spectrum that shows a plethora of
discrete features that are strongly dependent on orbital phase. Most
famously, it represents the first case were P-Cygni profiles were
detected in an \x\ spectrum \citep{brandt2000,schulz2002}. These are
thought to be associated with a high-velocity outflow from an
equatorial accretion disk wind.

Based on the discovery of Type I~\x\ bursts
\citep{tennant1986,linares2010}, the compact object is believed to be
a neutron star. Using the relativistic precession model
\citep{stella1999}, \citet{boutloukos2006} obtain a mass of
\mns~$=2.2\pm0.3$~\msun.

In none of the 5 observations in our sample analyzed here is the
\feka\ line detected. For the \bne\ model, the \zfe\ values are
therefore determined by the continuum only. In cases such as this in
which the emission line is weak or undetected, the line width and
shift had to be fixed (at $100$~\kmps\ FWHM and 0.0, respectively).  In
\cirxIg\ an additional Gaussian emission-line component was included at
6.67~keV due to an excess in the data relative to the basic model.

In addition, observations 5478 and 706 \citep{schulz2002} show clear
P-Cygni features from He-like and H-like ions of several elements, 
and these require dedicated, more complex modeling.
Such modeling requires that the emission and absorption feature be
fitted simultaneously, something that is beyond the scope of the
models used in this paper.

Finally, as mentioned earlier, observation 1905 may suffer from pile-up.

\subsection{\cygxI}\label{sec-cygxi}
\cygxI\ is a Galactic black hole HMXB known for its persistent
activity in the X-ray band. From spectroscopic studies of the
supergiant companion, the mass of the black hole is estimated at
10.1~\msun\ \citep{herrero1995}. Using an estimate for the radius of
the secondary, as well as improved distance estimates,
\citet{orosz2011} find a mass of $14.8\pm1.0$~\msun.

This HMXB has been extensively studied. It is usually found in either
a low/hard or high/soft state. The first is characterized by low
luminosity and a hard power-law continuum, while the second is
dominated by high luminosity, dominated by soft \x\ emission. The
obs. IDs of the HETG observations studied in the present work have
already been associated with distinct states in the literature
\citep[e.g.][]{miskovicova2016,grinberg2013}.  However, most works
focus on the ionized wind and broad \feka\ line, rather than the
narrow \feka\ line from neutral matter.
These works have used several \x\ telescopes to study the underlying
geometry of the system's accreting black hole environment
\citep[e.g.][]{nowak2011} and have also exploited the proximity of the
broad \feka\ line emitting region to the accreting black hole to infer high,
and even maximal, spin values
\citep{duro2011,gou2011,fabian2012,miller2012,gou2014,tomsick2014,parker2015,duro2016,walton2016}.
On the other hand, the
only results on the narrow
\feka\ line emission are based on \chandra\ obs.~IDs 2415 and 3815, both taken
in continuous clocking (CC) mode. Thus \citet{miller2002} use
obs.~ID. 2415 to estimate a centroid of $6.415 \pm 0.007$ keV, also
obtaining FWHM~$=$~\aer{80}{+28}{-19}~eV (or
\aer{3740}{+1310}{-890}~\kmps) and EW~$=$~\aer{16}{+3}{-2}~eV. For this
obs.~ID. \ctt\ obtain a centroid of $6.4115\pm0.0305$~keV and EW~$=20.82
\pm 2.05$~eV. For obs.~ID. 3815 they obtain $6.3966\pm0.0059$~keV and
EW~$=19.07 \pm 2.06$~eV.

We present results for 15 observations, and among these the
\feka\ line is not detected in 7 (according to our criterion
of $\Delta C\ge6.63$ for a detection).  We fit the \bne\ model to all of
the observations whether or not the \feka\ line was detected because
the absence of the line can still potentially constrain the parameters
of the \bne\ model. We find that in 10 of the observations the Fe
relative abundance in the \bne\ model fits reached its maximum table
limit of 10~\zfesun. In these cases the column density required to account for
the continuum shape is insufficient to produce enough \feka\ line flux
even for such an inflated Fe abundance.  In 3 of the remaining
observations only a lower limit on the Fe abundance could be
obtained. In the other 2 observations both lower and upper bounds on %
the Fe abundance were obtained. In one of these a supersolar Fe abundance
was required and in the other case the Fe abundance was approximately
solar.

As with all the sources in the sample, the \myt\ model was fitted to
those observations in which the \feka\ line is detected. A range in 
line-of-sight column density, \nhz, is found, from only an upper limit
of $10^{20} \rm \ cm^{-2}$, to a value of $\sim 2.5 \times 10^{23} \rm
\ cm^{-2}$.
In several cases, the rising spectrum towards low energies was
incompatible with the ``standard'' Galactic absorption value, so we
attempted letting Galactic absorption be a free parameter. However, in
practice there is a degeneracy between it and the line-of-sight column
density of the model. Since the fit is insensitive to the extra column
density parameter, in such cases we fix the component to a low value
of $10^{-18}$~\cunits, unless indicated otherwise (see
\tr{tab-sample}). In 5 of the 8 observations with a \feka\ line
detection the line width could not be constrained and had to be fixed
for the fits (at a FWHM of 100~\kmps).  In 2 of the remaining
observations (obs.~IDs~3814 and 3815), the \feka\ line was resolved,
and yielded robust lower and upper bounds, but in the other
observation (obs.~ID~9847), the \feka\ line was unresolved, with only
an upper limit on the FWHM of 4765~\kmps\ (see \tr{tab-feka} and
Appendix~\ref{app-con}). Below are some specific details about the
individual fits.

\subsubsection{\cygxIa-A}

For \bne, the \feka\ line is formally detected (\delc~$=37.58$).  However,
the column density of this source is so low, that even with
\zfe~$=$~10~\zfesun, the fit is poor, and we fix the \feka\ line FWHM at
100~\kmps.  \nhgal\ is also poorly constrained and formally fixed at
$10^{-18}$~\cunits.
For \myt, both $z$ and \sigl\ are poorly constrained, the fit is unstable
and tends to a best-fit where the continuum, rather than the line,
is fitted, likely due to noisy features on the low-energy side of the
line. We thus freeze the \feka\ line FWHM at 100~\kmps.

\subsubsection{\cygxIb-B}
The \feka\ line is not detected and only results for the \bne\ model
are shown. We fix the \feka\ line FWHM at 100~\kmps. The best fit
reaches \zfe~=10~\zfesun\ with a very low column density, \nhsph, and a strong
soft power-law component. \nhgal\ is poorly constrained and formally
fixed at $10^{-18}$~\cunits.

\subsubsection{\cygxIc-C}
Similar to \cygxIb.

\subsubsection{\cygxId-D}
Similar to \cygxIb.

\subsubsection{\cygxIe-E}
Similar to \cygxIb.
As mentioned earlier, this observation may be piled-up.

\subsubsection{\cygxIf-F}
Although a \delc\ of $8.26$ suggests a marginal detection, this may
well be due to a bump at the position of \feka\ emission at 6.4~keV.
Accordingly, for both models the line is not well constrained, and we
have to fix \sigl~$=$~100~\kmps\ and $z=0$. Only a lower limit can be
obtained for \zfe. 
No \mytz\ component is used
in the \myt\ model.  \nhgal\ is fixed at $10^{-18}$~\cunits\ to
stabilize the fit.

\subsubsection{\cygxIg-G}
This is a CC mode observation.  In the \bne\ fit \zfe~=10~\zfesun\ as
the very low column density, \nhsph, cannot produce enough flux for
the \feka\ line.  An additional soft power-law continuum component is
required. \nhgal\ is poorly constrained and formally fixed at
$10^{-18}$~\cunits\ for both the \bne\ and \myt\ models.  In the
\myt\ fit no additional power-law continuum is required and only an
upper limit on \nhz\ is obtained. The \feka\ line width had to be
fixed at a FWHM of $100$~\kmps\ in order to prevent the fits and error
analysis becoming unstable.

\subsubsection{\cygxIh-H}
Similar to \cygxIb.
As mentioned earlier, this observation may be piled-up.

\subsubsection{\cygxIi-I}
Similar to \cygxIb.
As mentioned earlier, this observation may be piled-up.

\subsubsection{\cygxIj-J}
Similar to \cygxIb.
As mentioned earlier, this observation may be piled-up.

\subsubsection{\cygxIk-K}
Both the \bne\ and \myt\ models require a soft power law to fit the
soft continuum upturn. The \feka\ line width could not be constrained
so the FWHM was fixed at 100~\kmps\ but $z$ was allowed to float.
Although the \feka\ line is weak, the \bne\ model requires a column
density that is too small to account for the \feka\ line flux so the
\bne\ best-fit leads to a supersolar abundance. However, the
solar-abundance \myt\ fit did not require an unusually high value of
$A_{S}$.  The \myt\ model also has a low \nhs~$\sim 0.1-0.2$
solution. As before we choose the high-\nhs\ solution as more
conservative.

\subsubsection{\cygxIm-M}
For this observation we exclude prominent emission features in the
$2.6-2.65$ keV energy band, as well as likely P-Cygni absorption in
the $6.6-6.8$ keV band.  The \bne\ fit gives a small column density
that drives the Fe abundance up to the maximum limit of 10. From the
\myt\ fit the \feka\ line is resolved, with a FWHM of $\sim
2500$~\kmps.

\subsubsection{\cygxIn-N}
Portions of the data with prominent absorption features in the energy
ranges $2.6-2.64$, $6.6-6.75$, $6.51-6.56$, and $6.95-7.0$~keV are
excluded, as well as a complex of absorption and emission in the
$7.45-7.8$~keV band.  Similar to obs.~ID~11044, the low-energy side of
the 6.4 keV line is noisy, so
we are again forced to freeze FWHM~$=$~100~\kmps.

\subsubsection{\cygxIo-O}
This obs.~ID. is very similar to \cygxIn.  Once more we exclude
prominent absorption features in the energy ranges $2.62-2.66$,
$7.72-7.9$, $6.9-7.06$, and $6.66-6.74$ keV.
For the \bne\ model only a lower limit on the Fe abundance is
obtained.  The \feka\ is unresolved in the \myt\ fit.

\subsubsection{\cygxIp-P}
This is a CC mode observation.  We exclude portions of the data with
prominent and complex absorption/emission features between
$2.45-2.48$, $2.61-2.63$, $4.32-4.4$, and $6.48-7.0$ keV.
The \bne\ model is forced to a highly super-solar value of $\sim
8$~\zfesun, for which only a lower limit can be estimated, as a result
of attempting to fit the \feka\ line with a column density that is too
small for solar abundance.

For both models, an additional very soft power law continuum is
required and hits the upper allowed bound of 10 for the slope, but
this soft continuum component only dominates a narrow energy range in
the spectrum at the low end of the bandpass.

\subsection{\cygxIII}
\cygxIII's ``microquasar'' is a HMXB with a Wolf-Rayet companion that
has a dense stellar wind, giving rise to a very rich and variable
emission-line spectrum \citep{paerels2000,vilhu2009}, but the wind
also prevents optical photons from being observed.  The system's
inclination and mass function thus remain highly uncertain, as does
the nature of the compact object. \citet{stark2003} fit the three most
prominent lines in \chandra-HETG spectra taken at different phases to
construct a mass-function for the compact object. They combine it with
the infrared mass-function of \citet{hanson2000} for the donor to
establish an upper limit for the compact object mass of 3.6~\msun.  On
the other hand, \citet{vilhu2009} construct photoionization models and
compare with \chandra-HETG observations to establish a mass-function
for the \fetwosix\ emission line. Combining this with the infrared
mass-function for the compact object of \citet{hanson2000}, they
calculate possible ranges for the compact object mass.  Depending on
inclination, these are in the range $2.8-8$~\msun\ ($i=30^{\circ}$) or
$1.0-3.2$~\msun\ ($i=60^{\circ}$).  From these estimates, we choose
the two extremes, namely \mcoa~$=$~1.0~\msun\ and \mcob~$=$~8.0~\msun.

We choose to analyze only \cygxIIIa, as \cygxIIIb\ and \cygxIIIc\ are
very strongly dominated by spectral features from non-neutral
material.  The \cygxIIIa\ observation also has features from ionized
Fe but there are two discrete lines from \fetwofive\ and \fetwosix, as
opposed to a blended complex as in the other 2 observations. Thus, in
addition to the baseline \bne\ and \myt\ models fitted to the
\cygxIIIa\ spectrum, two Gaussian emission-line components are
included (with line width and centroid energy free) at $\sim 6.65$~keV
and $\sim 6.97$~keV.  In the \bne\ fit the Fe abundance reaches its
maximum table value of 10~\zfesun\ and still cannot fit the
\feka\ line flux so the uniform, spherical, solar-abundance model can
be rejected (see \tr{tab-trans}). In the \myt\ fit, a line-of-sight
column density of \nhz~$\sim 2.5 \times 10^{22}$~\cunits\ is required
in addition to the Compton-thick reflector (see \tr{tab-myun}). The
\feka\ line is resolved, with a FWHM~$\sim 3800$~\kmps\ (see
\fr{fig-cont} in Appendix~\ref{app-con}, and \tr{tab-feka}).

\subsection{\gamcas}
This is a prototypical Be XRB, where a B-type star's circumstellar
disk gives rise to strong hydrogen emission lines, which together with
the stellar continuum define this class of objects.  Further, with an
X-ray luminosity more than an order of magnitude brighter than
``classical'' Be stars, \gamcas\ is the prototype of its own class of
``\gamcas\ analogs''.

\citet{harmanec2000} use \ha\ spectra to derive radial velocities and
deduce that the compact object has a mass in the range
$0.7-1.9$~\msun.  Based on a revised velocity solution and most
probable parameter values, \citet{smith2012} estimate a compact object
mass of $0.8\pm0.4$~\msun. We thus adopt the two extreme values of 0.4
and 1.9~\msun\ for our purposes.

Our analysis presents results for one HETG observation of \gamcas.
For the \feka\ line in the spectrum from this same obs.~ID., \ctt\ find
a line flux of $7.1\pm2.7 \times 10^{-5}$ \phunits,
an equivalent width of $39.74\pm15.04$~eV and a line
centroid of $1.9368\pm0.0001$~\AA\ or 
$6.4015\pm0.0003$~keV.
On
the other hand, \citet{smith2004} obtain results separately for the
K$\alpha_1$ and K$\alpha_2$ components: 
centroids at 1.9358 and 1.9398~\AA\ 
(6.4048~keV and 6.3916~keV),
and ``$\sigma_{\lambda}$'' values
of 3.1~m\AA\ (10.2~eV) for both components (corresponding to a FWHM of
$\sim 1125$~\kmps).

We add two Gaussian components to the \bne\ and \myt\ models in order
to fit \fetwofive(r) at $\sim 6.68$ keV and \fetwosix\ $\rm
Ly\alpha$ at $\sim 6.95$ keV.  In the \bne\ fit the Fe abundance
reaches its maximum table value of 10~\zfesun\ and still cannot fit the
\feka\ line flux so the uniform, spherical, solar-abundance model can
be rejected (see \tr{tab-trans}). In the \myt\ fit, the line-of-sight
column density is negligible, and only an upper limit is obtained (see
\tr{tab-myun}).  In the \bne\ fit we had to fix the FWHM of the line
broadening function at 100~\kmps\ otherwise the fit becomes
unstable. In contrast, the line width was constrained in the \myt\ fit
but it is unresolved and we obtained only an upper limit on the FWHM,
of $\sim 2850$~\kmps\ (see \tr{tab-feka}). The \feka\ line flux from
the \myt\ fit is consistent with the value obtained by \ctt\ using
phenomenological models.

\subsection{\gxIII}\label{sec-gxIII}
This X-ray pulsar and HMXB features a prominent \feka\ emission line,
and associated Compton shoulder, clearly detected in the HETG spectra.
This is one case where the \bne\ model with almost solar Fe abundance
can fit more than one observation (two in this case).  From the
B-hypergiant companion's radial velocity curve established with
high-resolution optical spectra \citet{kaper2006} estimate a
neutron-star mass of $1.85\pm0.6$ to $2.4\pm 0.7$, depending on the
companion radius.

We present results for two HETG observations, \gxIIIb\ and \gxIIIc.
\citet{watanabe2003} study \gxIIIc\ (``pre-periastron'' phase). They
derive EW values for the \feka\ line (including Compton shoulder) of
$643\pm20$ and $486\pm18$ eV for the first and second halves of the
observation, respectively. For a range in the power-law continuum
photon index of $\Gamma = 1.0 - 1.5$ they estimate metal abundances
$0.65-0.90$ of the cosmic values.  For the same observation \ctt\ obtain
a line flux of $\sim 32 \times 10^{-4}$ \phunits,
EW~$=113.54\pm4.08$~eV, for a central wavelength of
$1.9384\pm0.0002$~\AA\ or $6.3962\pm0.0007$~keV. For \gxIIIb\ \ctt\
obtain a line flux of $\sim 78 \times 10^{-4}$ \phunits,
EW~$=282.68\pm3.30$~eV for a central wavelength of
$1.9388\pm0.0002$~\AA\ or $6.3949\pm0.0007$~keV.
Neither the \citet{watanabe2003} nor the \ctt\ study attempted to
quantify the velocity width of the \feka\ line.

In our analysis of \gxIIIb, the energy regions $2.9-3.8$, $3.66-3.72$,
and $5.4-5.46$ keV are excluded due to multiple emission/absorption
features.  There is a prominent Ni~K$\alpha$ line at $\sim 7.47$~keV
in both observations and this line is already included in the
\bne\ table model, but for the \myt\ fits an additional Gaussian
component is added to the model.  At energies $\approxlt 3.6$ keV the
spectrum has a very low signal-to-noise ratio so the data in this
energy range provide little constraint to the fit.  The \bne\ model
gives an Fe abundance of 0.86 and 0.80 solar for \gxIIIb\ and
\gxIIIc, respectively.
For \gxIIIb, the \myt\ model significantly under-predicts the flux of
the \feka\ line (by $\sim 30\%$). This is not the case for \gxIIIc.
The \myt\ fit for \gxIIIb\ is also unsual in that it is one of the few
spectra in our sample of 56 observations that is able to constrain the
global column density, \nhs.  This is due to the high quality of the
data in the \feka\ line Compton shoulder, because the shape of the
shoulder and its magnitude relative to the line core are sensitive to
\nhs. On the other hand, for \gxIIIc, \nhs\ was not constrained and
was fixed at $10^{25}$~\cunits, the value used for most of the other
56 observations in the sample.

Our \feka\ line fluxes for both observations are in reasonable
agreement with the corresponding values in \ctt\ (see \tr{tab-feka}).
From our analysis, the \feka\ line is resolved in both observations,
with FWHM $\sim 400-600$~\kmps\ consistent with historical
measurements, but is smaller than most of the widths in the other
sources in the sample in which the \feka\ line is resolved (see
\fr{fig-cont}, and \tr{tab-feka}).

Using \xmm\ data, \citet{fuerst2011} measure an unabsorbed $2-10$
keV continuum flux ranging from \ten{1.350}{-8} to \ten{2.151}{-8}
\funits, which overlaps with our results from \myt\ that are in the
range between \ten{3.09}{-9} to \ten{1.48}{-8} \funits.  Similarly,
over the same energy range, using \suzaku\ data, \citet{suchy2012}
measure an unabsorbed continuum flux ranging from
$\sim$\ten{1.6}{-9} to \ten{3.6}{-9} \funits.

The very large \feka\ equivalent width for \gxIIIb\ is consistent
with the direct continuum being significantly depressed by the
almost Compton-thick \myt\ \nhz\ value along the line-of-sight. If
the line-of-sight extinction is caused by clumps, it is possible for
the continuum shape and line EW to undergo significant fluctuations.
\citet{suchy2012} carried out phase resolved analysis with
\suzaku\ that indicated high, strongly variable \nh\ (although that
analysis could not distinguish between in and out-of-the
line-of-sight column densities) due to a clumpy stellar wind from
the hypergiant companion Wray 977. \citet{mukherjee2004} based their
analysis on {\it Rossi X-ray Timing Explorer} data. They suggested a
model with ``stagnating'' clumps of matter and strong
inhomogeneities in the stellar wind, due to a combined low
stellar-wind terminal velocity and large mass-loss rate \citep[see
  also][]{leahy2008,manousakis2012}. High anisotropies in the
distribution of neutral matter (i.e. clumps) are also reported by
\citet{islam2014} in monitoring work with the {\it Monitor of All
  Sky X-ray Image}. 
  
\subsection{\gxI}
This object is a rare accreting \x\ pulsar with a low-mass companion.
It is the prototype of the small subclass of symbiotic XRBs.  There is
no direct mass determination for the neutron star in
literature. Following \citet{hinkle2006}, we adopt a value of
1.35~\msun\ as a canonical result for binary radio pulsars
\citep{thorsett1999}.

We present the results for 2 HETG observations of \gxI.  A hard
power-law continuum and a prominent \feka\ emission line characterize
these observations. In addition to the \feka\ line, both of the
observations in the present study have an additional emission line at
\simn6.95~keV, likely due to \fetwosix\ $\rm Ly\alpha$, that is fitted
by adding an additional Gaussian component to the models.

Spectral fits with the uniform spherical model (\bne) for both
observations require a super-solar Fe abundance. The flux in the
\feka\ line is much larger in \gxIb\ than it is in \gxIa, and the EW
is also larger in \gxIb\ than it is in \gxIa\ (see \tr{tab-feka}). The
\bne\ fits to both observations yield a similar column density,
\nhsph, to account for the continuum, so the very different
\feka\ line fluxes result in the Fe abundance for the fit to \gxIb\ to
increase to 10~\zfesun, the maximum limit of the model. This is still
insufficient to account for the \feka\ line flux in \gxIb. This
contrasts with the Fe abundance of $\sim 3$~\zfesun\ required for
\gxIa\ (see \tr{tab-trans}). Thus, the uniform, solar-abundance
spherical model is rejected on the basis of requiring a significantly
variable Fe abundance between observations.
We note that \citet{yoshida2017} find an iron abundance of $\sim 80\%$ solar
based on an independent analysis with {\it Suzaku}.

The \myt\ model, which has solar abundances, can account for both
observations. The \myt\ fits have the column density of the X-ray
reprocessor fixed at $10^{25}$~\cunits\ (Compton-thick) but both of
the \myt\ fits also have Compton-thin solutions. As for most of the
other sources in our sample, we give results for the more conservative
Compton-thick solutions (see \tr{tab-myun}). The \myt\ fits yield a
line-of-sight column density, \nhz, that is $\sim 2.8$ and $4.1 \times
10^{22}$~\cunits\ for \gxIa\ and \gxIb\ respectively (see
\tr{tab-myun}).

The \feka\ line flux in \gxIb\ is an order of magnitude larger than it
is in \gxIa\ (see \tr{tab-feka}), but the continuum flux in \gxIb\ is
only a factor of $\sim 3$ larger than it is in \gxIa\ (see
\tr{tab-flux}). There must be changes in the global column density
and/or covering factor between the two observations, but data with a
wider bandpass extending to higher energies are required to further
constrain models. The \feka\ line is unresolved in both observations,
and the largest upper limit on the FWHM is 1490~\kmps\ (see
\tr{tab-feka}).

Using \gxIa\ and simultaneous {\it RXTE} PCA data, \citet{paul2005}
obtain a line center of $6.400\pm 0.005$ keV and equivalent width of
\aer{71}{+33}{-16} eV, using simple phenomenological modeling.
For the same observation, \ctt\ obtain a centroid
$1.9376\pm0.0013$~\AA\ or $6.3988\pm0.0043$~keV, an \feka\ line flux
of $\sim 1.6\pm 0.4 \times 10^{-4}$ \phunits,
and EW~=$72.81\pm16.96$ eV. Our results are consistent with these
empirical measurements for obs.~ID.~2710 (see \tr{tab-feka}).
Historical results are not available for \gxIb.

\subsection{\herxI}\label{sec-herxi}
This eclipsing LMXB has a sizable accretion disk.
\citet{reynolds1997} used optical spectroscopy to establish a neutron
star mass of $1.5\pm0.3$~\msun, challenging earlier results of
significantly lower masses (1\msun, or lower). More recently,
\citet{leahy2014} model high-state eclipses observed with {\it RXTE}
to measure the radius and atmospheric scale height of HZ Her, the
stellar companion of the neutron star. They also fit stellar
atmosphere models and calculate stellar evolution models for a range
of allowed masses from orbital parameters and allowed metallicities
from optical spectra. These different models appear to be in agreement
with a narrow range in companion mass, corresponding to a neutron star
mass range between 1.3 and 1.7\msun.  We adopt these two extreme
values for our purposes.

Using {\it Suzaku} broadband spectra, \citet{asami2014} investigate
a broad $4-9$ keV hump. They consider either an ionized partial covering
model or an additional broad line at 6.5~keV but are unable to distinguish
between these alternatives. \citet{fuerst2013} and \citet{wolff2016}
fit both a narrow and a broad line to \nustar+\suzaku\ and \nustar-only
spectra, respectively. Once again the broad line is centered around
$\sim 6.5$ keV. According to these authors, the broad line is
most likely due to unresolved contributions from several ionization
stages.

Our sample consists of 10 HETG observations of \herxI\ (see
\tr{tab-sample}). \citet{jimenez-garate2005} study
obs.~ID. 2749. Although they focus on photoionized transitions, they
also model separately Fe K$\alpha_1$ and $\alpha_2$, finding a Doppler
width $<680$~\kmps\ for each.

\citet{ji2009} study obs.~IDs 2749, 3821, 3822, 4585, 6149 and
6150. For the \feka\ line they find centroids between $1.938\pm0.001$
and $1.940\pm0.001$ \AA\ or $6.398\pm0.003$ to $6.391\pm0.003$~keV,
and equivalent widths between \aer{81}{+15}{-14} and
\aer{635}{+59}{-57} eV.  From the same observations, \ctt\ find
\feka\ centroids between $1.9372\pm0.0037$ and
$1.9399\pm0.0031$~\AA\ ($6.4002\pm0.0122$ to $6.3913\pm0.0102$~keV),
line fluxes in the range $\sim 1.3$ to $5.5 \times 10^{-4}$~\phunits,
and EWs in the range $51.58 \pm 19.46$
and $512.88 \pm 68.98$~\AA.

The line-of-sight column density was less than the formal Galactic
value in all except 2 observations for both models, and a further 2
observations for the \bne\ model, so for these \nhgal\ was fixed at a
low value, $10^{18}$~\cunits, and any remaining Galactic absorption is
then degenerate with the relevant column density parameter in the
model (i.e., \nhsph\ for \bne\ and \nhz\ for \myt). For the remaining
observations the line-of-sight column density was greater than the
formal Galactic absorption so \nhgal\ was fixed at the formal value
(see \tr{tab-sample}).

For some of the observations a second power-law continuum in the
\bne\ fit dominates the continuum compared to the actual absorbed
continuum from the spherical matter distribution. Such a solution is
attempting to produce sufficient flux in the \feka\ line, yet
producing no line-of-sight extinction to reconcile a continuum that is
rising at low energies.  Physically, this is not a self-consistent
solution if both power-law continua are produced at the center of the
spherical matter distribution because covering of the X-ray source in
the \bne\ model is uniform and not patchy. It is possible that the
second power-law continuum in these cases is produced outside of the
spherical matter distribution. Even if that is the case, the scenario
that we are explicitly testing is that of a simple geometry of a
single X-ray source surrounded by a uniform, spherical distribution of
matter. In that respect the test fails in these cases, which will be
pointed out in the details for the individual fits given below.

In 4 of the 10 observations, the \feka\ line is resolved, and it is
unresolved in another 4 observations. In the remaining 2 observations
the line width could not be constrained and had to be fixed.  The
\feka\ line fluxes we obtain for the observations studied by \ctt\ all
agree (within statistical errors) with the \ctt\ values, which were
obtained from simplified ad hoc phenomenological modeling using a
fixed narrow line width. Case-by-case details on these and other
pertinent measurements are given in the individual fitting
descriptions below. Overall, considering that a large range in the Fe
abundance for the \bne\ model is required among the 10 observations
($\sim 0.5$ to the model limit of 10 times the solar value), a
uniform, solar abundance spherical matter distribution can be ruled
out for \herxI.

\subsubsection{\herxIa-A}
The signal-to-noise ratio of the spectrum for this observation is low
so that the FWHM of the \feka\ line had to be fixed at 100~\kmps.  In
the \bne\ fit the soft X-ray spectrum below $\sim 5$~keV is dominated
by a second power-law continuum because it is not attenuated. The Fe
abundance is $\sim 0.5$ solar. The \myt\ fit gives an unusually high
value of $A_{S} \sim 25$ (see \tr{tab-myun}).

\subsubsection{\herxIb-B}

The \feka\ line in this observation is weak and the line FWHM had to
be fixed at 100~\kmps. The \bne\ fit is forced to the maximum allowed
Fe abundance of 10~\zfesun.  The EW of the \feka\ line from the
\myt\ fit is only $\sim 30$~eV (in the \bne\ fit the \feka\ line model
flux falls short of the data even for the maximum Fe abundance).
There are some residuals at energies higher than that of the
\feka\ line but the signal-to-noise ratio is poor so the features are
not fitted.

\subsubsection{\herxIc-C}
The spectrum for this observation has an extremely low signal-to-noise
ratio.  For the \bne\ fit, the Fe abundance is sub-solar by a factor
of $\sim 4$ (see \tr{tab-trans}).  The spectrum continues to rise at
low energies, so that below $\sim 5$~keV the spectrum is dominated by
a second power-law continuum, so this is another case for which a
simple uniform spherical matter distribution is incompatible with the
data. The \feka\ line is unresolved and only an upper limit on the
FWHM ($\sim 3700$~\kmps) could be obtained (see \tr{tab-feka}).

\subsubsection{\herxId-D}\label{sec-herxid}

This spectrum of the source has one of the most prominent
\feka\ lines among the 10 observations, as borne out by a very large
EW of $\sim 500$~eV (see \tr{tab-feka}). In the spectral fitting
analysis, an additional Gaussian component is included to model an
\fetwosix\ $\rm Ly\alpha$ line at $\sim 6.96$~keV.

In the \bne\ fit, the Fe abundance is driven to its maximum value of
10~\zfesun\ because the fitted column density cannot produce enough \feka\ line
flux. Also, the spectrum below $\sim 5$~keV is dominated by a second
power-law continuum.

The \myt\ fit results correspond to a Compton-thick reflector with
\nhs\ fixed at $10^{25}$~\cunits\ but a Compton-thin solution also
exists. However, the Compton-thick solution is chosen as it is more
conservative. A very high value of $A_{S} \sim 65$ is required, as
would be expected for a reflection-dominated spectrum.
We note
that the ratio of maximum to minimum observed continuum flux among
the 10 observations is $\sim 45$, while the intrinsic continuum flux
varies by two orders of magnitude (see \tr{tab-flux} columns 4 and 6).
The \feka\ line is unresolved in this observation, with
FWHM~$<1195$~\kmps.

The very large \feka\ equivalent width for this specific observation
is also consistent with the results of \citet[][$\sim 600$ eV]{ji2009}
and \ctt\ ($\sim 500$ eV). \citet{ji2009} analyze several observations
in different phases and confirm that this observation is in the
``low'' (luminosity) state \citep[see also][]{jimenez-garate2005},
while \ctt\ state this is an eclipse. These imply a suppressed direct
continuum, which, together with a large $A_S$ \myt\ solution is
consistent with a strongly reflection dominated configuration. We note
though that the Compton-thin \myt\ \nhz\ solution is not consistent
with this interpretation.  However, it is possible that \nhz\ is
actually so large that it is not detected at all in the HEG bandpass.
Such a solution is completely degenerate with respect to the other
model parameters, as far as the HEG data are concerned.
\citet{zane2004} carried out phase-resolved analysis with \xmm, also
obtaining EW values up to $\sim 600$ eV. They suggest an origin for
the Fe fluorescent emission at the A/F-type companion, HZ Her. The
exceptional strength of the line would then be consistent with the
companion's high mass, which at 2.3 \msun\ is much larger than in
other accretion disk sources. However, an origin at the accretion disk
wind is also possible, as there is evidence of gas outflowing from HZ
Her \citep{anderson1994,boroson2001}.

\subsubsection{\herxIe-E}

The \bne\ fit gives a super-solar Fe abundance with large statistical
errors (the range is $\sim 1-9.3$ solar, see \tr{tab-trans}).  The spectrum
below $\sim 5$~keV is again dominated by a second power-law
continuum. The \myt\ fit is similar to \herxId, requiring a large
value of $A_{S} \sim 50$. However, this time the \feka\ line appears
to be resolved, with FWHM~$\sim 1500$~\kmps\ (see \tr{tab-feka}).

\subsubsection{\herxIf-F}

For this observation an additional Gaussian component is included in
the models to account for an emission line at $\sim 6.86$~keV, which
is likely due to either redshifted \fetwosix\ $\rm Ly\alpha$ or
blueshifted \fetwofive(r).

The \bne\ fit requires a sub-solar Fe abundance ($0.71\pm0.10$ solar)
and the continuum below $\sim 4$~keV is dominated by a second
power-law component.  The \myt\ fit requires a significant
line-of-sight column density in addition to the Compton-thick
reflector (see \tr{tab-feka}).
The \feka\ line appears resolved, with FWHM~$\sim 1400$~\kmps\ (see \tr{tab-feka}).

\subsubsection{\herxIg-G}

For this observation the \bne\ fit actually yields a solar Fe
abundance, meaning that the column density of the uniform, spherical
matter distribution, \nhsph, that is required to model the continuum,
correctly predicts the \feka\ line flux. However the entire continuum
in the fitted bandpass is dominated by a second power-law continuum
that is essentially unobscured, so the overall model is not
self-consistent. The \myt\ fit is able to fit the spectrum with a
combination of a direct power-law continuum and a Compton-thick
reflection spectrum with a value of $A_{S} \sim 3.4$ that is not
extreme. The \feka\ line is resolved and has a rather high FWHM of
$\sim 4200^{+2060}_{-1860}$~\kmps\ (see \tr{tab-feka}).  There appears
to be a weak excess of data above the model between the peak of the
\feka\ line and $\sim 7$~keV, which could be due to one or more
emission lines from ionized Fe. However, the statistical significance
is not high enough to warrant including additional emission lines in
the models.

\subsubsection{\herxIh-H}

The \bne\ model fit gives a sub-solar Fe abundance ($\sim 0.6$ solar) and
requires a very high photon index, \gammasph, of the primary power-law
continuum, reaching the maximum table model value of 3.0 (see
\tr{tab-trans}). Below $\sim 5$~keV, the spectrum is dominated by the
second power-law continuum so this spherical model configuration is
not self-consistent and therefore can be rejected. The \myt\ fit is
characterized by a dominance of a zeroth-order continuum with heavy
extinction, with \nhz~$=7.9 \times 10^{23}$~\cunits. The reflected
continuum is weaker in comparison to this. The \feka\ line is
unresolved in this observation, with FWHM~$<2675$~\kmps\ (see
\tr{tab-feka}).

\subsubsection{\herxIi-I}

The \bne\ fit is similar to that for \herxIh\ in the sense that the
primary power-law photon index is very steep and the spectrum in the
entire fitted energy band is dominated by the second power-law
continuum. This spherical matter distribution configuration is
therefore not self-consistent.  The Fe abundance is $\sim 1.24$ solar but
the model is rejected because of the dominance of an essentially
unabsorbed continuum.  The \myt\ fit does not have a zeroth-order
attenuated continuum that was present in \herxIh, only the
Compton-thick reflection component is required. Also, in this
observation the \feka\ line is resolved, with FWHM~$\sim 2600$~\kmps.

\subsubsection{\herxIj-J}

The \bne\ model fit gives a sub-solar Fe abundance of $\sim 0.6$ solar (see
\tr{tab-trans}).  The \myt\ model fit requires a significant
line-of-sight absorption column density (\nhz $\sim 5.3 \times
10^{23}$~\cunits), in addition to the Compton-thick reflection
component (see \tr{tab-myun}).  A second power-law continuum is
required for both fits. The \feka\ line is unresolved, with an upper
limit on the FWHM of $2014$~\kmps.

%LMC X-4 result summary:
%obs 2-10 keV flux (myt~bne):, Fe abun, A_S, EW feka
%A	9.33e-11	0.55	~5	68	
%B	0.48e-11	10	126	702
%C	1.51e-11	0.45	~1	129
%A/B flux ratio ~19

\subsection{\lmcxIV}\label{sec-lmcxIV}

\lmcxIV\ is another well-studied HMXB.  The most recent determination
of the neutron star mass is the one by \citet{falanga2015}, with
\mns~$=$~$1.57\pm0.11$~\msun\ (see \ref{subsec_fuxvii}).  We also use
the earlier result of \citet{vandermeer2007},
$1.25\pm0.11$~\msun. These authors use high-resolution echelle optical
spectra with VLT/UVES to improve significantly on earlier radial
velocity measurements.

The 3 HETG observations of this HMXB that we study here show large
changes in continuum flux and spectral shape, as well as significant
changes in the prominence of the \feka\ line relative to the
continuum. The signal-to-noise ratio of \lmcxIVa\ and \lmcxIVb\ is
poor but sufficient to attempt fits with both the \bne\ and
\myt\ models.

\citet{neilsen2009} average the same 3 HETG observations. They
estimate a \feka\ line centroid of $1.941\pm0.002$~\AA\ or
$6.388\pm0.007$~keV and an equivalent width of $130\pm30$~eV.  For
obs. IDs 9571, 9573 and 9574 \ctt\ find \feka\ centroids of
$1.9374\pm0.0054$~\AA\ or $6.3995\pm0.0178$~keV,
$1.9422\pm0.0056$~\AA\ or $6.3837\pm0.0184$~keV and
$1.9409\pm0.0035$~\AA\ or $6.3880\pm0.0115$~keV, respectively. They estimate
corresponding EW values of $73.71\pm21.95$, $890.00\pm267.00$ and
$243.03\pm37.74$~eV.

Our fits with the \bne\ model give an Fe abundance of $\sim 0.5$ solar for
the first and third observations (obs.~IDs 9571 and 9574
respectively).  However, for the second observation the Fe abundance
reaches the maximum value of 10~\zfesun. The variable Fe abundance between
observations rules out the simple uniform, solar-abundance spherical model. However, the first observation alone makes the uniform spherical model unlikely
because the continuum in the entire fitted bandpass is dominated by
the second power-law continuum, resulting in a scenario that is not
self-consistent.

The \myt\ fits (see \tr{tab-myun}, \tr{tab-feka}, and \tr{tab-flux}),
reveal that the variability in the X-ray spectrum can be interpreted
as variability in the relative proportions of the direct and reflected
spectra (the latter including the \feka\ line emission).  The observed
continuum flux drops by a factor of $\sim 19$ between the first 2
observations, from $\sim 9.3 \times 10^{-11}$ to $\sim 0.48 \times
10^{-11} \rm \ erg \ cm^{-2} \ s^{-1}$. This is accompanied by the
X-ray spectrum in the second observation becoming
reflection-dominated, as evidenced by the very large EW of the
\feka\ line and the very high value of $A_{S}$.
In the third observation the continuum flux
increased by a factor of $\sim 3$, but still remained a factor of
$\sim 6$ below the flux in the first observation. The spectrum is no
longer reflection-dominated and has signatures of continuum components
that are direct, transmitted, and reflected. Consistent with this
change in spectral shape, the EW of the \feka\ line and the value of
$A_{S}$ both decrease between the second and third observations.  In
the first observation the \feka\ line is unresolved, and in the third
observation the line width was fixed (at FWHM~$=100 \rm \ km
\ s^{-1}$) in order to obtain a stable fit. In the second observation
the \feka\ line was resolved with FWHM~$\sim 2740$~\kmps.

Recently, \citet{shtykovsky2017} used \nustar\ data covering about
half of the system's orbital cycle to carry out tomography of the
narrow \feka\ line. Based on the phase shift between the \feka\ EW and
the pulsing profile, they suggest that the line originates on the
outer edge of the accretion disk \citep[see also][]{neilsen2009}, and
most likely either in the inflowing accretion stream that flows
through the inner Lagrangian point and falls on the outer accretion
disk edge or in the area where the flow interacts with the outer edge
of the accretion disk (the so-called hot spot). At the same time,
\lmcxIV\ is known for its so-called ``superorbital '' period and
variability, with the source intensity changing more than $\sim 50$
times during $\simeq 30.5$ days.  This is believed to be due to a
tilted and precessing accretion disk \citep{lang1981, heemskerk1989},
thus obscuring the direct X-ray continuum from the source, and
favoring a high EW measurement.  As \citet{neilsen2009} note (see also
the observations dates in \tr{tab-sample}), in fact, observations
9573, 9574, and 9571 are in this order and represent a transition from
a ``low'' to a ``high'' state, with 9573 then likely representing the
strongest suppressed continuum conditions. We note that the
Compton-thin \myt\ \nhz\ solution quoted is not consistent with this
interpretation, but it is possible that \nhz\ is actually so large
that it is not detected at all in the HEG bandpass.  Such a solution
is completely degenerate with respect to the other model parameters,
as far as the HEG data are concerned.
  
We note that our \feka\ line fluxes and EW values are consistent with
the corresponding values obtained by \ctt\ that are based on simple
phenomenological models. In the spectral fits to \lmcxIVa\ we had to
include an additional narrow Gaussian emission-line component to model
an excess at $\sim 6.65$~keV, likely due to emission from \fetwofive.

Finally, we note that \citet{hung2010} use \suzaku\ and the {\it Rossi
  X-ray Timing Explorer}, and report a separate, Doppler broadened
\feka\ line originating in the inner accretion disk.

\subsection{\oaoXVI}\label{sec-oaoXVI}

This HMXB is one of the few eclipsing X-ray pulsars known, with a
supergiant companion to a neutron star.  The most recent mass
estimates for the neutron star compact object are by
\citet{falanga2015} at $1.74\pm0.30$~\msun\ (see \ref{subsec_fuxvii})
and \citet{mason2012} at $1.42\pm0.26$~\msun, using refined
near-infrared radial velocity measurements of the donor star, only
detected in the infrared.

In the short exposure with obs.~ID. 1947, \citet{chakrabarty2002}
clearly detect \feka\ line emission at 6.4 keV, and estimate EW~$\sim
111$~eV. For the same observation, \ctt\ obtain a \feka\ line centroid of
$1.9366\pm0.0037$~\AA\ or $6.4021\pm0.0122$~keV and
EW~$=$~$147.09\pm56.71$.  Otherwise, there are, surprisingly, no results
in the literature for the other observation in our sample (obs.~ID
12460).

In our analysis, we fit both of the HETG observations. The spectrum of
the first observation (obs.~ID 12460) has most of the signal in the
\feka\ line, its EW being so large that the spectrum is clearly
reflection-dominated. Not surprisingly, the \bne\ fit to obs.~ID 12460
drives the Fe abundance to the maximum value of 10~\zfesun, and an
additional power-law continuum is required that dominates the
continuum below $\sim 4$~keV. The \myt\ fit accounts for the
reflection-dominated spectrum in terms of the direct continuum being
suppressed by a large line-of-sight column density, \nhz~$>7.2 \times
10^{23}$~\cunits\ (see \tr{tab-myun}) and the value $A_{S} \sim 1.2$
confirms that it is not the intrinsic continuum that has decreased
relative to the reflection spectrum.  The EW of the \feka\ line is
very large, $\sim 2$~keV (see \tr{tab-feka}).  The \myt\ fit still
under-predicts the \feka\ line flux by $\sim 15\%$, but the Fe
abundance is fixed at the solar value in this model, so the line flux
could easily be accommodated with a modest deviation from solar Fe
abundance.  Due to the low signal-to-noise ratio of the data, the
limitation of the bandpass, and complexity of the model, the
parameters \gammas, \nhs, \gammaz\ and $A_S$ had to be frozen in order
to obtain the constraint on \nhz.  Note that an additional Gaussian
component had to be included for both the \bne\ and \myt\ models due
to excess line-like emission at $\sim 6.68$~keV, likely due to
\fetwofive(r).

\citet{pradhan2014} interpret the strong spectral and X-ray intensity
variations of their phase-resolved \suzaku\ observations as evidence
of a highly inhomogeneous, clumpy stellar wind from the supergiant
Ofpe/WN9-type companion, which belongs to a class characterized by
exceptionally intense stellar winds with low terminal velocities and
high mass loss rates \citep{martins2007,mason2012}. They estimate
clump masses of the order of $\sim 3\times 10^{24}$ g, and further
suggest that this object could belong to a distinct class between
Supergiant Fast X-ray Transients (SFXTs), which are known to show
irregular outbursts lasting from minutes to hours, and ``normal''
HMXBs.  The exceptional EW of this observation could thus be
associated to a strong suppression of the direct continuum due to
massive clumps.

In the second observation (obs.~ID 1947), the \feka\ line is much less
prominent compared to the continuum, than it was in the first
observation. The \bne\ fit gives an Fe abundance of $\sim 0.5$ solar.  The
\myt\ fit does not require the large line-of-sight extinction that the
first observation did, and consistent with this, the EW of the
\feka\ line is only $\sim 170$~eV. However, it appears that the
\feka\ line flux is similar for the two observations (see
\tr{tab-feka}), while the observed continuum flux increases by an
order of magnitude, going from the first to second observation (see
\tr{tab-flux}). However, the {\it intrinsic} continuum flux {\it
  decreases} going from the first to second observation, indicating
the importance of the role of variable line-of-sight extinction in
these observations of \oaoXVI. In conclusion, the uniform, solar-abundance spherical
model can be rejected on the basis of the large variations in Fe
abundance required between observations.

The \feka\ line is unresolved in both observations, the upper limits
on the FWHM being $\sim 1560$ and $\sim 2370$~\kmps\ for the first and
second observations, respectively.

\subsection{\velxI}
This is an archetypal eclipsing HMXB, consisting of a pulsing neutron
star and supergiant companion.  There are several estimates for the
neutron star's mass in this system. The most recent is by
\citet{falanga2015}, with \mco~$=$~$2.12\pm0.16$~\msun\ (see
\ref{subsec_fuxvii}).  Further, \citet{rawls2011} improve on earlier
estimates of the X-ray eclipse duration by means of an optimized
numerical code for Roche geometry, leading to a neutron star mass of
$1.77 \pm 0.08$~\msun.  We use these two results for our purposes.  It
is worth noting that these are both consistent with those of
\citet{quaintrell2003} who obtain $1.88\pm0.13$ or
$2.27\pm0.17$\msun\ for inclinations of 90 and 70.1 degrees,
respectively.

In our sample there are 5 observations of \velxI\ (see
\tr{tab-sample}).  Using obs.~ID. 102, \citet{schulz2002c} measure the
centroid of the \feka\ line to be $1.937\pm0.001$~\AA\ or
$6.401\pm0.003$~keV, with a flux of $10.4\pm1.1 \ \times \ 10^{-5}$
\phunits. It is stated by \citet{schulz2002c} that energy shifts and
broadening are negligible but no quantitative analysis is provided.
\citet{goldstein2004} use the 3 obs.~IDs 1926, 1928
and 1927 to study the \feka\ line as a function of orbital phases 0,
0.25 and 0.5, finding velocity shifts of \aer{210}{+89}{-87},
\aer{-67}{+67}{-68} and \aer{119}{+48}{-48}~\kmps,
respectively. \citet{goldstein2004} measure a large range in the
\feka\ line flux of $\sim 18.1\pm2.6$ to $374 \pm 27
\ \times\ 10^{-5}$ \phunits, corresponding to a ratio of $\sim 20.7$
between the maximum and minimum flux.  They interpret the fluorescent
lines as indicators of cooler clumps in the otherwise hot,
photoionized stellar wind.

For the same 3 observations and orbital phases as in
\citet{goldstein2004}, \citet{watanabe2006} find line centroids of
$6.3958\pm 0.0022$, \aer{6.3992}{+0.0018}{-0.0005} and
\aer{6.3965}{+0.0011}{-0.0012}~keV, and corresponding Gaussian widths
of \aer{7.2}{+3.9}{-5.3}, \aer{0.0}{+7.4}{-0.0}
(value quoted as shown in their paper)
and
\aer{11.0}{+1.8}{-1.9}~eV, respectively. These widths are equivalent
to FWHM values of $\sim 790$, $<820$, and $\sim 1215$~\kmps\ for
phases 0.0, 0.25, and 0.50 respectively. \citet{watanabe2006} deduce
that the fluorescent emission must be produced in three, or more,
distinct regions, namely the extended stellar wind, reflection off the
stellar photosphere, and in dense material partially covering, and
possibly in the accretion wake of, the neutron star. In the above
studies, \feka\ line parameters were derived using simple, ad hoc,
phenomenological models, although \citet{watanabe2006} applied, in 
addition, a simple clumpy wind model. The latter was used to deduce
that a column density of $\sim 1.7 \times 10^{23}$~\cunits\ for the
material producing the \feka\ line is consistent with the data but no
constraints on the allowed range were provided.  The model provided
only a simple treatment of the \feka\ line, and did not include the
Compton shoulder.

\ctt\ study obs.~IDs 102, 1926, 1927 and 1928 and obtain centroids of
the \feka\ line in the range $6.3936\pm0.0096$~keV to
$6.3992\pm0.0013$~keV.
Using a phenomenological model only, they measured ranges in the
\feka\ line flux and EW of $\sim 15$ to $340 \ \times 10^{-5}$ \phunits\
and $\sim 54$ to $932$~eV,
respectively. Thus, the \feka\ line flux varied by a factor of $>22$,
and the EW by a factor of $>17$.

In our analysis of 5 HETG observations of \velxI\ (the 4 observations
mentioned above, and obs.~ID.~14654), the \bne\ fits require an Fe
abundance that varies between $\sim 0.3$ solar to the maximum table
value of 10~\zfesun\ (see \tr{tab-trans}). Thus, the simple, uniform,
solar-abundance spherical model is ruled out.  The \myt\ fits reveal
spectra with a large dynamic range in the magnitude of the reflection
spectrum and \feka\ line relative to the direct continuum (see
\tr{tab-myun}), while the observed continuum flux varies by a factor
of $\sim 220$ (see \tr{tab-flux}).  The \feka\ line flux varies
between $\sim 15$ to $400 \ \times 10^{-5}$ \phunits, and the EW
varies between $\sim 57$ to $927$~eV (see \tr{tab-feka}). The
\feka\ line is unresolved in all of the observations, the maximum
upper limit on the FWHM being $1525$~\kmps. More details for each of
the 5 observations are provided below (refer to the spectral plots
for \velxI\
in Appendix~\ref{app-ufda}, and contour plots
in Appendix~\ref{app-con}).

We note that \citet{martinez-nunez2014} used \xmm\ and different
modeling to obtain an unabsorbed continuum flux in the $0.6-10$ keV
band in the range $\sim$\ten{3}{-9} to $\sim$\ten{4}{-8} \funits. This
compares to a range of $\sim$\ten{9}{-12} to $\sim$\ten{3}{-9}
\funits\ obtained for our observations with \myt. However, their flux
is over a more extended energy range, while the \xmm\ data include a
giant flare and no eclipse, whereas our data also include the eclipse
phase.

Two of the observations show exceptionally large EW values, and
are discussed in detail below (\velxIa-A and \velxIc-C).

\subsubsection{\velxIa-A}\label{sec-velxIa}

The bulk of the signal in the spectrum is in the \feka\ line, against
a relatively weak continuum. The Fe abundance in the \bne\ model is
driven to the maximum value of 10~\zfesun, but still the model is
unable to produce the required flux in the \feka\ line (see
\tr{tab-trans}). The \myt\ model fit is one of the few cases in the
entire XRB sample for which the column density \nhs\ can be
constrained, and the fit gives $2.1^{+0.6}_{-0.8} \times
10^{23}$~\cunits\ (see \tr{tab-myun}). A line-of-sight column density
of \nhz~$\sim 1.4 \times 10^{24}$~\cunits\ is also required,
substantially suppressing the direct continuum.  The EW of the
\feka\ line is consequently large, $\sim 530$~eV, characteristic of a
reflection-dominated spectrum.  This large value agrees within the
errors with the value of $\sim 630$ eV obtained with the
phenomenological modeling of \ctt. The \velxI\ neutron star is
essentially embedded in the dense stellar wind of its B0.5Ib-type
supergiant companion, leading to strong variability for this HMXB
\citep{kreykenbohm2008}. \citet{pradhan2014} consider this HMXB to be
very similar in this respect to \oaoXVI, as a possible link between
SFXTs and standard HMXBs (see \scr{sec-oaoXVI}). The interpretation
for the large EW might then be similarly involving a clumpy stellar
wind, as also modeled by \citet{watanabe2006}.
  
The \feka\ line is unresolved, with
FWHM~$<1525$~\kmps.  A second power-law continuum is required in the
fits with both of the models, and dominates the continuum at low
energies.

\subsubsection{\velxIb-B}

The continuum flux is substantially larger (by a factor of $\sim 60$)
compared to the previous observation and the \feka\ line does not
dominate the spectrum. The \bne\ fit gives a sub-solar Fe abundance of
$\sim 0.34$ (see \tr{tab-trans}). The \myt\ fit shows that the
\feka\ line flux increases by a factor of $\sim 7$ compared to the
previous observation but due to the larger increase in the continuum,
the EW of the line is only $\sim 57$~eV (see \tr{tab-feka}).  Also,
the line-of-sight column density decreases by an order of magnitude to
$\sim 1.6 \times 10^{23}$~\cunits.  The column density of the
reflector, \nhs, had to be frozen at $10^{25}$~\cunits\ as it could
not be constrained.  The \feka\ line is unresolved, with
FWHM~$<940$~\kmps.

\subsubsection{\velxIc-C}\label{sec-velxIc} % 0 = Eclipse 1926

In this observation, the continuum has dropped back down to $\sim
20\%$ below the level it was in the first observation (A), or a factor
of $\sim 80$ lower than in observation B.  Thus, the \feka\ line is
even more prominent than it was in observation A. As would be
expected, the \bne\ fit then cannot produce sufficient \feka\ line
flux even for the maximum Fe abundance of 10~\zfesun\ (see \tr{tab-trans}), and
the \myt\ fit gives a large \feka\ line EW of $\sim 927$~eV,
characteristic of a reflection-dominated spectrum (see \tr{tab-myun}).
This value also agrees with the one reported by \ctt.  A
  reflection dominated spectrum is also corroborated by the very large
  value of $A_{S}$ of $\sim 120$, but the implication is that the
  apparently Compton-thin \nhz\ value is not the line-of-sight column
  density of the reflector, with the true value being so large that
  its effect cannot be detected in the HEG bandpass.  The EW
  discussion for \velxIa-A is also relevant here. In addition, this
  observation corresponds to an eclipse phase \citep{goldstein2004},
  further suppressing the direct continuum.  
The \feka\ line flux is also reduced compared to the previous
observation, but only by a factor $\sim 5$. The line-of-sight column
density, \nhz\ is driven to zero, but could be so large
(i.e. $10^{25}$~\cunits\ or more) that it is not detectable in the HEG
bandpass.
Note that the FWHM of the \feka\ line was fixed at $100$~\kmps\ in
order to achieve a stable fit and error analysis.

\subsubsection{\velxId-D}

In this observation the observed continuum flux has increased by a
factor of $\sim 147$ compared to the previous observation and the
spectrum is no longer reflection-dominated. The \bne\ model yields a
fit with an Fe abundance that is only $\sim 10\%$ below solar (see
\tr{tab-trans}).  The \myt\ fit shows that the \feka\ line flux
increases by a factor of $\sim 19$ compared to the previous
observation.  The EW of the line is now only $\sim 122$~eV (see
\tr{tab-feka}).  Also, the line-of-sight column density is only
moderately less than that in the previous observation, at $\sim 9
\times 10^{22}$~\cunits.  The column density of the reflector, \nhs,
had to be frozen at $10^{25}$~\cunits\ as it could not be
constrained. Note that the FWHM of the \feka\ line was fixed at
$100$~\kmps\ in order to achieve a stable fit and error analysis.

\subsubsection{\velxIe-E}

In this observation, the observed continuum flux has increased by a
factor of $\sim 220$ compared to observation C, which is the most
reflection-dominated spectrum from the 5 observations. The \bne\ fit
to observation E requires an additional soft power-law continuum, and
a super-solar Fe abundance of $\sim 3.1$ in order to produce
sufficient flux in the \feka\ line considering the small column
density, \nhsph, of $\sim 1.5 \times 10^{22}$~\cunits\ (see
\tr{tab-trans}).  The \myt\ fit accordingly gives a relatively small
value for the EW of the \feka\ line of $\sim 64$~eV. The flux in the
\feka\ line is only a factor of $\sim 12$ higher than it was in
observation C, when the EW of the \feka\ line was nearly a keV. The
\feka\ line is unresolved, with FWHM~$<789$~\kmps.

\section{\\Plots of fitted spectra for individual observations}\label{app-ufda}
For each observation we show the counts spectra,
zooming into the region $5.8-7.6$ keV,
and unfolded photon
spectra
over the full fitting region.
In separate panels, these spectra are overlaid with the
fitted \bne\ and \myt\ models.  The counts spectra are identified with
the label ``Data and Folded Model'' above the plot, and the photon
spectra are identified with the label ``Unfolded Spectrum'' above the
plot.

In the case of photon spectra, the red line shows the total model (all
components).  The blue line is the \bne\ model (for spherical model
fits where there are also other components) or the direct continuum
component, \mytz\ (for \myt\ fits).  The purple line represents the
scattered continuum component, \myts, for the \myt\ model. The second
power law, when used, is shown as an orange line both for the
\bne\ and the \myt\ models. Extra emission lines modeled by Gaussians
appear in magenta.
  
\begin{figure*}
  \includegraphics[width=\textwidth]{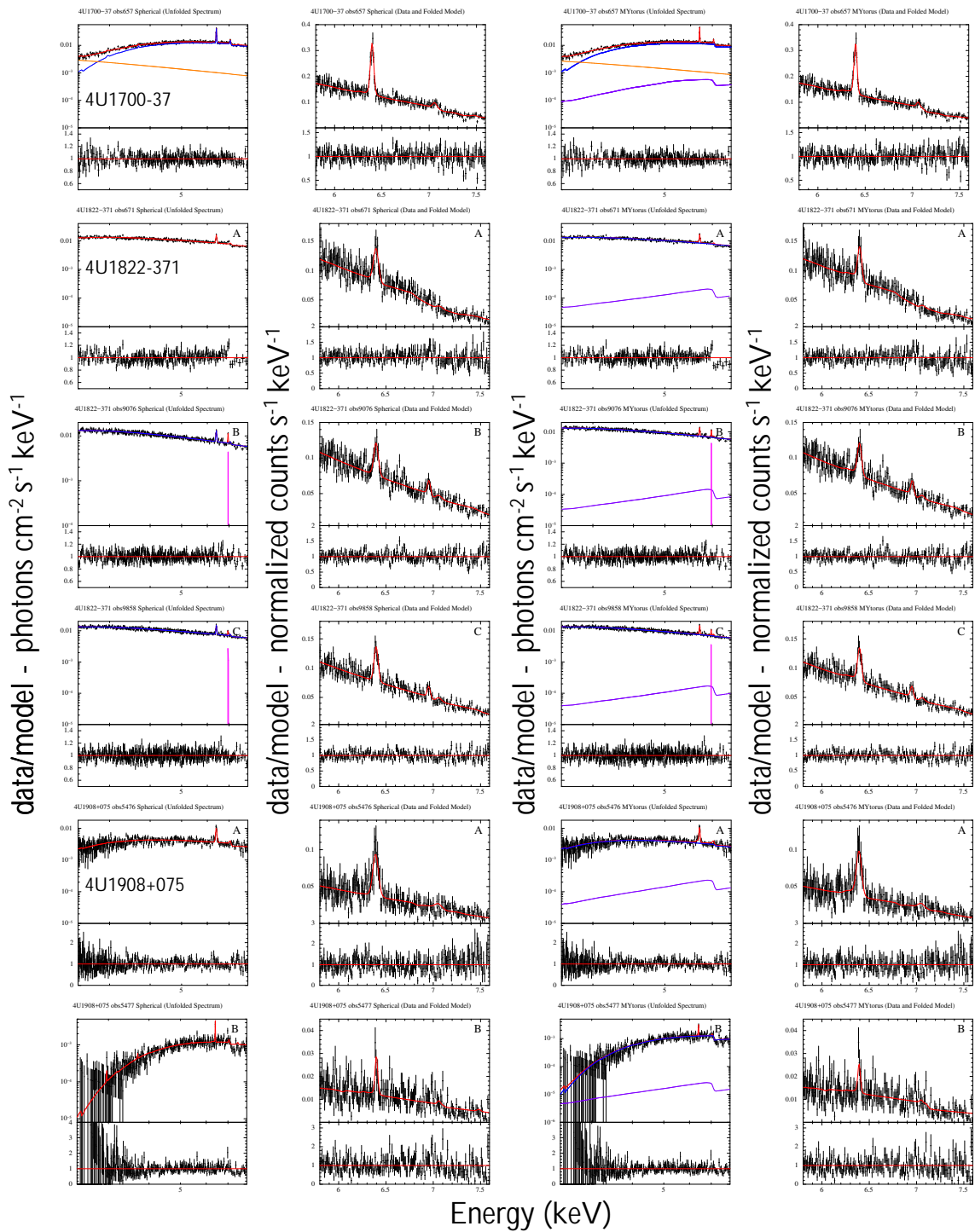}
  \vspace{-2cm}
  \caption{Results
    for observations are presented in the order of \tr{tab-sample},
    from left to right and top to bottom. Each row shows results for a single
    observation. The two leftmost panels show \bne\ model results, and the two
    rightmost panels \myt\ model results. In each panel pair, the left panel
    shows unfolded spectra over the full fitting region, and the right panel
    counts spectra in the vicinity of the \feka\ line. Data/model ratios
    are shown at the bottom of each panel. To facilitate locating
    specific observations, the object name is shown in the panel for
    the first observation for a given object.}
  \label{fig-ufda}
  \end{figure*}
\begin{figure*}
  \addtocounter{figure}{-1}
  \includegraphics[width=\textwidth]{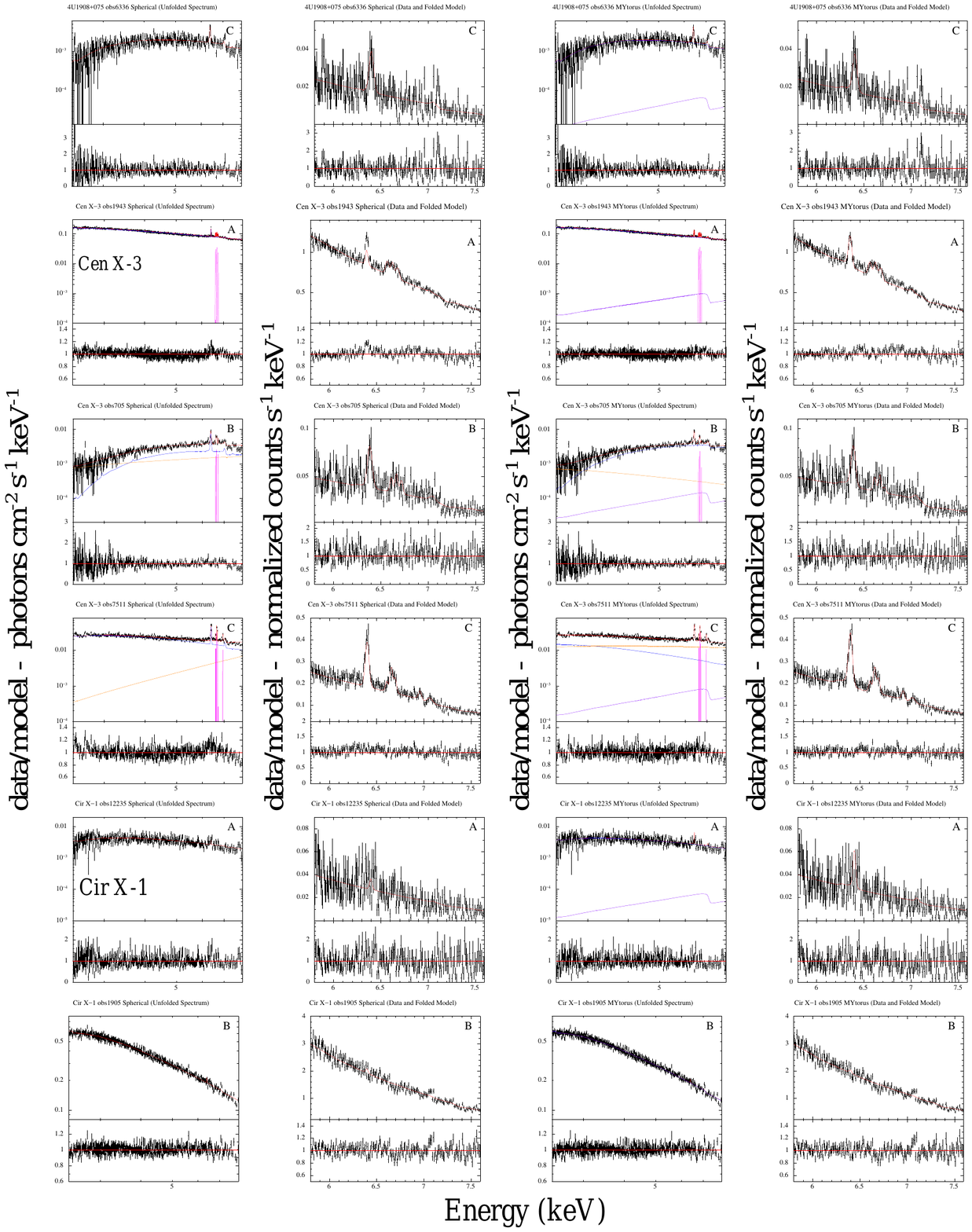}
  \vspace{-2cm}
  \caption{\it Continued.}
  \end{figure*}
\begin{figure*}
  \addtocounter{figure}{-1}
  \includegraphics[width=\textwidth]{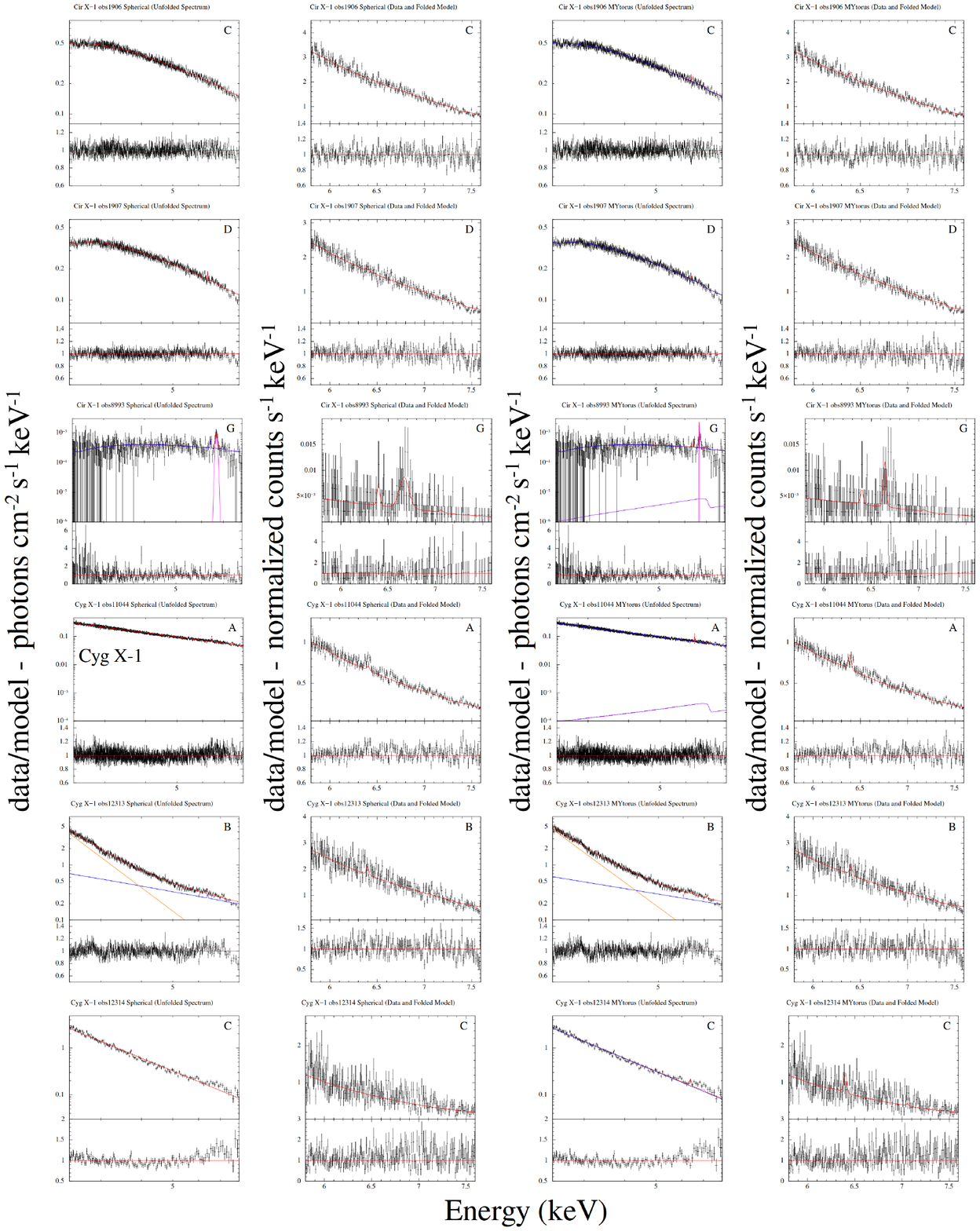}
  \vspace{-2cm}
  \caption{\it Continued.}
  \end{figure*}
\begin{figure*}
  \addtocounter{figure}{-1}
  \includegraphics[width=\textwidth]{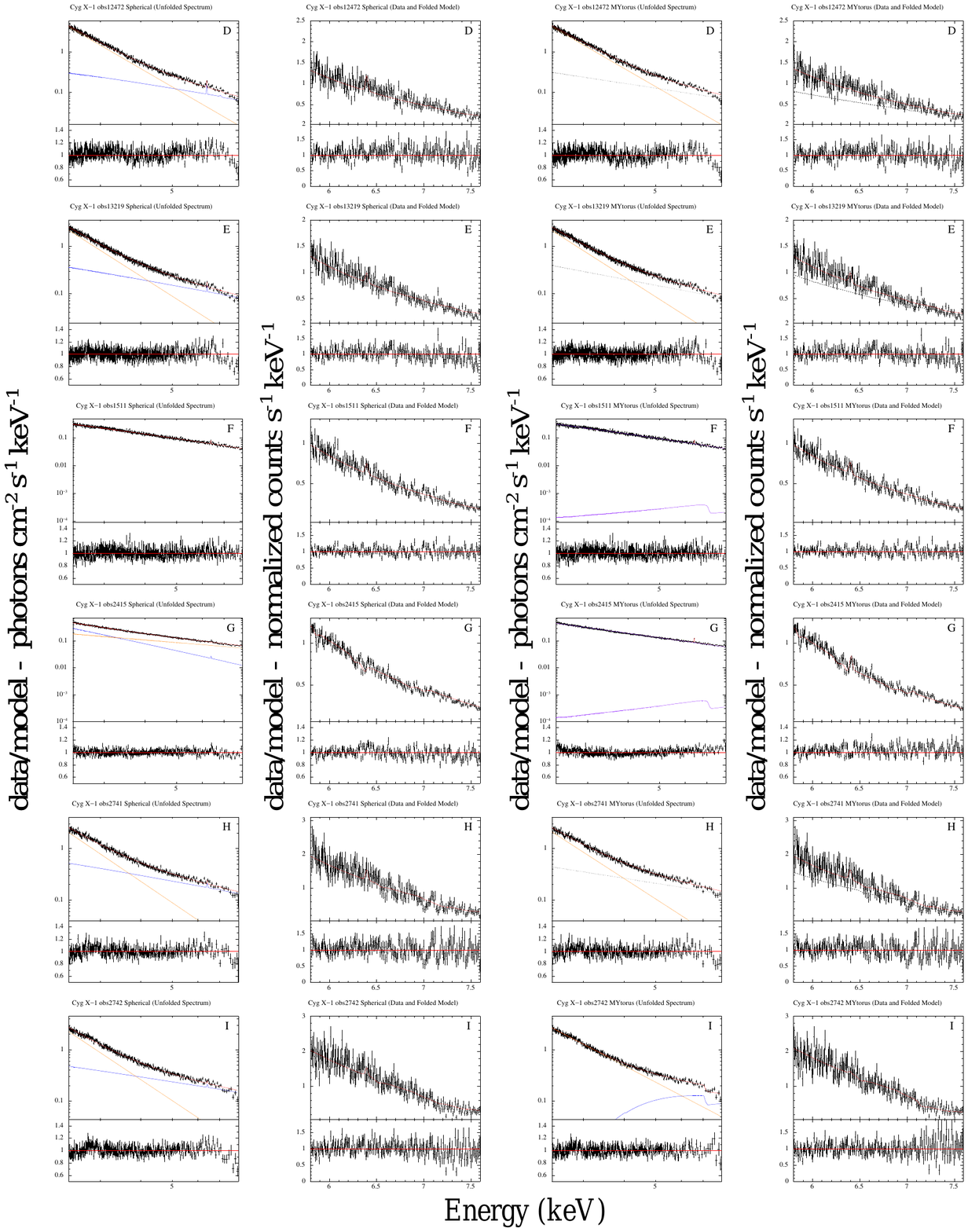}
  \vspace{-2cm}
  \caption{\it Continued.}
  \end{figure*}
\begin{figure*}
  \addtocounter{figure}{-1}
  \includegraphics[width=\textwidth]{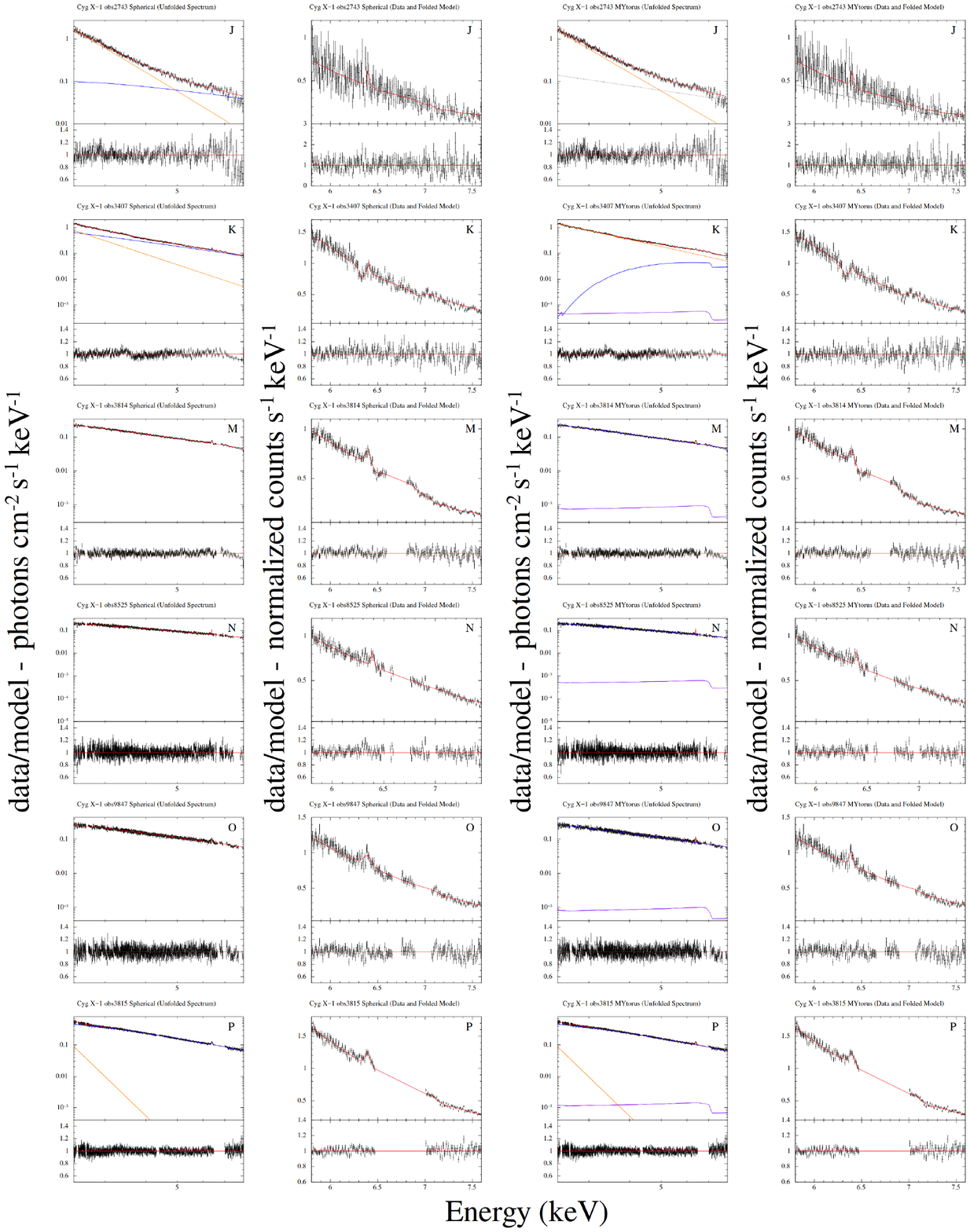}
  \vspace{-2cm}
  \caption{\it Continued.}
  \end{figure*}
\begin{figure*}
  \addtocounter{figure}{-1}
  \includegraphics[width=\textwidth]{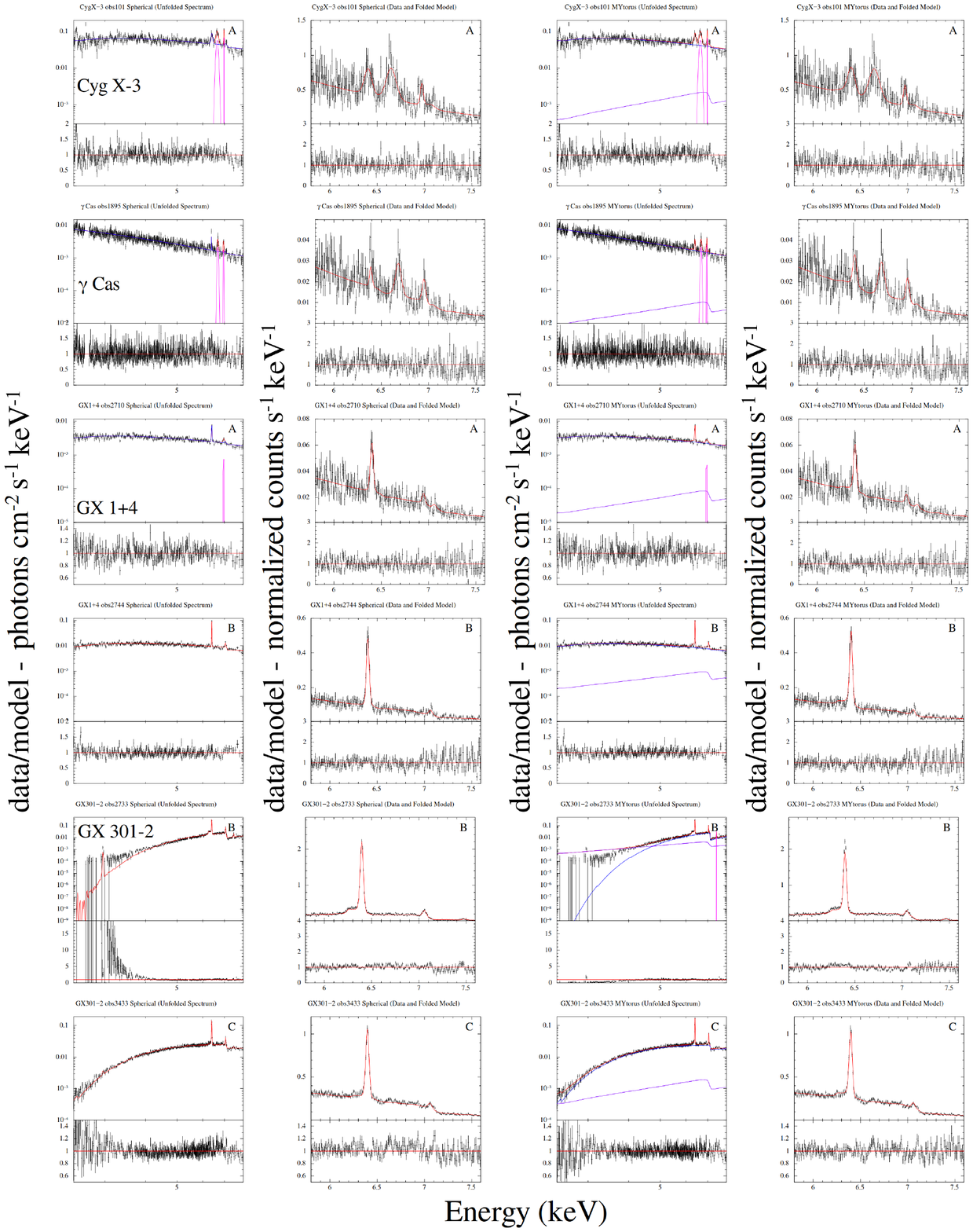}
  \vspace{-2cm}
  \caption{\it Continued.}
  \end{figure*}
\begin{figure*}
  \addtocounter{figure}{-1}
  \includegraphics[width=\textwidth]{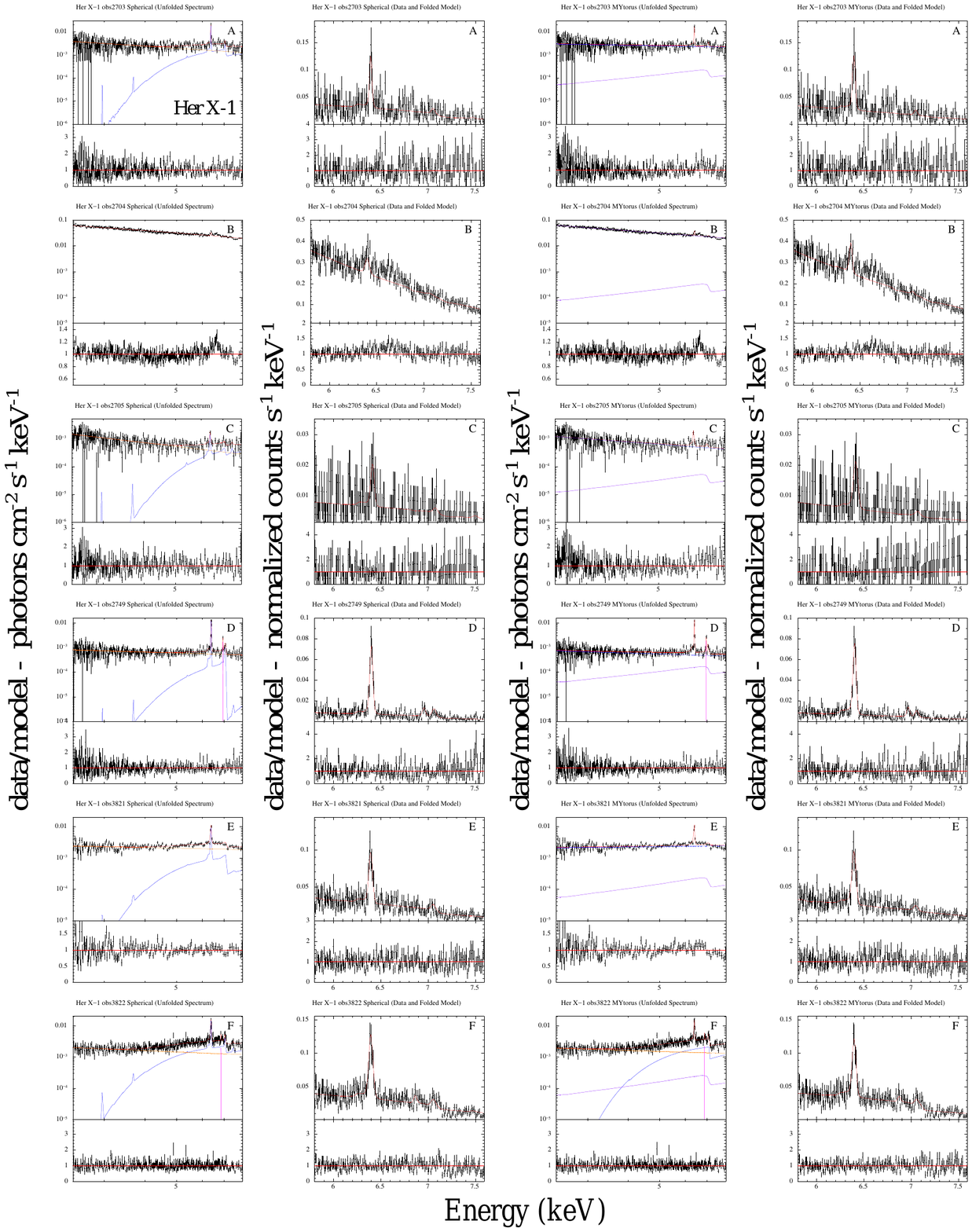}
  \vspace{-2cm}
  \caption{\it Continued.}
  \end{figure*}
\begin{figure*}
  \addtocounter{figure}{-1}
  \includegraphics[width=\textwidth]{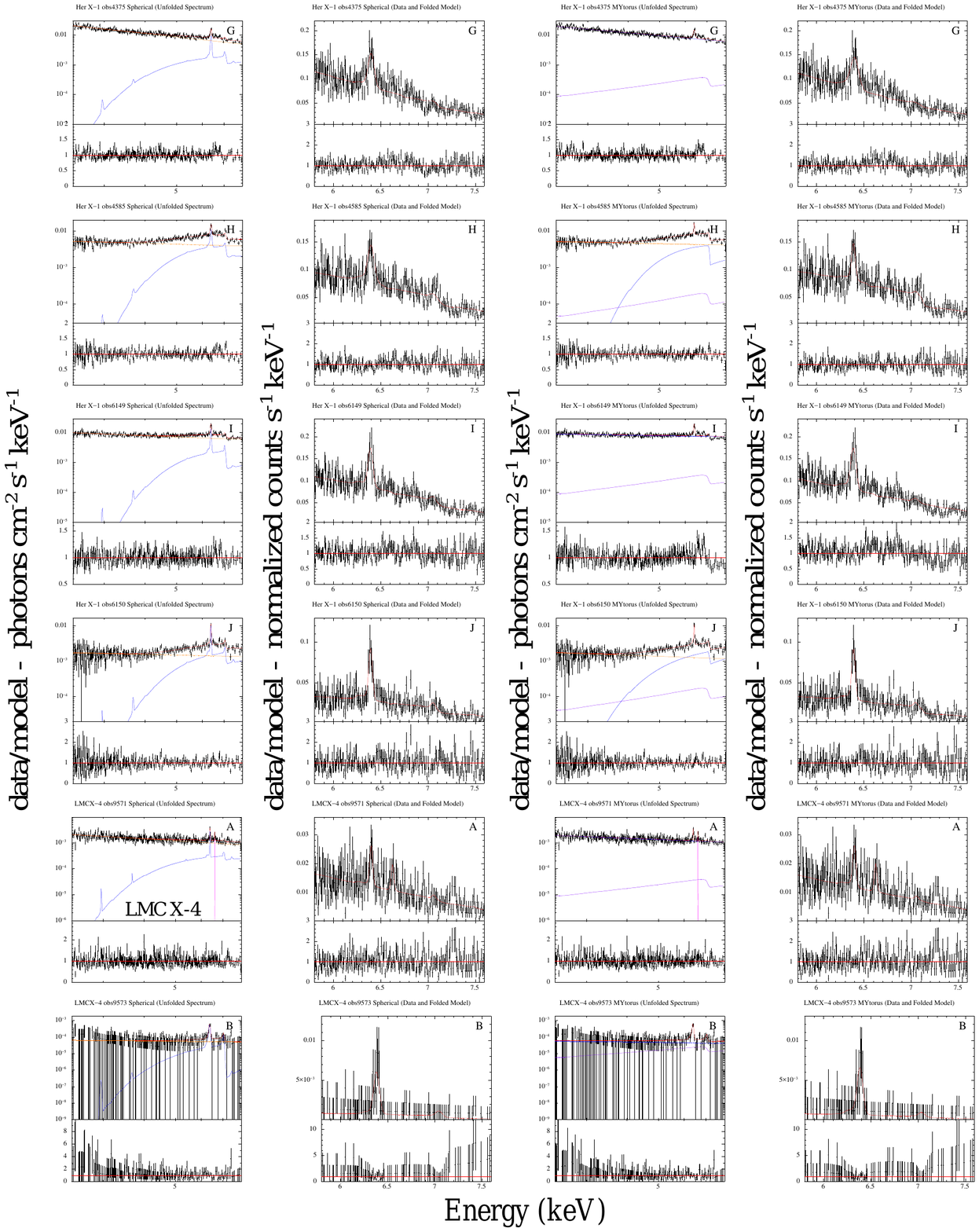}
  \vspace{-2cm}
  \caption{\it Continued.}
  \end{figure*}
\begin{figure*}
  \addtocounter{figure}{-1}
  \includegraphics[width=\textwidth]{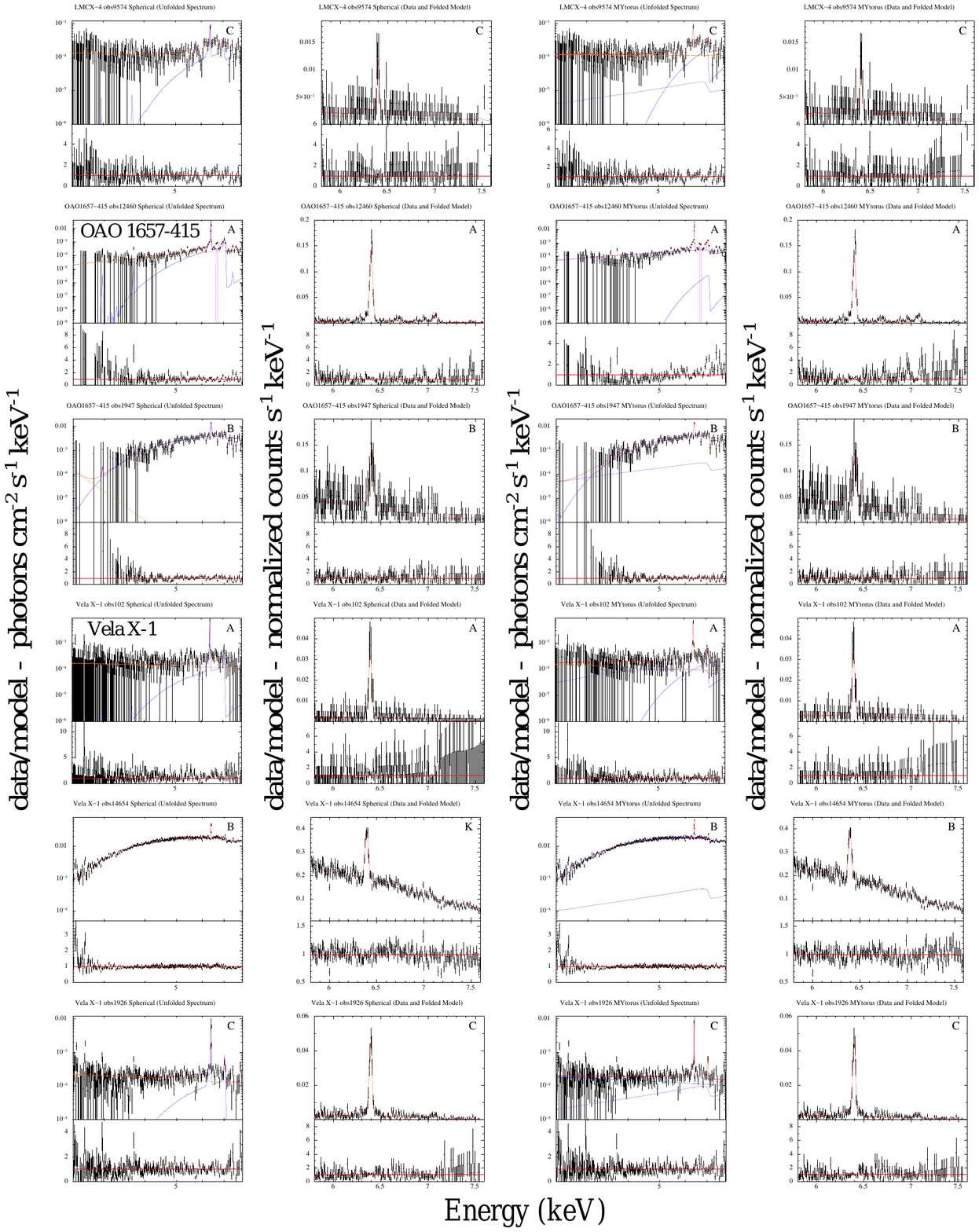}
  \vspace{-2cm}
  \caption{\it Continued.}
  \end{figure*}
 \begin{figure*}
  \addtocounter{figure}{-1}
  \includegraphics[width=\textwidth]{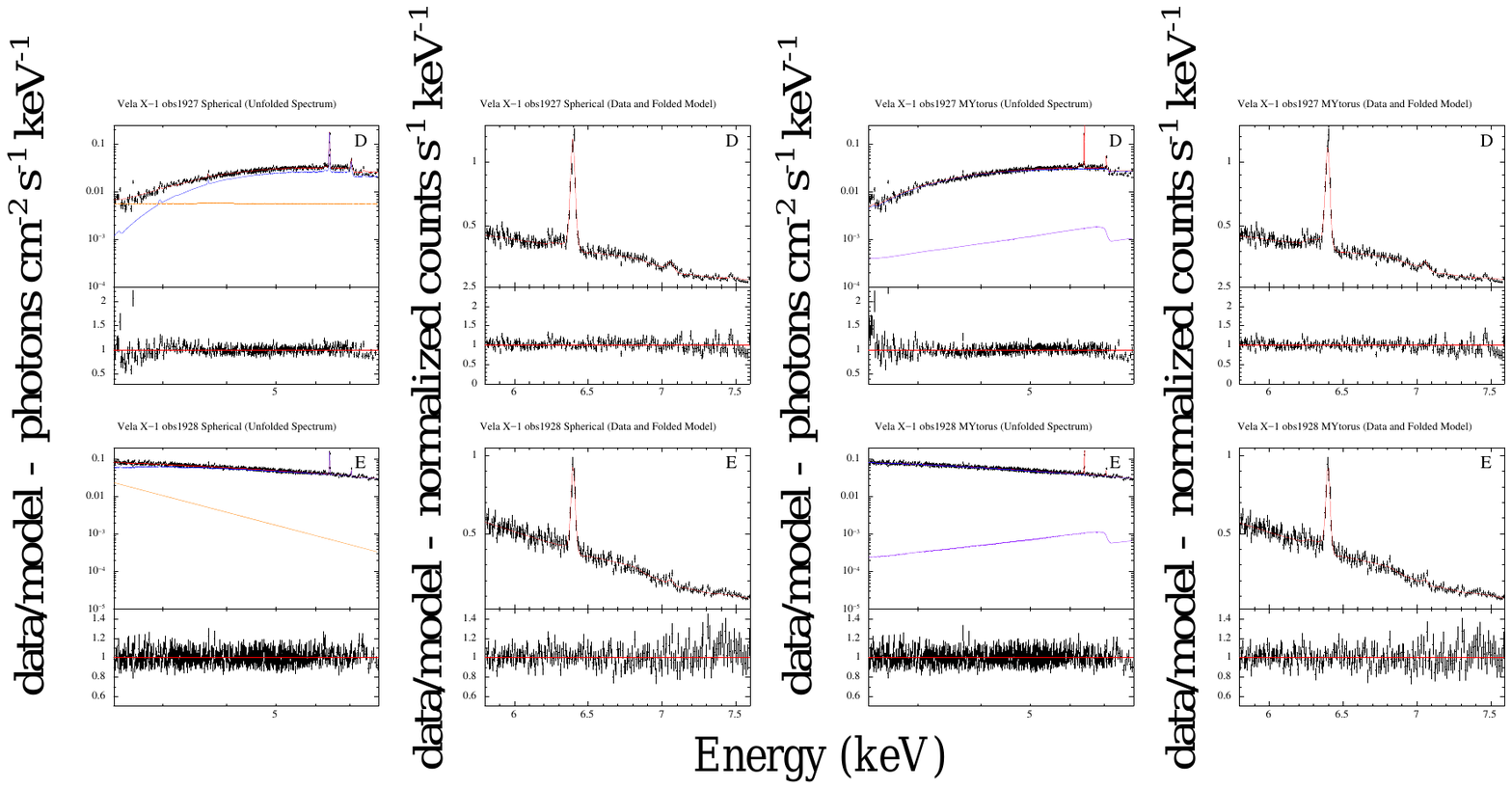}
  \vspace{-2cm}
  \caption{\it Continued.}
  \end{figure*}
\clearpage

\section{\\Contour plots of FWHM vs.~\textit{N}$_{\textbf{H,S}}$ for \feka\ detections.}\label{app-con}
We plot FWHM vs.~\nhs\ contours for \myt\ fits at join confidence
levels corresponding to 68\%\ ({\it black curves}), 90\%\ ({\it red
  curves}), and 99\%\ ({\it blue curves}).

%%%%%%%%%%%%%%%%%%%%%%%%%%%%%%%%%%%%%%%%%%%%%%%%%%
\begin{figure*}
  \includegraphics[width=\textwidth]{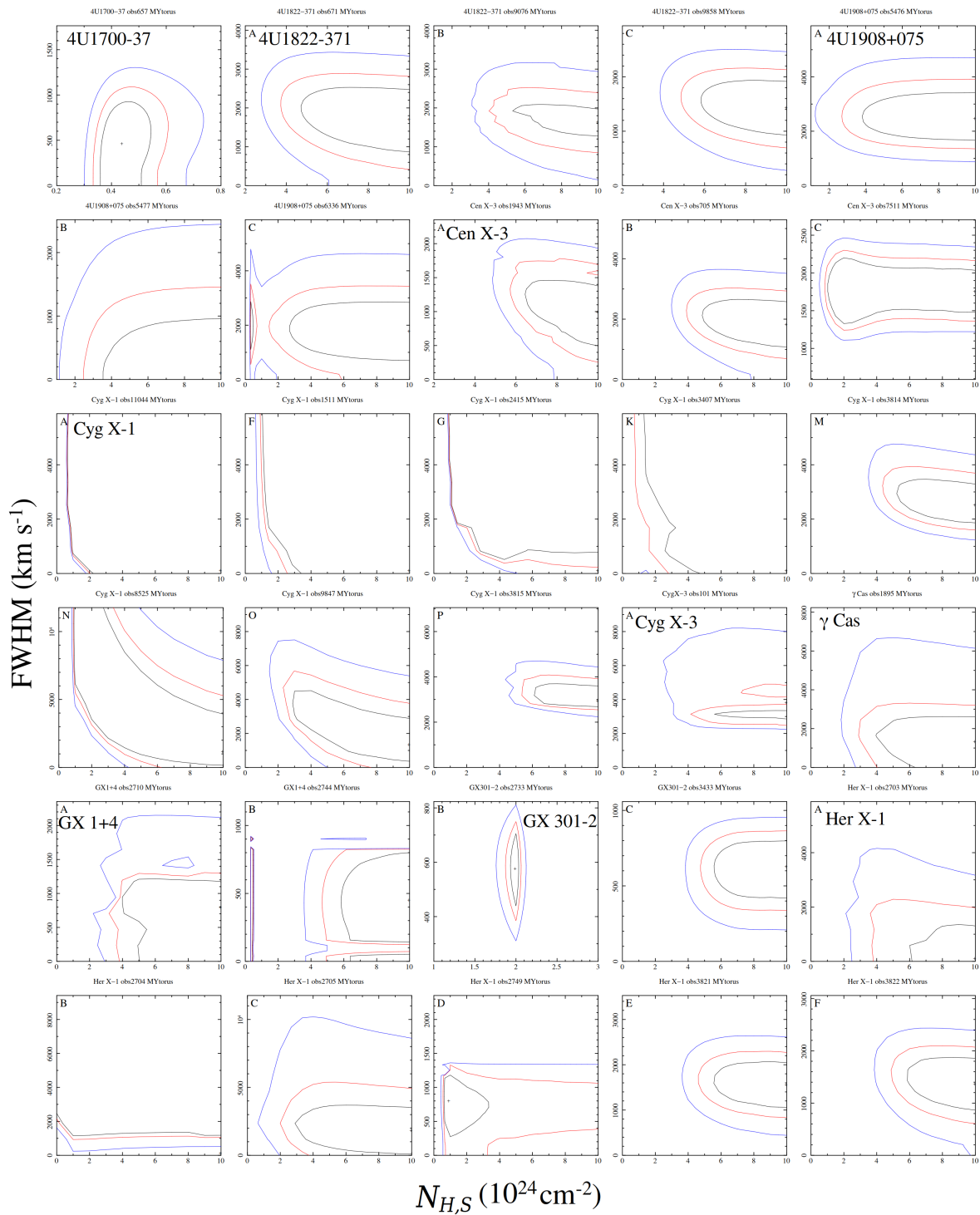}
  \vspace{-2cm}
  \caption{For each observation with a significant \feka\ line
    detection we show two-parameter contours of FWHM vs. \nhs\ from
    the \myt\ fits. Results
    for observations are presented in the order of \tr{tab-sample},
    from left to right and top to bottom. To facilitate locating
    specific observations, the object name is shown in the panel for
    the first observation for a given object.}
  \label{fig-cont}
  \end{figure*}
\begin{figure*}
  \addtocounter{figure}{-1}
  \includegraphics[width=\textwidth]{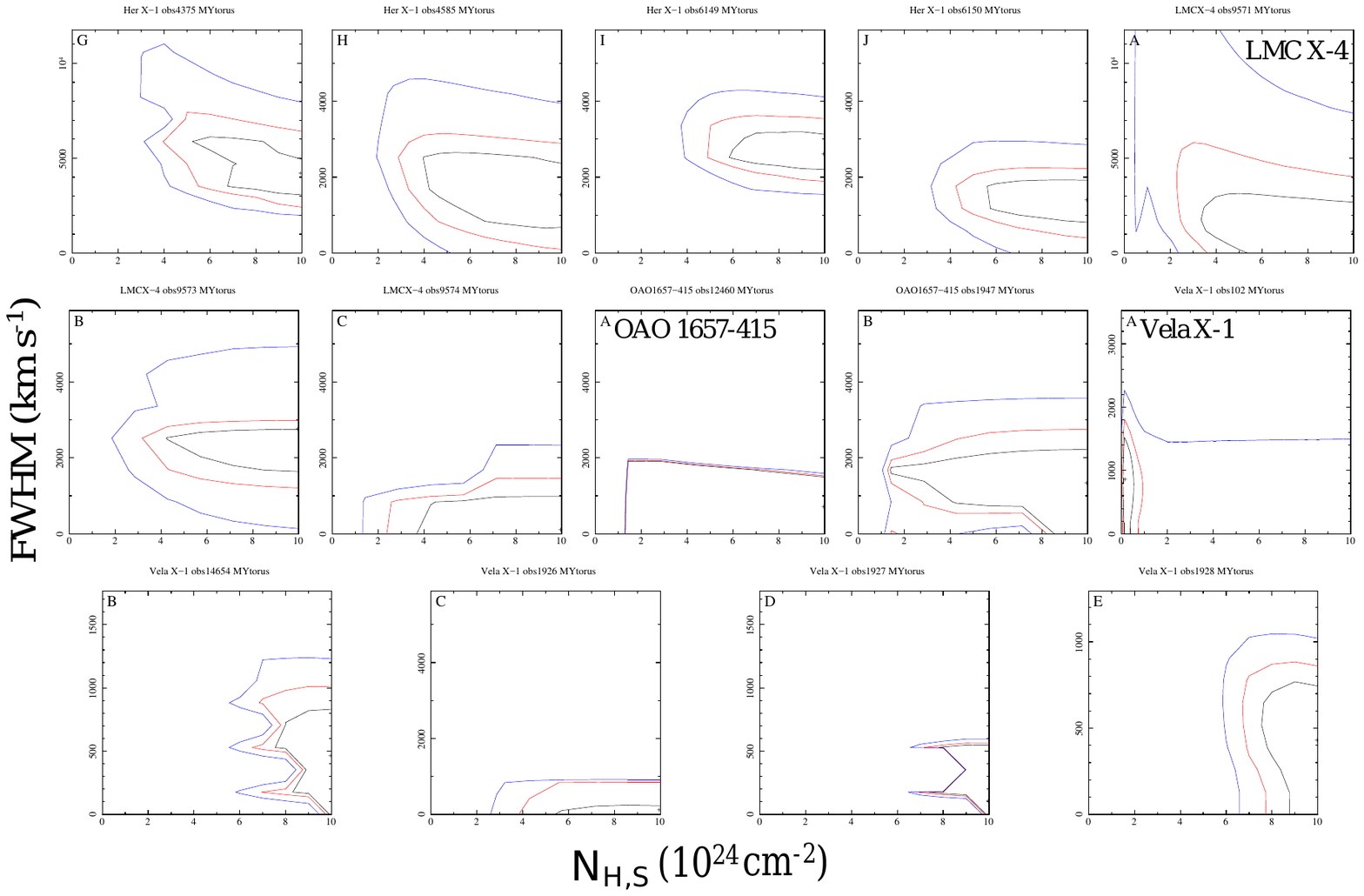}
  \vspace{-11.5cm}
  \caption{\it Continued.}
  \end{figure*}

\clearpage
\bibliographystyle{likeapj}
\bibliography{masterbib}   

\end{document}

%% file: newcomm.tex
\newcommand{\fuXVII}{4U1700$-$37}
\newcommand{\fuXVIII}{4U1822$-$371}
\newcommand{\fuXIX}{4U1908$-$075}
\newcommand{\cenxIII}{Cen~X$-$3}
\newcommand{\cirxI}{Cir~X$-$1}
\newcommand{\cygxI}{Cyg~X$-$1}
\newcommand{\cygxIII}{Cyg~X$-$3}
\newcommand{\gamcas}{$\gamma$~Cas}
\newcommand{\gxI}{GX~1+4}
\newcommand{\gxIII}{GX~301$-$2}
\newcommand{\herxI}{Her~X$-$1}
\newcommand{\lmcxIV}{LMCX$-$4}
\newcommand{\oaoXVI}{OAO~1657$-$415}
\newcommand{\velxI}{Vela~X$-$1}

\newcommand{\fuXIXbt}{4U1908$+$075 (obs.~ID.~5477)}

\newcommand{\gxIIIbt}{GX~301$-$2 (obs.~ID.~2733)}
\newcommand{\gxIIIct}{GX~301$-$2 (obs.~ID.~3433)}

\newcommand{\herxIdt}{Her~X$-$1 (obs.~ID.~2749)}

\newcommand{\lmcxIVbt}{LMC~X$-$4 (obs.~ID.~9573)}
\newcommand{\lmcxIVct}{LMC~X$-$4 (obs.~ID.~9574)}
\newcommand{\oaoXVIat}{OAO~1657$-$415 (obs.~ID.~12460)}

\newcommand{\velxIat}{Vela~X$-$1 (obs.~ID.~102)}

\newcommand{\velxIct}{Vela~X$-$1 (obs.~ID.~1926)}

\newcommand{\fuXVIIa}{obs.~ID.~657}

\newcommand{\fuXIXa}{obs.~ID.~5476}

\newcommand{\cenxIIIa}{obs.~ID.~1943}
\newcommand{\cenxIIIb}{obs.~ID.~705}
\newcommand{\cenxIIIc}{obs.~ID.~7511}

\newcommand{\cirxIg}{obs.~ID.~8993}

\newcommand{\cygxIa}{obs.~ID.~11044}
\newcommand{\cygxIb}{obs.~ID.~12313}
\newcommand{\cygxIc}{obs.~ID.~12314}
\newcommand{\cygxId}{obs.~ID.~12472}
\newcommand{\cygxIe}{obs.~ID.~13219}
\newcommand{\cygxIf}{obs.~ID.~1511}
\newcommand{\cygxIg}{obs.~ID.~2415}
\newcommand{\cygxIh}{obs.~ID.~2741}
\newcommand{\cygxIi}{obs.~ID.~2742}
\newcommand{\cygxIj}{obs.~ID.~2743}
\newcommand{\cygxIk}{obs.~ID.~3407}

\newcommand{\cygxIm}{obs.~ID.~3814}
\newcommand{\cygxIn}{obs.~ID.~8525}
\newcommand{\cygxIo}{obs.~ID.~9847}
\newcommand{\cygxIp}{obs.~ID.~3815}
\newcommand{\cygxIIIa}{obs.~ID.~101}
\newcommand{\cygxIIIb}{obs.~ID.~6601}
\newcommand{\cygxIIIc}{obs.~ID.~7268}

\newcommand{\gxIIIb}{obs.~ID.~2733}
\newcommand{\gxIIIc}{obs.~ID.~3433}
\newcommand{\gxIa}{obs.~ID.~2710}
\newcommand{\gxIb}{obs.~ID.~2744}
\newcommand{\herxIa}{obs.~ID.~2703}
\newcommand{\herxIb}{obs.~ID.~2704}
\newcommand{\herxIc}{obs.~ID.~2705}
\newcommand{\herxId}{obs.~ID.~2749}
\newcommand{\herxIe}{obs.~ID.~3821}
\newcommand{\herxIf}{obs.~ID.~3822}
\newcommand{\herxIg}{obs.~ID.~4375}
\newcommand{\herxIh}{obs.~ID.~4585}
\newcommand{\herxIi}{obs.~ID.~6149}
\newcommand{\herxIj}{obs.~ID.~6150}
\newcommand{\lmcxIVa}{obs.~ID.~9571}
\newcommand{\lmcxIVb}{obs.~ID.~9573}

\newcommand{\velxIa}{obs.~ID.~102}
\newcommand{\velxIb}{obs.~ID.~14654}
\newcommand{\velxIc}{obs.~ID.~1926}
\newcommand{\velxId}{obs.~ID.~1927}
\newcommand{\velxIe}{obs.~ID.~1928}

\newcommand{\cunits}{cm$^{-2}$}
\newcommand{\funits}{erg~cm$^{-2}$~s$^{-1}$}

\newcommand{\kmps}{km s$^{-1}$}

\newcommand{\lunits}{erg~s$^{-1}$}

\newcommand{\mco}{$M_{\rm CO}$}
\newcommand{\mcoa}{$M_{\rm CO,1}$}
\newcommand{\mcob}{$M_{\rm CO,2}$}

\newcommand{\mns}{$M_{\rm NS}$}

\newcommand{\msun}{$M_{\odot}$}

\newcommand{\phunits}{photons~cm$^{-2}$~s$^{-1}$}
\newcommand{\ps}{s$^{-1}$}

\newcommand{\delc}{$\Delta C$}

\newcommand{\gammasph}{$\Gamma_{\rm sph}$}
\newcommand{\normgammasph}{${N_{\rm sph}}$}
\newcommand{\normgammaz}{${N_{\rm Z}}$}
\newcommand{\normgammas}{${N_{\rm S}}$}
\newcommand{\gammas}{$\Gamma_{\rm S}$}
\newcommand{\gammasoft}{$\Gamma_{\rm soft}$}
\newcommand{\gammaz}{$\Gamma_{\rm Z}$}

\newcommand{\nh}{$N_{\rm H}$}

\newcommand{\nhz}{$N_{\rm H,Z}$} 
\newcommand{\nhs}{$N_{\rm H,S}$} 
 
\newcommand{\nhsph}{$N_{\rm H,sph}$}

\newcommand{\nhgal}{$N_{\rm H}^{\rm gal}$}

\newcommand{\sigl}{$\sigma_L$}

\newcommand{\x}{X-ray}

\newcommand{\lcobs}{$L_{\rm 2-10, c, obs}$}
\newcommand{\lcintr}{$L_{\rm 2-10, c, intr}$}

\newcommand{\ftot}{$f_{\rm 2-10, tot}$}
\newcommand{\fcobs}{$f_{\rm 2-10, c, obs}$}
\newcommand{\fcintr}{$f_{\rm 2-10, c, intr}$}
\newcommand{\ffeka}{$f_{\rm Fe K\alpha}$}

\newcommand{\thobss}{$\theta_{\rm obs, S}$}

\newcommand{\zs}{$z_{\rm S}$}
\newcommand{\zsph}{$z_{\rm sph}$}

\newcommand{\zfe}{$A_{\rm Fe}$}
\newcommand{\zfeten}{$A_{\rm Fe} = 10 \,\, A_{\rm Fe,\odot}$}
\newcommand{\zm}{$A_{\rm M}$}
\newcommand{\zfesun}{$A_{\rm Fe,\odot}$}

\newcommand{\feka}{Fe K$\alpha$}
\newcommand{\fekb}{Fe K$\beta$}

\newcommand{\fetwofive}{Fe{\sc \,xxv}}
\newcommand{\fetwosix}{Fe{\sc \,xxvi}}
\newcommand{\ha}{H$\alpha$}

\newcommand{\chandra}{{\it Chandra}}
\newcommand{\nustar}{{\it NuSTAR}}
\newcommand{\suzaku}{{\it Suzaku}}

\newcommand{\xmm}{{\it XMM-Newton}}

\newcommand{\myt}{{\sc mytorus}}

\newcommand{\myty}{{\sc myt}}
\newcommand{\myts}{{\sc my}torus{\sc s}}
\newcommand{\mytl}{{\sc my}torus{\sc l}}
\newcommand{\mytz}{{\sc my}torus{\sc z}}

\newcommand{\xspec}{{\sc xspec}}

\newcommand{\fr}{Figure~\ref}
\newcommand{\scr}{Section~\ref}
\newcommand{\tr}{Table~\ref}
\newcommand{\fspace}{\hspace{-2pt}} %For main text after footnote.
\newcommand{\exi}{\begin{equation}}
\newcommand{\exo}{\end{equation}}

\newcommand{\aer}[3]{$#1^{#2}_{#3}$} %1.2^+0.1_-0.2
\newcommand{\ten}[2]{$#1\times 10^{#2}$} %{some number} x 10^{some power}
\newcommand{\up}[1]{$^{#1}$}

\def\spose#1{\hbox to 0pt{#1\hss}} 
\def\approxlt{\mathrel{\spose{\lower 3pt\hbox{$\sim$}}
        \raise 2.0pt\hbox{$<$}}}
\def\approxgt{\mathrel{\spose{\lower 3pt\hbox{$\sim$}}
        \raise 2.0pt\hbox{$>$}}}

\newcommand{\simn}{$\sim$~}

%% file: tab_sample.tex
\startlongtable
%\begin{deluxetable*}{ C{2cm} c ccc ccc cc c}
\begin{deluxetable*}{ c c ccc ccc cc c}
  \tablecaption{\chandra/HETG Galactic XRB sample\label{tab-sample}}
\tabletypesize{\scriptsize}
\tablehead{
\colhead{Source} & 
\multicolumn{2}{c}{obs.} &
\colhead{Type} &
\colhead{Date/Time} &
\colhead{Exposure} &
\colhead{Fitting} &
\colhead{Count Rate} &
\multicolumn{2}{c}{\nhgal} &
\colhead{\delc}
\\
\colhead{}&
\colhead{}&
\colhead{}& %for letter
\colhead{}&
\colhead{}&
\colhead{Time} &
\colhead{Band} &
\colhead{($2.0-8.0$~keV)}&
\colhead{\bnet} & \colhead{\myt} &
\\
\colhead{}&
\colhead{}&
\colhead{}& %for letter
\colhead{}&
\colhead{}&
\colhead{(ks)}&
\colhead{(keV)}&
\colhead{(\ps)}&
\multicolumn{2}{c}{(10\up{21}~\cunits)}&
}
\colnumbers
\startdata
4U1700$-$37 & 657 & A & HMXB/NS? & 2000-08-22T11:35:29 & 42.4 & $2.4-8.0$ & $1.5927\pm0.0061$ & \multicolumn{2}{c}{\aer{5.74}{\quad}{\quad}} & 462.52\\ 
4U1822$-$371$^1$ & 671 & A & LMXB/NS & 2000-08-23T16:20:37 & 39.4 & $2.4-8.0$ & $1.5985\pm0.0064$ & \multicolumn{2}{c}{\aer{1.04}{\quad}{\quad}} & 89.86\\ 
4U1822$-$371$^1$ & 9076 & B & LMXB/NS & 2008-05-20T22:46:21 & 62.2 & $2.4-8.0$ & $1.5271\pm0.0050$ & \multicolumn{2}{c}{\aer{1.04}{\quad}{\quad}} & 83.86\\ 
4U1822$-$371$^1$ & 9858 & C & LMXB/NS & 2008-05-23T13:14:05 & 80.3 & $2.4-8.0$ & $1.5574\pm0.0044$ & \multicolumn{2}{c}{\aer{1.04}{\quad}{\quad}} & 157.30\\ 
4U1908$+$075$^2$ & 5476 & A & HMXB/NS & 2005-06-27T18:24:12 & 18.7 & $2.4-8.0$ & $0.5585\pm0.0055$ & \multicolumn{2}{c}{\aer{12.1}{\quad}{\quad}} & 85.57\\ 
4U1908$+$075$^2$ & 5477 & B & HMXB/NS & 2005-11-26T12:28:41 & 49.0 & $2.4-8.0$ & $0.1074\pm0.0015$ & \multicolumn{2}{c}{\aer{12.1}{\quad}{\quad}} & 22.83\\ 
4U1908$+$075$^2$ & 6336 & C & HMXB/NS & 2005-07-11T10:53:57 & 31.4 & $2.4-8.0$ & $0.2419\pm0.0028$ & \multicolumn{2}{c}{\aer{12.1}{\quad}{\quad}} & 35.03\\ 
Cen~X$-$3 & 1943 & A & HMXB/NS & 2000-12-30T00:13:30 & 45.3 & $2.4-8.0$ & $17.4900\pm0.0196$ & \multicolumn{2}{c}{\aer{9.95}{\quad}{\quad}} & 292.35\\ 
Cen~X$-$3 & 705 & B & HMXB/NS & 2000-03-05T08:10:14 & 39.2 & $2.4-8.0$ & $0.4137\pm0.0032$ & \multicolumn{2}{c}{\aer{9.95}{\quad}{\quad}} & 48.25\\ 
Cen~X$-$3 & 7511 & C & HMXB/NS & 2007-09-12T11:07:00 & 39.4 & $2.4-8.0$ & $3.3428\pm0.0092$ & \multicolumn{2}{c}{\aer{9.95}{\quad}{\quad}} & 513.38\\ 
Cir~X$-$1 & 12235 & A & LMXB/NS & 2010-07-04T05:05:10 & 19.4 & $2.4-8.0$ & $0.5238\pm0.0052$ & \multicolumn{2}{c}{\aer{15.9}{\quad}{\quad}} & 1.16\\ 
Cir~X$-$1 & 1905 & B & LMXB/NS & 2001-08-01T09:28:05 & 7.9 & $2.4-8.0$ & $53.7788\pm0.0825$ & \multicolumn{2}{c}{\aer{15.9}{\quad}{\quad}} & 0.00\\ 
Cir~X$-$1 & 1906 & C & LMXB/NS & 2001-08-05T17:33:09 & 7.9 & $2.4-8.0$ & $51.4620\pm0.0808$ & \multicolumn{2}{c}{\aer{15.9}{\quad}{\quad}} & 0.00\\ 
Cir~X$-$1 & 1907 & D & LMXB/NS & 2001-08-09T14:28:36 & 8.0 & $2.4-8.0$ & $38.0570\pm0.0691$ & \multicolumn{2}{c}{\aer{15.9}{\quad}{\quad}} & 1.62\\ 
Cir~X$-$1 & 8993 & G & LMXB/NS & 2008-07-16T07:59:00 & 32.7 & $2.4-8.0$ & $0.0538\pm0.0013$ & \multicolumn{2}{c}{\aer{15.9}{\quad}{\quad}} & 5.18\\ 
Cyg~X$-$1 & 11044 & A & HMXB/BH & 2010-01-14T02:49:28 & 29.4 & $2.4-8.0$ & $20.6219\pm0.0265$ & \aer{0.001}{f}{\quad} & \aer{0.001}{f}{\quad} & 37.58\\ 
Cyg~X$-$1 & 12313 & B & HMXB/BH & 2010-07-22T16:22:27 & 2.1 & $2.4-8.0$ & $65.5693\pm0.1748$ & \aer{0.001}{f}{\quad} & \aer{\ldots}{\quad}{\quad} & 0.15\\ 
Cyg~X$-$1 & 12314 & C & HMXB/BH & 2010-07-24T17:21:43 & 0.9 & $2.4-8.0$ & $47.3176\pm0.2296$ & \aer{0.001}{f}{\quad} & \aer{\ldots}{\quad}{\quad} & 0.76\\ 
Cyg~X$-$1 & 12472 & D & HMXB/BH & 2011-01-06T13:47:10 & 3.3 & $2.4-8.0$ & $61.9591\pm0.1369$ & \aer{0.001}{f}{\quad} & \aer{\ldots}{\quad}{\quad} & 6.09\\ 
Cyg~X$-$1 & 13219 & E & HMXB/BH & 2011-02-05T06:35:01 & 4.4 & $2.4-8.0$ & $56.9174\pm0.1135$ & \aer{0.001}{f}{\quad} & \aer{\ldots}{\quad}{\quad} & 2.59\\ 
Cyg~X$-$1 & 1511 & F & HMXB/BH & 2000-01-12T08:15:16 & 12.6 & $2.4-8.0$ & $22.1196\pm0.0419$ & \aer{0.001}{f}{\quad} & \aer{0.001}{f}{\quad} & 8.26\\ 
Cyg~X$-$1 & 2415 & G & HMXB/BH & 2001-01-04T06:03:47 & 30.0 & $2.4-8.0$ & $28.8914\pm0.0310$ & \aer{0.001}{f}{\quad} & \aer{0.001}{f}{\quad} & 27.17\\ 
Cyg~X$-$1 & 2741 & H & HMXB/BH & 2002-01-28T05:34:31 & 1.9 & $2.4-8.0$ & $62.7450\pm0.1824$ & \aer{0.001}{f}{\quad} & \aer{\ldots}{\quad}{\quad} & 1.01\\ 
Cyg~X$-$1 & 2742 & I & HMXB/BH & 2002-01-30T01:23:31 & 1.9 & $2.4-8.0$ & $65.1236\pm0.1865$ & \aer{0.001}{f}{\quad} & \aer{\ldots}{\quad}{\quad} & 0.23\\ 
Cyg~X$-$1 & 2743 & J & HMXB/BH & 2002-04-13T20:53:02 & 2.4 & $2.4-8.0$ & $36.4287\pm0.1226$ & \aer{7.210}{f}{\quad} & \aer{\ldots}{\quad}{\quad} & 2.28\\ 
Cyg~X$-$1 & 3407 & K & HMXB/BH & 2001-10-28T16:14:56 & 16.1 & $2.4-8.0$ & $45.6119\pm0.0533$ & \aer{0.001}{f}{\quad} & \aer{0.001}{f}{\quad} & 8.70\\ 
Cyg~X$-$1 & 3814 & M & HMXB/BH & 2003-04-19T16:47:31 & 47.2 & $2.4-8.0$ & $17.5420\pm0.0193$ & \aer{0.001}{f}{\quad} & \aer{0.001}{f}{\quad} & 59.70\\ 
Cyg~X$-$1 & 8525 & N & HMXB/BH & 2008-04-18T18:09:48 & 29.4 & $2.4-8.0$ & $16.9730\pm0.0240$ & \aer{0.001}{f}{\quad} & \aer{0.001}{f}{\quad} & 24.71\\ 
Cyg~X$-$1 & 9847 & O & HMXB/BH & 2008-04-19T14:44:56 & 18.9 & $2.4-8.0$ & $21.4173\pm0.0337$ & \aer{0.001}{f}{\quad} & \aer{0.001}{f}{\quad} & 29.63\\ 
Cyg~X$-$1 & 3815 & P & HMXB/BH & 2003-03-04T15:46:06 & 57.0 & $2.4-8.0$ & $36.1843\pm0.0255$ & \aer{0.001}{f}{\quad} & \aer{0.001}{f}{\quad} & 117.62\\ 
Cyg~X$-$3 & 101 & A & HMXB/? & 1999-10-19T23:52:10 & 1.9 & $2.4-8.0$ & $8.4046\pm0.0657$ & \multicolumn{2}{c}{\aer{11.7}{\quad}{\quad}} & 22.80\\ 
$\gamma$~Cas & 1895 & A & HMXB/NS & 2001-08-10T09:21:57 & 51.2 & $2.4-8.0$ & $0.5643\pm0.0033$ & \multicolumn{2}{c}{\aer{4.12}{\quad}{\quad}} & 20.92\\ 
GX~301$-$2$^3$ & 2733 & B & HMXB/NS & 2002-01-13T09:00:24 & 39.2 & $2.4-8.0$ & $1.1393\pm0.0054$ & \multicolumn{2}{c}{\aer{13.0}{\quad}{\quad}} & 16006.26\\ 
GX~301$-$2$^3$ & 3433 & C & HMXB/NS & 2002-02-03T12:34:10 & 59.0 & $2.4-8.0$ & $2.2827\pm0.0062$ & \multicolumn{2}{c}{\aer{13.0}{\quad}{\quad}} & 3975.42\\ 
GX~1$+$4 & 2710 & A & LMXB/NS & 2002-08-05T21:34:10 & 56.6 & $2.4-8.0$ & $0.4505\pm0.0028$ & \multicolumn{2}{c}{\aer{3.07}{\quad}{\quad}} & 126.19\\ 
GX~1$+$4 & 2744 & B & LMXB/NS & 2002-04-26T23:01:38 & 20.3 & $2.4-8.0$ & $1.6385\pm0.0090$ & \multicolumn{2}{c}{\aer{3.07}{\quad}{\quad}} & 1028.75\\ 
Her~X$-$1 & 2703 & A & LMXB/NS & 2002-06-29T03:00:01 & 9.1 & $2.4-8.0$ & $0.4268\pm0.0068$ & \aer{0.169}{f}{\quad} & \aer{0.001}{f}{\quad} & 67.77\\ 
Her~X$-$1 & 2704 & B & LMXB/NS & 2002-07-05T17:45:04 & 18.7 & $2.4-8.0$ & $5.5899\pm0.0173$ & \aer{0.001}{f}{\quad} & \aer{0.001}{f}{\quad} & 37.32\\ 
Her~X$-$1 & 2705 & C & LMXB/NS & 2002-07-18T15:26:29 & 19.0 & $2.4-8.0$ & $0.1226\pm0.0025$ & \aer{0.001}{f}{\quad} & \aer{0.001}{f}{\quad} & 18.12\\ 
Her~X$-$1 & 2749 & D & LMXB/NS & 2002-05-05T10:15:53 & 49.4 & $2.4-8.0$ & $0.1102\pm0.0015$ & \aer{0.001}{f}{\quad} & \aer{0.001}{f}{\quad} & 589.35\\ 
Her~X$-$1 & 3821 & E & LMXB/NS & 2003-08-24T21:16:19 & 29.6 & $2.4-8.0$ & $0.3712\pm0.0035$ & \aer{0.001}{f}{\quad} & \aer{0.001}{f}{\quad} & 162.70\\ 
Her~X$-$1 & 3822 & F & LMXB/NS & 2003-12-09T23:42:48 & 29.7 & $2.4-8.0$ & $0.3707\pm0.0035$ & \aer{0.001}{f}{\quad} & \aer{0.001}{f}{\quad} & 217.72\\ 
Her~X$-$1 & 4375 & G & LMXB/NS & 2002-11-03T07:07:34 & 20.3 & $2.4-8.0$ & $1.7699\pm0.0093$ & \aer{0.169}{f}{\quad} & \aer{0.001}{f}{\quad} & 67.90\\ 
Her~X$-$1 & 4585 & H & LMXB/NS & 2004-11-26T06:09:20 & 19.8 & $2.4-8.0$ & $0.8919\pm0.0067$ & \multicolumn{2}{c}{\aer{0.169}{\quad}{\quad}} & 43.36\\ 
Her~X$-$1 & 6149 & I & LMXB/NS & 2004-11-29T15:07:41 & 21.8 & $2.4-8.0$ & $1.2689\pm0.0076$ & \multicolumn{2}{c}{\aer{0.169}{\quad}{\quad}} & 77.65\\ 
Her~X$-$1 & 6150 & J & LMXB/NS & 2004-12-01T08:53:43 & 21.7 & $2.4-8.0$ & $0.3133\pm0.0038$ & \aer{0.001}{f}{\quad} & \aer{0.001}{f}{\quad} & 94.77\\ 
LMC~X$-$4 & 9571 & A & HMXB/NS & 2007-09-01T16:19:49 & 46.5 & $2.4-8.0$ & $0.2181\pm0.0022$ & \multicolumn{2}{c}{\aer{1.17}{\quad}{\quad}} & 29.16\\ 
LMC~X$-$4 & 9573 & B & HMXB/NS & 2007-08-30T01:53:25 & 50.3 & $2.4-8.0$ & $0.0106\pm0.0005$ & \multicolumn{2}{c}{\aer{1.17}{\quad}{\quad}} & 56.56\\ 
LMC~X$-$4 & 9574 & C & HMXB/NS & 2007-08-31T10:50:21 & 43.5 & $2.4-8.0$ & $0.0225\pm0.0007$ & \multicolumn{2}{c}{\aer{1.17}{\quad}{\quad}} & 41.70\\ 
OAO~1657$-$415 & 12460 & A & HMXB/NS & 2011-05-17T13:29:11 & 48.9 & $2.4-8.0$ & $0.0397\pm0.0009$ & \multicolumn{2}{c}{\aer{15.0}{\quad}{\quad}} & 2395.52\\ 
OAO~1657$-$415 & 1947 & B & HMXB/NS & 2001-02-10T19:12:29 & 5.1 & $2.4-8.0$ & $0.2573\pm0.0071$ & \multicolumn{2}{c}{\aer{15.0}{\quad}{\quad}} & 42.86\\ 
Vela~X$-$1 & 102 & A & HMXB/NS & 2000-04-13T09:57:52 & 28.0 & $2.4-8.0$ & $0.0350\pm0.0011$ & \multicolumn{2}{c}{\aer{4.14}{\quad}{\quad}} & 154.90\\ 
Vela~X$-$1 & 14654 & B & HMXB/NS & 2013-07-30T16:54:07 & 45.9 & $2.4-8.0$ & $1.8966\pm0.0064$ & \multicolumn{2}{c}{\aer{4.14}{\quad}{\quad}} & 393.02\\ 
Vela~X$-$1 & 1926 & C & HMXB/NS & 2001-02-11T21:20:17 & 83.1 & $2.4-8.0$ & $0.0343\pm0.0006$ & \multicolumn{2}{c}{\aer{4.14}{\quad}{\quad}} & 789.41\\ 
Vela~X$-$1 & 1927 & D & HMXB/NS & 2001-02-07T09:57:17 & 29.4 & $2.4-8.0$ & $3.5611\pm0.0110$ & \multicolumn{2}{c}{\aer{4.14}{\quad}{\quad}} & 1794.20\\ 
Vela~X$-$1 & 1928 & E & HMXB/NS & 2001-02-05T05:29:55 & 29.6 & $2.4-8.0$ & $8.2778\pm0.0167$ & \multicolumn{2}{c}{\aer{4.14}{\quad}{\quad}} & 727.24\\ 
\enddata
\tablecomments{
Columns (1), (2), and (3) give the XRB system name, \chandra-HETG observation
ID, and associated alphabetical label used in this paper.
Column (4) distinguishes between High- vs. Low-Mass X-ray binaries, and
neutron-star vs. black-hole accretors.
Column (5) is the observation start time (UTC) recorded in 
the {\sc date-obs} header keyword. 
Column (6) is the exposure time (header keyword {\sc exposure})
corrected for all effects, including spatial ones such as vignetting.
Column (8) gives the count rate in the fitting band.
Columns (9) and (10)
give either the \citep[tabulated,][weighted average]{kalberla2005} 
foreground Galactic hydrogen column 
density used or fitted values for the two models in this study.
For completeness and clarity we note that for the
two objects with mostly low and untabulated values,
\cygxI\ and \herxI, the tabulated values are
\ten{7.21}{21} and \ten{0.169}{21}~\cunits, respectively.
However, in line with all tables, no \myt\ values are given
for non-detections. The superscript $^f$ indicates a fixed value.
%$^a$ indicates that no errors were calculated because the spherical
%fit reached \zfe~=~10\zsun.
Finally, column (11) gives our quantitative measure of \feka\ line
detectability in terms of the change in the $C$-statistic value between
a fit with a Gaussian fixed at an energy of 6.4 keV and FWHM of 100~\kmps\
and a fit lacking such a component (\scr{sec-detectability}).
$^1$: Also known as X1822$-$371; $^2$: Also known as X1908$+$075; $^3$: Also
known as 4U1223$-$62.
}  
\end{deluxetable*}

%% file: fig-NHZFe.tex
\begin{figure}
\hspace{-20pt}  
\includegraphics[scale=0.2,clip=true]{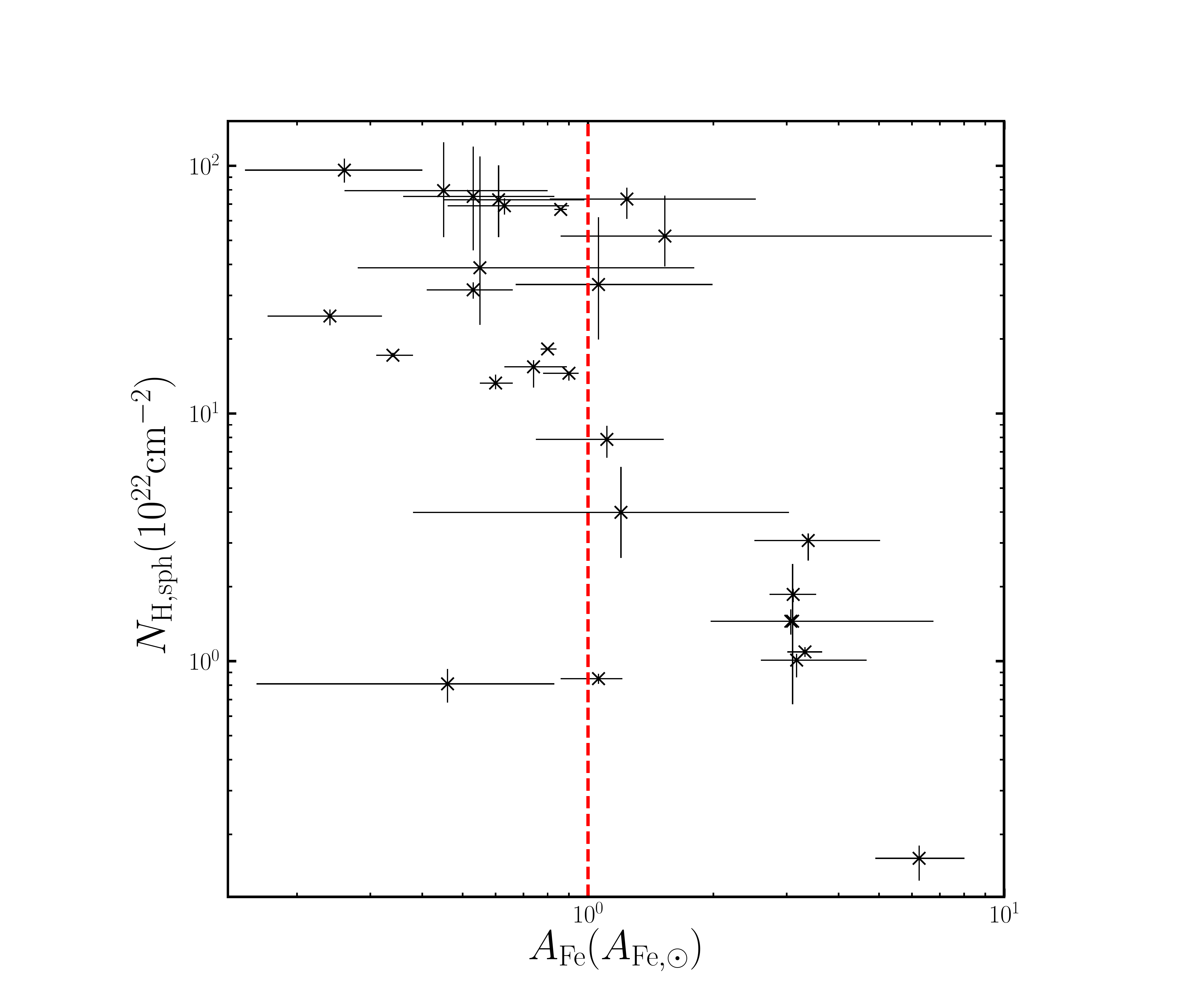}
\caption{\nhsph\ against \zfe\ for \bne\ model. Only points for which
\zfe\ was not a lower or upper limit, nor reaching \zfeten\ are
shown. The vertical dotted line marks \zfe~=~\zfesun.}
\label{fig-NHZFe}
\end{figure}

%% file: fig-compA.tex
\begin{figure*}
  \vspace{-2.5cm}
  \hspace{-2cm}
  \includegraphics[scale=1,clip=true]{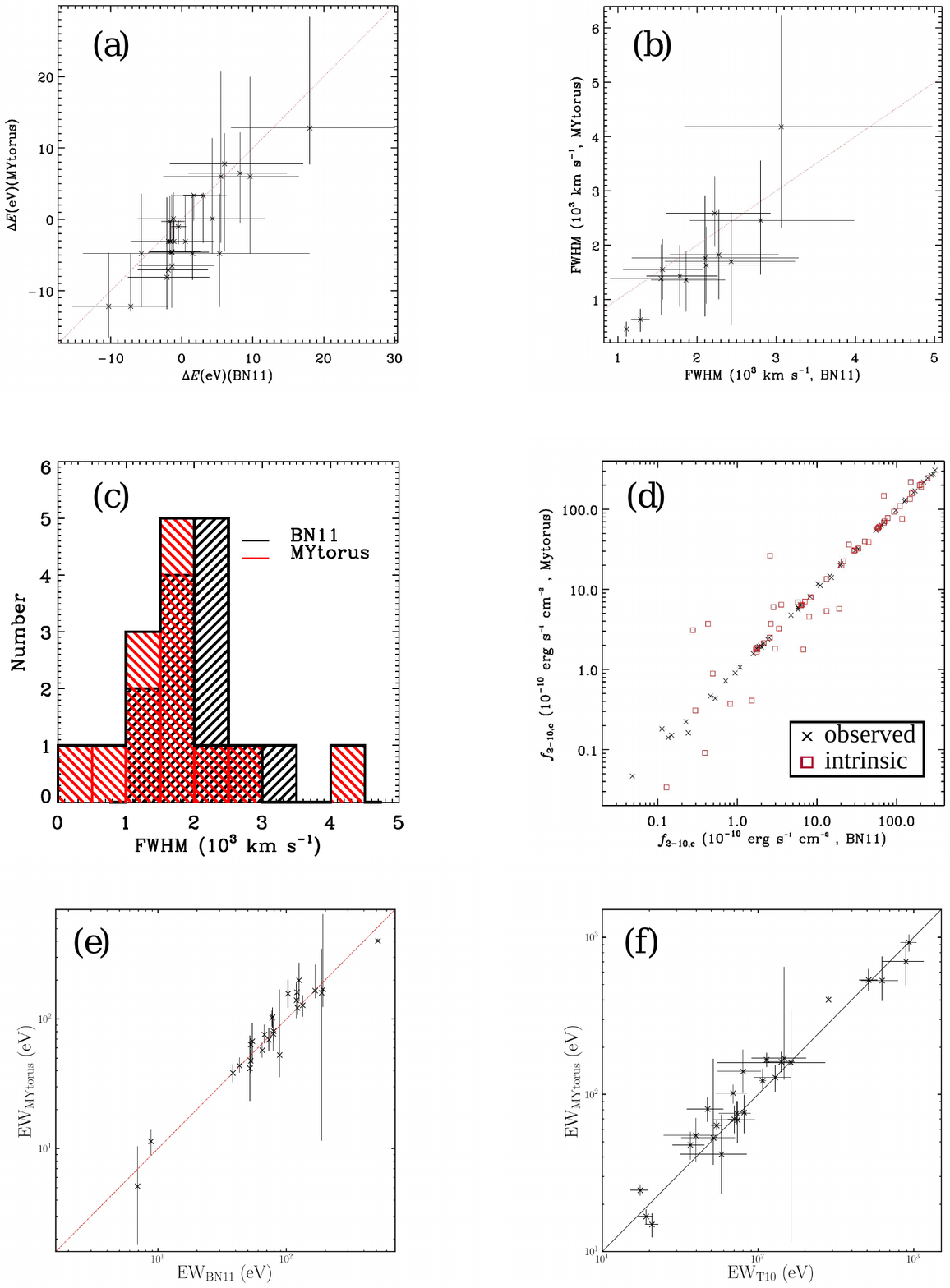}
   \vspace{-4.5cm}
 
\caption{
Comparisons of different fitted models. Panels (a), (b),
(d), (e), and (f) plot fitted values for \myt\ against the
\bne\ spherical model (or against \ctt, panel f) for \feka\ line
energy shifts, \feka\ line FWHM, continuum fluxes, and \feka\ line
EW, respectively. Panel (c) shows FWHM distributions for the
\bne\ (black) and \myt\ (red) models. Only line results for which
the line was detected at $\Delta C\ge6.63$ and \zfe\ was neither a
lower limit nor reached \zfeten\ are shown. In panels (e) and (f)
the point for \lmcxIVct\ is not shown, as the \myt\ result has a
very large upper 90\%\ uncertainty (\tr{tab-feka}), making it
essentially unconstrained.
}
\label{fig-comp}
\end{figure*}

%% file: taball_trans.tex
\startlongtable
\begin{deluxetable*}{ C{2cm} cccc cccc c}
  \tablecaption{Fitting results for the spherical model of
  \citet[][\bnet]{brightman2011}\label{tab-trans}}
\tabletypesize{\scriptsize}
\tablehead{
\colhead{Source} & 
\multicolumn{2}{c}{obs.} &
\colhead{$C$-stat(d.o.f)} &
\colhead{goodness} &
\colhead{$\Gamma_{\rm sph}$}&
\colhead{\zfe} &
\colhead{$\Gamma_{\rm soft}$}&
\colhead{$N_{\rm soft}/N_{\rm sph}$}&
\colhead{\nhsph} 
\\
\colhead{}&
\colhead{}&
\colhead{}& %for letter
\colhead{}&
\colhead{\%} &
\colhead{}&
\colhead{(\zfesun)}&
\colhead{}&
\colhead{(log)}&
\colhead{($10^{22}$~\cunits})
}
\colnumbers
\startdata
4U1700$-$37 & 657 & A & 1468.75(1439) & 48.7 & \aer{1.21}{+0.11}{-0.07} & \aer{0.60}{+0.06}{-0.05} & \aer{1.21}{t}{} & \aer{-1.17}{+0.33}{-0.90} & \aer{13.27}{+1.07}{-0.75}\\ 
4U1822$-$371 & 671 & A & 1508.48(1440) & 86.0 & \aer{1.00}{f}{} & \aer{3.07}{+1.17}{-0.67} & \aer{\ldots}{}{} & \aer{\ldots}{}{} & \aer{1.45}{+0.17}{-0.17}\\ 
4U1822$-$371 & 9076 & B & 1494.82(1438) & 81.2 & \aer{<1.06}{c}{} & \aer{3.17}{+1.50}{-0.57} & \aer{\ldots}{}{} & \aer{\ldots}{}{} & \aer{1.01}{+0.06}{-0.15}\\ 
4U1822$-$371 & 9858 & C & 1567.34(1438) & 98.9 & \aer{<1.02}{c}{} & \aer{3.32}{+0.33}{-0.31} & \aer{\ldots}{}{} & \aer{\ldots}{}{} & \aer{1.09}{+0.05}{-0.05}\\ 
4U1908$+$075 & 5476 & A & 1522.94(1440) & 46.1 & \aer{<1.11}{c}{} & \aer{3.38}{+1.65}{-0.87} & \aer{\ldots}{}{} & \aer{\ldots}{}{} & \aer{3.07}{+0.21}{-0.52}\\ 
4U1908$+$075 & 5477 & B & 1312.14(1440) & 98.3 & \aer{1.40}{+0.24}{-0.24} & \aer{0.24}{+0.08}{-0.07} & \aer{\ldots}{}{} & \aer{\ldots}{}{} & \aer{24.75}{+1.57}{-2.05}\\ 
4U1908$+$075 & 6336 & C & 1572.95(1440) & 79.5 & \aer{1.39}{+0.19}{-0.21} & \aer{1.11}{+0.41}{-0.36} & \aer{\ldots}{}{} & \aer{\ldots}{}{} & \aer{7.86}{+1.06}{-1.24}\\ 
Cen~X$-$3 & 1943 & A & 2072.94(1435) & 100.0 & \aer{1.00}{a}{} & \aer{10.00}{a}{} & \aer{\ldots}{}{} & \aer{\ldots}{}{} & \aer{0.07}{a}{}\\ 
Cen~X$-$3 & 705 & B & 1587.92(1432) & 92.7 & \aer{<1.08}{c}{} & \aer{0.74}{+0.15}{-0.11} & \aer{-0.45}{+0.15}{-0.13} & \aer{-1.51}{+0.33}{-0.88} & \aer{15.43}{+0.99}{-2.72}\\ 
Cen~X$-$3 & 7511 & C & 1714.11(1434) & 100.0 & \aer{1.29}{a}{} & \aer{10.00}{a}{} & \aer{-2.28}{a}{} & \aer{-3.44}{a}{} & \aer{0.78}{a}{}\\ 
Cir~X$-$1 & 12235 & A & 1611.18(1442) & 84.6 & \aer{1.49}{+0.15}{-0.16} & \aer{<1.10}{}{} & \aer{\ldots}{}{} & \aer{\ldots}{}{} & \aer{3.15}{+0.79}{-0.87}\\ 
Cir~X$-$1 & 1905 & B & 1722.32(1442) & 100.0 & \aer{1.86}{+0.03}{-0.03} & \aer{0.46}{+0.37}{-0.30} & \aer{\ldots}{}{} & \aer{\ldots}{}{} & \aer{0.81}{+0.12}{-0.13}\\ 
Cir~X$-$1 & 1906 & C & 1523.77(1442) & 91.3 & \aer{1.56}{+0.02}{-0.02} & \aer{<0.23}{}{} & \aer{\ldots}{}{} & \aer{\ldots}{}{} & \aer{1.02}{+0.11}{-0.11}\\ 
Cir~X$-$1 & 1907 & D & 1660.01(1442) & 100.0 & \aer{1.59}{+0.03}{-0.02} & \aer{<0.38}{}{} & \aer{\ldots}{}{} & \aer{\ldots}{}{} & \aer{1.35}{+0.13}{-0.13}\\ 
Cir~X$-$1 & 8993 & G & 1557.56(1439) & 86.2 & \aer{<1.65}{c}{} & \aer{1.20}{+1.84}{-0.82} & \aer{\ldots}{}{} & \aer{\ldots}{}{} & \aer{3.99}{+2.10}{-1.38}\\ 
Cyg~X$-$1 & 11044 & A & 1760.80(1441) & 100.0 & \aer{1.57}{a}{} & \aer{10.00}{a}{} & \aer{\ldots}{}{} & \aer{\ldots}{}{} & \aer{0.03}{a}{}\\ 
Cyg~X$-$1 & 12313 & B & 1554.00(1440) & 96.8 & \aer{1.10}{a}{} & \aer{10.00}{a}{} & \aer{4.42}{a}{} & \aer{1.97}{a}{} & \aer{0.07}{a}{}\\ 
Cyg~X$-$1 & 12314 & C & 1595.79(1442) & 99.9 & \aer{2.88}{a}{} & \aer{10.00}{a}{} & \aer{\ldots}{}{} & \aer{\ldots}{}{} & \aer{0.01}{a}{}\\ 
Cyg~X$-$1 & 12472 & D & 1733.94(1440) & 100.0 & \aer{1.40}{a}{} & \aer{10.00}{a}{} & \aer{4.69}{a}{} & \aer{2.32}{a}{} & \aer{0.25}{a}{}\\ 
Cyg~X$-$1 & 13219 & E & 1556.91(1440) & 97.8 & \aer{1.31}{a}{} & \aer{10.00}{a}{} & \aer{4.40}{a}{} & \aer{1.93}{a}{} & \aer{0.11}{a}{}\\ 
Cyg~X$-$1 & 1511 & F & 1580.01(1442) & 99.0 & \aer{1.77}{+0.02}{-0.03} & \aer{>2.15}{}{} & \aer{\ldots}{}{} & \aer{\ldots}{}{} & \aer{0.22}{+0.10}{-0.14}\\ 
Cyg~X$-$1 & 2415 & G & 1704.36(1439) & 100.0 & \aer{2.71}{a}{} & \aer{10.00}{a}{} & \aer{1.01}{a}{} & \aer{-0.88}{a}{} & \aer{0.09}{a}{}\\ 
Cyg~X$-$1 & 2741 & H & 1504.55(1440) & 87.7 & \aer{1.17}{a}{} & \aer{10.00}{a}{} & \aer{4.27}{a}{} & \aer{1.74}{a}{} & \aer{0.10}{a}{}\\ 
Cyg~X$-$1 & 2742 & I & 1453.70(1440) & 41.5 & \aer{1.00}{a}{} & \aer{10.00}{a}{} & \aer{4.21}{a}{} & \aer{1.88}{a}{} & \aer{0.06}{a}{}\\ 
Cyg~X$-$1 & 2743 & J & 1446.95(1439) & 37.3 & \aer{1.00}{f}{} & \aer{10.00}{a}{} & \aer{4.73}{a}{} & \aer{2.58}{a}{} & \aer{0.17}{a}{}\\ 
Cyg~X$-$1 & 3407 & K & 1663.29(1439) & 100.0 & \aer{1.79}{+0.03}{-0.02} & \aer{6.24}{+1.78}{-1.34} & \aer{4.18}{+0.02}{-0.02} & \aer{0.96}{+0.30}{-1.78} & \aer{0.16}{+0.02}{-0.03}\\ 
Cyg~X$-$1 & 3814 & M & 1474.62(1380) & 93.6 & \aer{1.39}{a}{} & \aer{10.00}{a}{} & \aer{\ldots}{}{} & \aer{\ldots}{}{} & \aer{0.16}{a}{}\\ 
Cyg~X$-$1 & 8525 & N & 1532.08(1350) & 99.9 & \aer{1.39}{+0.01}{-0.01} & \aer{1.06}{+0.15}{-0.20} & \aer{\ldots}{}{} & \aer{\ldots}{}{} & \aer{0.85}{+0.04}{-0.04}\\ 
Cyg~X$-$1 & 9847 & O & 1543.94(1368) & 100.0 & \aer{1.36}{+0.02}{-0.02} & \aer{>3.94}{}{} & \aer{\ldots}{}{} & \aer{\ldots}{}{} & \aer{0.24}{+0.15}{-0.07}\\ 
Cyg~X$-$1 & 3815 & P & 1932.04(1315) & 100.0 & \aer{1.97}{+0.02}{-0.01} & \aer{>3.25}{}{} & \aer{>9.72}{}{} & \aer{2.15}{+0.32}{-1.37} & \aer{0.22}{+0.23}{-0.04}\\ 
Cyg~X$-$3 & 101 & A & 1689.23(1436) & 100.0 & \aer{1.00}{f}{} & \aer{10.00}{a}{} & \aer{\ldots}{}{} & \aer{\ldots}{}{} & \aer{0.78}{a}{}\\ 
$\gamma$~Cas & 1895 & A & 1459.02(1377) & 77.7 & \aer{1.78}{a}{} & \aer{10.00}{a}{} & \aer{\ldots}{}{} & \aer{\ldots}{}{} & \aer{0.25}{a}{}\\ 
GX~301$-$2 & 2733 & B & 1849.36(1406) & 100.0 & \aer{1.20}{+0.08}{-0.08} & \aer{0.86}{+0.03}{-0.03} & \aer{\ldots}{}{} & \aer{\ldots}{}{} & \aer{66.68}{+1.19}{-0.87}\\ 
GX~301$-$2 & 3433 & C & 1570.46(1424) & 99.6 & \aer{<1.01}{c}{} & \aer{0.80}{+0.04}{-0.03} & \aer{\ldots}{}{} & \aer{\ldots}{}{} & \aer{18.22}{+0.16}{-0.16}\\ 
GX~1$+$4 & 2710 & A & 1518.76(1438) & 87.5 & \aer{<1.09}{c}{} & \aer{3.11}{+0.42}{-0.38} & \aer{\ldots}{}{} & \aer{\ldots}{}{} & \aer{1.86}{+0.10}{-0.13}\\ 
GX~1$+$4 & 2744 & B & 1433.02(1440) & 40.0 & \aer{1.05}{a}{} & \aer{10.00}{a}{} & \aer{\ldots}{}{} & \aer{\ldots}{}{} & \aer{1.56}{a}{}\\ 
Her~X$-$1 & 2703 & A & 1647.84(1439) & 81.0 & \aer{<3.02}{c}{} & \aer{0.53}{+0.30}{-0.17} & \aer{0.92}{+0.34}{-0.29} & \aer{-1.47}{+0.42}{-0.32} & \aer{75.23}{+44.36}{-29.67}\\ 
Her~X$-$1 & 2704 & B & 1557.74(1440) & 98.8 & \aer{1.00}{a}{} & \aer{10.00}{a}{} & \aer{\ldots}{}{} & \aer{\ldots}{}{} & \aer{0.15}{a}{}\\ 
Her~X$-$1 & 2705 & C & 1615.42(1438) & 38.9 & \aer{1.26}{+0.09}{-0.09} & \aer{0.26}{+0.14}{-0.11} & \aer{1.37}{+0.16}{-0.18} & \aer{-0.59}{+0.35}{-0.79} & \aer{96.06}{+10.90}{-10.56}\\ 
Her~X$-$1 & 2749 & D & 1527.76(1436) & 33.8 & \aer{1.00}{a}{} & \aer{10.00}{a}{} & \aer{0.40}{a}{} & \aer{-0.75}{a}{} & \aer{37.47}{a}{}\\ 
Her~X$-$1 & 3821 & E & 1539.99(1438) & 77.7 & \aer{<2.16}{c}{} & \aer{1.53}{+7.80}{-0.67} & \aer{0.23}{+0.28}{-0.18} & \aer{-1.45}{+0.72}{-0.05} & \aer{52.05}{+23.82}{-12.82}\\ 
Her~X$-$1 & 3822 & F & 1547.62(1436) & 43.4 & \aer{1.00}{f}{} & \aer{0.71}{+0.10}{-0.10} & \aer{0.43}{+0.27}{-0.10} & \aer{-0.98}{+0.35}{-1.02} & \aer{>36.40}{}{}\\ 
Her~X$-$1 & 4375 & G & 1476.10(1438) & 57.2 & \aer{1.00}{f}{} & \aer{1.06}{+0.93}{-0.39} & \aer{1.19}{+0.26}{-0.14} & \aer{0.42}{+0.36}{-0.60} & \aer{33.17}{+28.84}{-13.26}\\ 
Her~X$-$1 & 4585 & H & 1496.73(1438) & 52.4 & \aer{3.00}{f}{} & \aer{0.63}{+0.27}{-0.17} & \aer{0.26}{+0.19}{-0.18} & \aer{-2.78}{+0.34}{-0.83} & \aer{68.99}{+5.01}{-5.42}\\ 
Her~X$-$1 & 6149 & I & 1600.65(1438) & 96.5 & \aer{>2.55}{}{} & \aer{1.24}{+1.29}{-0.43} & \aer{0.44}{+0.11}{-0.11} & \aer{-2.31}{+0.32}{-0.59} & \aer{73.44}{+8.21}{-12.42}\\ 
Her~X$-$1 & 6150 & J & 1555.47(1438) & 46.5 & \aer{>1.39}{}{} & \aer{0.61}{+0.37}{-0.16} & \aer{0.22}{+0.31}{-0.30} & \aer{-2.38}{+0.38}{-0.52} & \aer{72.95}{+27.55}{-21.43}\\ 
LMC~X$-$4 & 9571 & A & 1583.06(1438) & 86.9 & \aer{1.00}{f}{} & \aer{0.55}{+1.25}{-0.27} & \aer{0.79}{+0.43}{-0.24} & \aer{0.04}{+0.44}{-0.28} & \aer{38.77}{+70.41}{-16.00}\\ 
LMC~X$-$4 & 9573 & B & 1069.11(1439) & 22.2 & \aer{1.00}{a}{} & \aer{10.00}{a}{} & \aer{0.23}{a}{} & \aer{-0.98}{a}{} & \aer{32.83}{a}{}\\ 
LMC~X$-$4 & 9574 & C & 1297.27(1439) & 23.4 & \aer{1.00}{f}{} & \aer{0.45}{+0.35}{-0.19} & \aer{0.33}{+0.74}{-0.50} & \aer{-1.25}{+0.58}{-0.21} & \aer{79.42}{+45.22}{-27.85}\\ 
OAO~1657$-$415 & 12460 & A & 1154.82(1438) & 80.2 & \aer{1.00}{a}{} & \aer{10.00}{a}{} & \aer{-1.86}{a}{} & \aer{-3.52}{a}{} & \aer{51.88}{a}{}\\ 
OAO~1657$-$415 & 1947 & B & 1006.02(1438) & 55.5 & \aer{<1.22}{c}{} & \aer{0.53}{+0.13}{-0.12} & \aer{>9.19}{}{} & \aer{1.50}{+0.50}{-0.17} & \aer{31.57}{+2.28}{-2.48}\\ 
Vela~X$-$1 & 102 & A & 505.41(1438) & 100.0 & \aer{1.00}{a}{} & \aer{10.00}{a}{} & \aer{0.50}{a}{} & \aer{-0.86}{a}{} & \aer{28.24}{a}{}\\ 
Vela~X$-$1 & 14654 & B & 1757.05(1440) & 100.0 & \aer{1.20}{+0.07}{-0.04} & \aer{0.34}{+0.04}{-0.03} & \aer{\ldots}{}{} & \aer{\ldots}{}{} & \aer{17.19}{+0.41}{-0.31}\\ 
Vela~X$-$1 & 1926 & C & 1831.44(1439) & 100.0 & \aer{1.00}{a}{} & \aer{10.00}{a}{} & \aer{0.59}{a}{} & \aer{-0.86}{a}{} & \aer{30.44}{a}{}\\ 
Vela~X$-$1 & 1927 & D & 1635.06(1438) & 100.0 & \aer{<1.66}{c}{} & \aer{0.90}{+0.05}{-0.12} & \aer{0.09}{+0.06}{-0.26} & \aer{-1.53}{+0.37}{-1.29} & \aer{14.54}{+0.25}{-0.96}\\ 
Vela~X$-$1 & 1928 & E & 1452.55(1439) & 55.8 & \aer{1.00}{f}{} & \aer{3.10}{+3.66}{-1.13} & \aer{3.55}{+1.59}{-0.65} & \aer{0.31}{+0.45}{-0.46} & \aer{1.45}{+1.02}{-0.78}\\ 
\enddata
\tablecomments{
Columns (1), (2), and (3) give the XRB system name, \chandra-HETG
observation ID, and associated alphabetical label used in this paper.
Columns (4) and (5) give the best-fit value of the $C$-statistic, and
associated degrees of freedom, and the output of the \xspec\ {\tt
goodness} command for 2000 random realizations of the data. Column (6)
gives the \bnet\ power-law continuum photon index; column (7) the \bnet\
iron abundance relative to solar; column (8) the photon index for any
additional, ``soft'' power law continuum. Column (9) gives the ratio of the
normalizations of the soft-to-\bnet\ power law continua.
Column (10) gives the
model's equivalent hydrogen column density.  The line shift and line
width parameters for each fit are shown in \tr{tab-feka}. $^{f}$:
Parameter frozen; $^{a}$: No errors calculated because upper model
limit of Fe abundance reached. d.o.f.: degrees of freedom;
%$^{b}$: Actual value is \ten{-3.6}{-3};
$^c$: The lower limit on \gammasph\
($=1.00$) is set by limitations of the model; $^{t}$: Parameter is
tied to \gammasph.}
\end{deluxetable*}

%% file: taball_myun.tex
\startlongtable
\begin{deluxetable*}{ C{2cm} cccc cccc cccc c}
\tablecaption{Fitting results for the uncoupled \myt\ model \citep{murphy2009}\label{tab-myun}}
%\vspace{-0.5cm}
\tablewidth{0pc}
\tabletypesize{\scriptsize}
\tablehead{
\colhead{Source} & 
\multicolumn{2}{c}{obs.} &
\colhead{$C$-stat(d.o.f)}&
\colhead{goodness} &
\colhead{$\Gamma_{\rm Z}$}&
\colhead{$\Gamma_{\rm S}$}&
\colhead{$\Gamma_{\rm soft}$}&
\colhead{$N_{\rm soft}/N_{\rm Z}$}&
\colhead{$A_S$} &
\colhead{\nhz} &
\colhead{\nhs} \\
\colhead{}& 
\colhead{}&
\colhead{}& %for letter
\colhead{}&
\colhead{\%} &
\colhead{}&
\colhead{}&
\colhead{}&
\colhead{(log)}&
\colhead{}&
\colhead{($10^{22}$~\cunits)} &
\colhead{($10^{22}$~\cunits)}
\\
\colhead{(1)}&
\multicolumn{2}{c}{(2)}&
\colhead{(3)}&
\colhead{(4)}&
\colhead{(5)}&
\colhead{(6)}&
\colhead{(7)}&
\colhead{(8)}&
\colhead{(9)}&
\colhead{(10)}&
\colhead{(11)}
}
\startdata
4U1700$-$37 & 657 & A & 1463.57(1437) & 46.4 & \aer{1.05}{f}{} & \aer{1.40}{f}{} & \aer{1.05}{t}{} & \aer{-1.16}{+0.33}{-0.88} & \aer{1.16}{+0.26}{-0.21} & \aer{13.02}{+1.00}{-1.01} & \aer{44}{f}{}\\ 
4U1822$-$371 & 671 & A & 1501.90(1438) & 78.5 & \aer{0.93}{+0.07}{-0.07} & \aer{1.40}{f}{} & \aer{\ldots}{}{} & \aer{\ldots}{}{} & \aer{1.76}{+0.38}{-0.34} & \aer{1.75}{+0.31}{-0.31} & \aer{1000}{f}{}\\ 
4U1822$-$371 & 9076 & B & 1495.11(1436) & 79.8 & \aer{1.06}{+0.02}{-0.02} & \aer{1.40}{f}{} & \aer{\ldots}{}{} & \aer{\ldots}{}{} & \aer{1.12}{+0.19}{-0.17} & \aer{1.66}{+0.11}{-0.08} & \aer{1000}{f}{}\\ 
4U1822$-$371 & 9858 & C & 1567.21(1436) & 98.3 & \aer{1.04}{+0.02}{-0.02} & \aer{1.40}{f}{} & \aer{\ldots}{}{} & \aer{\ldots}{}{} & \aer{1.33}{+0.20}{-0.16} & \aer{1.72}{+0.14}{-0.08} & \aer{1000}{f}{}\\ 
4U1908$+$075 & 5476 & A & 1517.84(1438) & 40.4 & \aer{1.10}{+0.17}{-0.16} & \aer{1.40}{f}{} & \aer{\ldots}{}{} & \aer{\ldots}{}{} & \aer{3.28}{+1.23}{-0.91} & \aer{4.77}{+0.85}{-0.84} & \aer{1000}{f}{}\\ 
4U1908$+$075 & 5477 & B & 1289.84(1438) & 94.7 & \aer{0.84}{+0.25}{-0.25} & \aer{1.40}{f}{} & \aer{\ldots}{}{} & \aer{\ldots}{}{} & \aer{1.07}{+0.84}{-0.47} & \aer{21.50}{+1.50}{-1.10} & \aer{1000}{f}{}\\ 
4U1908$+$075 & 6336 & C & 1571.48(1438) & 80.0 & \aer{1.39}{+0.08}{-0.08} & \aer{1.40}{f}{} & \aer{\ldots}{}{} & \aer{\ldots}{}{} & \aer{1.01}{+0.29}{-0.27} & \aer{8.89}{+0.44}{-0.43} & \aer{1000}{f}{}\\ 
Cen~X$-$3 & 1943 & A & 1780.34(1434) & 100.0 & \aer{1.05}{+0.01}{-0.01} & \aer{1.40}{f}{} & \aer{\ldots}{}{} & \aer{\ldots}{}{} & \aer{0.74}{+0.06}{-0.06} & \aer{0.09}{+0.03}{-0.03} & \aer{1000}{f}{}\\ 
Cen~X$-$3 & 705 & B & 1565.02(1430) & 84.4 & \aer{0.33}{+0.04}{-0.04} & \aer{1.40}{f}{} & \aer{1.09}{+0.13}{-0.13} & \aer{-0.57}{+0.33}{-0.95} & \aer{6.84}{+1.26}{-1.19} & \aer{12.41}{+0.69}{-0.67} & \aer{1000}{f}{}\\ 
Cen~X$-$3 & 7511 & C & 1610.44(1428) & 99.7 & \aer{1.59}{+0.08}{-0.08} & \aer{1.40}{f}{} & \aer{0.24}{f}{} & \aer{-0.75}{+0.32}{-1.20} & \aer{3.14}{+0.22}{-0.21} & \aer{1.34}{+0.15}{-0.15} & \aer{1000}{f}{}\\ 
%Cir~X$-$1 & 12235 & A & 1628.08(1443) & 98.9 & \aer{1.26}{+0.12}{-0.12} & \aer{1.40}{f}{} & \aer{\ldots}{}{} & \aer{\ldots}{}{} & \aer{1.00}{f}{} & \aer{2.50}{+0.57}{-0.60} & \aer{1000}{f}{}\\ 
%Cir~X$-$1 & 1905 & B & 1717.88(1442) & 100.0 & \aer{1.85}{+0.03}{-0.03} & \aer{2.60}{f}{} & \aer{\ldots}{}{} & \aer{\ldots}{}{} & \aer{0.30}{+0.29}{-0.29} & \aer{1.22}{+0.11}{-0.11} & \aer{1000}{f}{}\\ 
%Cir~X$-$1 & 1906 & C & 1518.60(1442) & 91.8 & \aer{1.54}{+0.03}{-0.02} & \aer{2.60}{f}{} & \aer{\ldots}{}{} & \aer{\ldots}{}{} & \aer{1.00}{f}{} & \aer{1.35}{+0.12}{-0.12} & \aer{1000}{f}{}\\ 
%Cir~X$-$1 & 1907 & D & 1649.85(1442) & 100.0 & \aer{1.58}{+0.03}{-0.03} & \aer{2.60}{f}{} & \aer{\ldots}{}{} & \aer{\ldots}{}{} & \aer{<0.94}{}{} & \aer{1.73}{+0.14}{-0.13} & \aer{1000}{f}{}\\ 
%Cir~X$-$1 & 8993 & G & 1583.74(1442) & 77.0 & \aer{1.04}{+0.31}{-0.35} & \aer{1.40}{f}{} & \aer{\ldots}{}{} & \aer{\ldots}{}{} & \aer{1.00}{f}{} & \aer{4.45}{+1.61}{-1.78} & \aer{1000}{f}{}\\ 
Cyg~X$-$1 & 11044 & A & 1692.59(1440) & 100.0 & \aer{1.55}{+0.01}{-0.01} & \aer{1.40}{f}{} & \aer{\ldots}{}{} & \aer{\ldots}{}{} & \aer{0.15}{+0.03}{-0.03} & \aer{<0.01}{}{} & \aer{1000}{f}{}\\ 
%Cyg~X$-$1 & 12313 & B & 1659.81(1531) & 98.8 & \aer{0.94}{+0.21}{-0.24} & \aer{2.60}{f}{} & \aer{4.19}{+0.21}{-0.18} & \aer{2.04}{+0.39}{-0.56} & \aer{7.11}{+7.48}{-4.52} & \aer{\ldots}{}{} & \aer{1000}{f}{}\\ 
%Cyg~X$-$1 & 12314 & C & 1579.28(1443) & 61.0 & \aer{2.91}{+0.04}{-0.04} & \aer{2.60}{f}{} & \aer{\ldots}{}{} & \aer{\ldots}{}{} & \aer{0.59}{+0.28}{-0.26} & \aer{\ldots}{}{} & \aer{1000}{f}{}\\ 
%Cyg~X$-$1 & 12472 & D & 1689.78(1442) & 100.0 & \aer{1.24}{+0.30}{-0.31} & \aer{2.60}{f}{} & \aer{4.67}{+0.16}{-0.13} & \aer{2.43}{+0.44}{-0.39} & \aer{1.00}{f}{} & \aer{\ldots}{}{} & \aer{1000}{f}{}\\ 
%Cyg~X$-$1 & 13219 & E & 1556.15(1441) & 97.9 & \aer{1.32}{+0.21}{-0.22} & \aer{2.60}{f}{} & \aer{4.41}{+0.20}{-0.18} & \aer{1.92}{+0.24}{-0.57} & \aer{4.06}{+3.82}{-2.34} & \aer{\ldots}{}{} & \aer{1000}{f}{}\\ 
Cyg~X$-$1 & 1511 & F & 1576.05(1438) & 98.8 & \aer{1.78}{+0.03}{-0.03} & \aer{1.80}{f}{} & \aer{\ldots}{}{} & \aer{\ldots}{}{} & \aer{0.21}{+0.08}{-0.07} & \aer{0.37}{+0.14}{-0.14} & \aer{1000}{f}{}\\ 
Cyg~X$-$1 & 2415 & G & 2051.46(1440) & 100.0 & \aer{1.70}{+0.01}{-0.01} & \aer{1.40}{f}{} & \aer{\ldots}{}{} & \aer{\ldots}{}{} & \aer{0.12}{+0.02}{-0.02} & \aer{<0.01}{}{} & \aer{1000}{f}{}\\ 
%Cyg~X$-$1 & 2741 & H & 1502.70(1441) & 76.8 & \aer{1.00}{+0.26}{-0.30} & \aer{2.60}{f}{} & \aer{4.07}{+0.31}{-0.26} & \aer{1.84}{+0.42}{-0.51} & \aer{7.55}{+9.96}{-5.02} & \aer{\ldots}{}{} & \aer{1000}{f}{}\\ 
%Cyg~X$-$1 & 2742 & I & 1402.17(1441) & 28.4 & \aer{1.24}{+0.37}{-0.33} & \aer{2.60}{f}{} & \aer{3.31}{+0.10}{-0.08} & \aer{0.52}{+0.31}{-1.60} & \aer{<3.75}{}{} & \aer{27.24}{+6.90}{-6.00} & \aer{1000}{f}{}\\ 
%Cyg~X$-$1 & 2743 & J & 1451.70(1441) & 71.8 & \aer{1.12}{+0.43}{-0.48} & \aer{2.60}{f}{} & \aer{4.47}{+0.27}{-0.21} & \aer{2.32}{+0.52}{-0.29} & \aer{10.00}{+23.70}{-6.76} & \aer{\ldots}{}{} & \aer{1000}{f}{}\\ 
Cyg~X$-$1 & 3407 & K & 1590.75(1436) & 99.9 & \aer{1.44}{+0.25}{-0.23} & \aer{2.60}{f}{} & \aer{2.77}{+0.06}{-0.04} & \aer{1.11}{+0.42}{-0.36} & \aer{1.94}{+2.02}{-1.25} & \aer{25.42}{+5.49}{-4.97} & \aer{1000}{f}{}\\ 
Cyg~X$-$1 & 3814 & M & 1471.92(1378) & 93.3 & \aer{1.41}{+0.02}{-0.02} & \aer{2.60}{f}{} & \aer{\ldots}{}{} & \aer{\ldots}{}{} & \aer{4.44}{+0.79}{-0.76} & \aer{0.43}{+0.09}{-0.09} & \aer{1000}{f}{}\\ 
Cyg~X$-$1 & 8525 & N & 1518.46(1348) & 99.8 & \aer{1.40}{+0.01}{-0.01} & \aer{2.60}{f}{} & \aer{\ldots}{}{} & \aer{\ldots}{}{} & \aer{2.94}{+0.64}{-0.64} & \aer{1.01}{+0.04}{-0.04} & \aer{1000}{f}{}\\ 
Cyg~X$-$1 & 9847 & O & 1549.62(1366) & 99.8 & \aer{1.37}{+0.03}{-0.03} & \aer{2.60}{f}{} & \aer{\ldots}{}{} & \aer{\ldots}{}{} & \aer{4.05}{+1.82}{-1.05} & \aer{0.58}{+0.12}{-0.12} & \aer{1000}{f}{}\\ 
Cyg~X$-$1 & 3815 & P & 1921.64(1314) & 100.0 & \aer{1.98}{+0.02}{-0.02} & \aer{2.60}{f}{} & \aer{>9.76}{}{} & \aer{2.13}{+0.32}{-1.36} & \aer{1.46}{+0.17}{-0.17} & \aer{1.72}{+0.13}{-0.12} & \aer{1000}{f}{}\\ 
Cyg~X$-$3 & 101 & A & 1692.00(1433) & 99.9 & \aer{1.18}{+0.05}{-0.05} & \aer{1.40}{f}{} & \aer{\ldots}{}{} & \aer{\ldots}{}{} & \aer{2.26}{+0.71}{-0.56} & \aer{2.47}{+0.23}{-0.23} & \aer{1000}{f}{}\\ 
$\gamma$~Cas & 1895 & A & 1442.21(1374) & 61.0 & \aer{1.72}{+0.04}{-0.04} & \aer{1.40}{f}{} & \aer{\ldots}{}{} & \aer{\ldots}{}{} & \aer{0.44}{+0.13}{-0.14} & \aer{<0.33}{}{} & \aer{1000}{f}{}\\ 
GX~301$-$2 & 2733 & B & 2373.86(1405) & 100.0 & \aer{0.37}{+0.04}{-0.01} & \aer{1.40}{f}{} & \aer{\ldots}{}{} & \aer{\ldots}{}{} & \aer{2.87}{+0.13}{-0.06} & \aer{91.79}{+1.08}{-0.91} & \aer{200}{+2}{-5}\\ 
GX~301$-$2 & 3433 & C & 1448.51(1422) & 22.3 & \aer{0.78}{+0.05}{-0.06} & \aer{1.40}{f}{} & \aer{\ldots}{}{} & \aer{\ldots}{}{} & \aer{4.90}{+0.52}{-0.52} & \aer{19.01}{+0.40}{-0.44} & \aer{1000}{f}{}\\ 
GX~1$+$4 & 2710 & A & 1524.86(1436) & 87.5 & \aer{1.09}{+0.04}{-0.04} & \aer{1.40}{f}{} & \aer{\ldots}{}{} & \aer{\ldots}{}{} & \aer{1.84}{+0.36}{-0.28} & \aer{2.79}{+0.19}{-0.19} & \aer{1000}{f}{}\\ 
GX~1$+$4 & 2744 & B & 1428.65(1438) & 26.9 & \aer{1.24}{+0.10}{-0.10} & \aer{1.40}{f}{} & \aer{\ldots}{}{} & \aer{\ldots}{}{} & \aer{4.22}{+0.82}{-0.69} & \aer{4.11}{+0.34}{-0.38} & \aer{1000}{f}{}\\ 
Her~X$-$1 & 2703 & A & 1684.08(1439) & 98.9 & \aer{0.26}{+0.11}{-0.11} & \aer{1.40}{f}{} & \aer{\ldots}{}{} & \aer{\ldots}{}{} & \aer{24.49}{+6.71}{-5.59} & \aer{<0.19}{}{} & \aer{1000}{f}{}\\ 
Her~X$-$1 & 2704 & B & 1516.57(1439) & 91.1 & \aer{0.94}{+0.02}{-0.02} & \aer{1.40}{f}{} & \aer{\ldots}{}{} & \aer{\ldots}{}{} & \aer{0.98}{+0.21}{-0.20} & \aer{<0.05}{}{} & \aer{1000}{f}{}\\ 
Her~X$-$1 & 2705 & C & 1640.76(1439) & 85.0 & \aer{0.87}{+0.14}{-0.14} & \aer{1.40}{f}{} & \aer{\ldots}{}{} & \aer{\ldots}{}{} & \aer{8.48}{+3.11}{-2.65} & \aer{<0.21}{}{} & \aer{1000}{f}{}\\ 
Her~X$-$1 & 2749 & D & 1534.06(1438) & 43.2 & \aer{0.43}{+0.11}{-0.11} & \aer{1.40}{f}{} & \aer{\ldots}{}{} & \aer{\ldots}{}{} & \aer{65.28}{+11.37}{-9.63} & \aer{<0.28}{}{} & \aer{1000}{f}{}\\ 
Her~X$-$1 & 3821 & E & 1570.57(1439) & 94.8 & \aer{-0.11}{+0.06}{-0.06} & \aer{1.40}{f}{} & \aer{\ldots}{}{} & \aer{\ldots}{}{} & \aer{50.04}{+7.96}{-7.11} & \aer{<0.39}{}{} & \aer{1000}{f}{}\\ 
Her~X$-$1 & 3822 & F & 1533.52(1436) & 33.5 & \aer{1.10}{+0.16}{-0.04} & \aer{1.40}{f}{} & \aer{0.31}{+0.15}{-0.19} & \aer{-1.38}{+0.48}{-0.04} & \aer{1.80}{+2.15}{-1.67} & \aer{56.38}{+9.37}{-6.47} & \aer{1000}{f}{}\\ 
Her~X$-$1 & 4375 & G & 1494.95(1439) & 75.0 & \aer{0.98}{+0.04}{-0.04} & \aer{1.40}{f}{} & \aer{\ldots}{}{} & \aer{\ldots}{}{} & \aer{3.40}{+0.61}{-0.62} & \aer{<0.07}{}{} & \aer{1000}{f}{}\\ 
Her~X$-$1 & 4585 & H & 1472.87(1436) & 35.2 & \aer{2.54}{+0.90}{-0.17} & \aer{1.40}{f}{} & \aer{0.18}{+0.25}{-0.19} & \aer{-2.71}{+0.73}{-0.06} & \aer{0.03}{+0.06}{-0.01} & \aer{78.86}{+8.23}{-24.40} & \aer{1000}{f}{}\\ 
Her~X$-$1 & 6149 & I & 1670.44(1439) & 99.8 & \aer{0.18}{+0.04}{-0.04} & \aer{1.40}{f}{} & \aer{\ldots}{}{} & \aer{\ldots}{}{} & \aer{14.88}{+2.00}{-2.12} & \aer{<0.11}{}{} & \aer{1000}{f}{}\\ 
Her~X$-$1 & 6150 & J & 1553.33(1436) & 41.5 & \aer{0.38}{+0.06}{-0.06} & \aer{1.40}{f}{} & \aer{0.32}{+0.10}{-0.11} & \aer{-0.71}{+0.32}{-1.05} & \aer{6.13}{+1.19}{-1.15} & \aer{52.94}{+4.00}{-4.08} & \aer{1000}{f}{}\\ 
LMC~X$-$4 & 9571 & A & 1588.69(1437) & 91.3 & \aer{0.52}{+0.06}{-0.06} & \aer{1.40}{f}{} & \aer{\ldots}{}{} & \aer{\ldots}{}{} & \aer{5.18}{+1.66}{-1.49} & \aer{<0.27}{}{} & \aer{1000}{f}{}\\ 
LMC~X$-$4 & 9573 & B & 1066.10(1438) & 22.4 & \aer{0.34}{+0.20}{-0.26} & \aer{1.40}{f}{} & \aer{\ldots}{}{} & \aer{\ldots}{}{} & \aer{126.40}{+42.45}{-37.35} & \aer{<1.79}{}{} & \aer{1000}{f}{}\\ 
LMC~X$-$4 & 9574 & C & 1295.39(1437) & 34.4 & \aer{0.29}{+0.19}{-0.16} & \aer{1.40}{f}{} & \aer{0.04}{+0.19}{-0.34} & \aer{-1.83}{+0.36}{-0.69} & \aer{0.92}{+339.67}{-0.85} & \aer{153.91}{+14.64}{-12.24} & \aer{1000}{f}{}\\ 
OAO~1657$-$415 & 12460 & A & 1341.48(1438) & 10.4 & \aer{0.66}{f}{} & \aer{1.40}{f}{} & \aer{\ldots}{}{} & \aer{\ldots}{}{} & \aer{1.18}{f}{} & \aer{>72.06}{}{} & \aer{1000}{f}{}\\ 
OAO~1657$-$415 & 1947 & B & 1004.77(1438) & 85.4 & \aer{0.70}{+0.68}{-0.15} & \aer{1.40}{f}{} & \aer{\ldots}{}{} & \aer{\ldots}{}{} & \aer{3.97}{+11.17}{-1.08} & \aer{33.52}{+6.29}{-2.67} & \aer{1000}{f}{}\\ 
Vela~X$-$1 & 102 & A & 1360.28(1436) & 88.0 & \aer{0.91}{+0.28}{-0.23} & \aer{1.40}{f}{} & \aer{0.58}{+2.00}{-0.24} & \aer{-1.87}{+0.36}{-0.49} & \aer{1.60}{+0.69}{-0.41} & \aer{141.47}{+139.82}{-14.98} & \aer{21}{+6}{-8}\\ 
Vela~X$-$1 & 14654 & B & 1715.27(1439) & 100.0 & \aer{0.88}{+0.06}{-0.06} & \aer{1.40}{f}{} & \aer{\ldots}{}{} & \aer{\ldots}{}{} & \aer{1.43}{+0.21}{-0.18} & \aer{15.71}{+0.41}{-0.41} & \aer{1000}{f}{}\\ 
Vela~X$-$1 & 1926 & C & 1842.27(1439) & 100.0 & \aer{0.57}{+0.09}{-0.10} & \aer{1.40}{f}{} & \aer{\ldots}{}{} & \aer{\ldots}{}{} & \aer{119.49}{+15.00}{-14.94} & \aer{<0.34}{}{} & \aer{1000}{f}{}\\ 
Vela~X$-$1 & 1927 & D & 1718.44(1438) & 100.0 & \aer{0.44}{+0.09}{-0.03} & \aer{1.40}{f}{} & \aer{\ldots}{}{} & \aer{\ldots}{}{} & \aer{8.67}{+1.65}{-1.04} & \aer{9.15}{+0.49}{-0.23} & \aer{1000}{f}{}\\ 
Vela~X$-$1 & 1928 & E & 1454.03(1439) & 53.9 & \aer{1.03}{+0.01}{-0.01} & \aer{1.40}{f}{} & \aer{\ldots}{}{} & \aer{\ldots}{}{} & \aer{1.82}{+0.14}{-0.13} & \aer{<0.74}{}{} & \aer{1000}{f}{}\\
\enddata
\tabletypesize{\small}
\tablecomments{\small
Columns (1) and (2) give the XRB system name, \chandra-HETG
observation ID, and associated alphabetical label used in this paper.
Columns (3) and (4) give the best-fit value of the $C$-statistic, with
associated degrees of freedom, and the output of the \xspec\ {\tt
goodness} command for 2000 random realizations of the data.  Column
(5) gives the photon index for the direct, line-of-sight, zeroth-order
continuum. Column (6) gives the photon index for the
Compton-scattered, or ``reflected'' continuum. Column (7) gives the
photon index for any additional, ``soft'' power law continuum, and
column (8) the ratio of the normalizations of this continuum to that
of the direct continuum. Column (9) gives the relative normalization
factor for the Compton-scattered continuum. Columns (10) and (11) give
the equivalent neutral hydrogen column densities of the direct and
scattered components, respectively.  The line shift and line width
parameters for each fit are shown in
\tr{tab-feka}. $^{f}$: Parameter frozen; $^{t}$: tied to
\gammaz; d.o.f.: degrees of freedom. As explained in the text,
the \myt\ model has only been applied to those observations
in \tr{tab-sample} that have a significant \feka\ line
detection ($\Delta C \ge 6.63$, \scr{sec-detectability}).
}

\end{deluxetable*}

%% file: taball_feka.tex
\begin{longrotatetable}
%\begin{deluxetable*}{C{1.4cm} cccc cccc cc c ccp{2pt}p{.1pt} c}\label{tab-feka}
\begin{deluxetable*}{C{1.4cm} cccc cccc cc@{} c @{}c@{}c@{}c@{}c@{} c@{}}
\tablecaption{Properties of Narrow \feka\ line\label{tab-feka}} 
%\tablewidth{0pc}
%\tablewidth{10pt}
\tabletypesize{\scriptsize}
\tablehead{
\colhead{Source} & 
\multicolumn{2}{c}{obs.} &
\multicolumn{2}{c}{$\Delta E$}&
\multicolumn{2}{c}{$\Delta v$}&
\multicolumn{2}{c}{$\sigma_L$}&
\multicolumn{2}{c}{FWHM}&
\multicolumn{2}{c}{residuals}&
%$\Delta C$&
\multicolumn{2}{c}{Flux}&
\multicolumn{2}{c}{EW}
\\
\colhead{}&
\colhead{}&
\colhead{}& %for letter
\multicolumn{2}{c}{(eV)}&
\multicolumn{2}{c}{(\kmps)}&
\multicolumn{2}{c}{(eV)}&
\multicolumn{2}{c}{($10^3$\kmps)}&
\multicolumn{2}{c}{(\%)}&
%&
\multicolumn{2}{c}{($10^{-4}$\phunits)}&
\multicolumn{2}{c}{(eV)}
\\
\colhead{}&
\colhead{}&
\colhead{}& %for letter
\colhead{\bnet}&\colhead{\myty}&
\colhead{\bnet}&\colhead{\myty}&
\colhead{\bnet}&\colhead{\myty}&
\colhead{\bnet}&\colhead{\myty}&
\colhead{\bnet}&\colhead{\myty}&
\colhead{\bnet}&\colhead{\myty}&
\colhead{\bnet}&\colhead{\myty}
\\
\colhead{(1)}&
\multicolumn{2}{c}{(2)}&
\colhead{(3)}&
\colhead{(4)}&
\colhead{(5)}&
\colhead{(6)}&
\colhead{(7)}&
\colhead{(8)}&
\colhead{(9)}&
\colhead{(10)}&
\colhead{(11)}&
\colhead{(12)}&
\colhead{(13)}&
\colhead{(14)}&
\colhead{(15)}&
\colhead{(16)}
}
\decimals
\startdata
4U1700$-$37 & 657 & A & \aer{-1.1}{+2.4}{-1.4} & \aer{-3.1}{+3.0}{-0.3} & \aer{52}{+67}{-112} & \aer{145}{+16}{-138} & \aer{10.9}{+2.5}{-2.6} & \aer{<8.3}{\quad}{\quad} & \aer{1.276}{+0.297}{-0.309} & \aer{<0.975}{\quad}{\quad} & 4.1 & 4.0 & 9.49 & \aer{9.33}{+2.11}{-1.73} & 72.7 & \aer{69.4}{+15.7}{-12.8}\\ 
4U1822$-$371 & 671 & A & \aer{-1.2}{+6.8}{-5.0} & \aer{0.1}{+3.7}{-4.7} & \aer{58}{+235}{-316} & \aer{-3}{+222}{-174} & \aer{20.7}{+6.9}{-7.1} & \aer{14.4}{+7.7}{-10.0} & \aer{2.431}{+0.807}{-0.836} & \aer{1.697}{+0.908}{-1.176} & 2.4 & 2.3 & 4.42 & \aer{4.17}{+0.89}{-0.81} & 52.7 & \aer{47.7}{+10.2}{-9.3}\\ 
4U1822$-$371 & 9076 & B & \aer{-1.9}{+5.9}{-5.3} & \aer{-3.1}{+6.6}{-4.5} & \aer{89}{+247}{-277} & \aer{144}{+213}{-311} & \aer{18.0}{+5.7}{-4.7} & \aer{13.9}{+6.1}{-6.0} & \aer{2.115}{+0.672}{-0.548} & \aer{1.630}{+0.714}{-0.706} & 1.4 & 1.4 & 2.85 & \aer{2.91}{+0.48}{-0.44} & 38.1 & \aer{38.3}{+6.4}{-5.8}\\ 
4U1822$-$371 & 9858 & C & \aer{-1.7}{+4.1}{-2.9} & \aer{-3.1}{+6.4}{-1.6} & \aer{78}{+136}{-192} & \aer{143}{+77}{-300} & \aer{15.2}{+4.1}{-3.6} & \aer{12.2}{+4.8}{-4.8} & \aer{1.782}{+0.480}{-0.420} & \aer{1.432}{+0.565}{-0.568} & 1.5 & 1.6 & 3.30 & \aer{3.43}{+0.52}{-0.40} & 43.0 & \aer{43.7}{+6.6}{-5.1}\\ 
4U1908$+$075 & 5476 & A & \aer{-5.7}{+6.1}{-8.2} & \aer{-4.8}{+8.4}{-7.5} & \aer{267}{+387}{-285} & \aer{223}{+352}{-396} & \aer{23.8}{+10.0}{-7.6} & \aer{20.9}{+9.4}{-8.5} & \aer{2.803}{+1.182}{-0.895} & \aer{2.458}{+1.100}{-1.001} & 3.0 & 3.2 & 4.32 & \aer{5.11}{+1.92}{-1.41} & 119.4 & \aer{140.0}{+52.6}{-38.6}\\ 
4U1908$+$075 & 5477 & B & \aer{1.5}{+4.9}{-6.0} & \aer{-4.8}{+8.3}{-3.7} & \aer{-73}{+282}{-230} & \aer{224}{+174}{-390} & \aer{<12.7}{\quad}{\quad} & \aer{<9.0}{\quad}{\quad} & \aer{<1.499}{\quad}{\quad} & \aer{<1.061}{\quad}{\quad} & 1.1 & 1.0 & 0.63 & \aer{0.54}{+0.42}{-0.24} & 51.9 & \aer{41.7}{+32.8}{-18.4}\\ 
4U1908$+$075 & 6336 & C & \aer{6.0}{+11.1}{-7.7} & \aer{7.8}{+4.3}{-12.3} & \aer{-282}{+360}{-520} & \aer{-363}{+575}{-199} & \aer{17.9}{+10.1}{-7.9} & \aer{15.0}{+9.8}{-9.2} & \aer{2.103}{+1.183}{-0.926} & \aer{1.766}{+1.147}{-1.082} & 0.9 & 0.9 & 1.33 & \aer{1.35}{+0.39}{-0.36} & 78.7 & \aer{76.8}{+22.4}{-20.3}\\ 
Cen~X$-$3 & 1943 & A & \aer{-1.4}{a}{\quad} & \aer{-4.4}{+4.4}{-0.1} & \aer{66}{a}{\quad} & \aer{207}{+6}{-207} & \aer{0.8}{a}{\quad} & \aer{8.4}{+3.5}{-4.5} & \aer{\ldots}{a}{\quad} & \aer{0.988}{+0.417}{-0.523} & 18.4 & 9.2 & 6.41 & \aer{20.13}{+1.71}{-1.66} & 7.8 & \aer{24.6}{+2.1}{-2.0}\\ 
Cen~X$-$3 & 705 & B & \aer{-10.3}{+6.6}{-5.1} & \aer{-12.2}{+7.5}{-4.4} & \aer{483}{+239}{-310} & \aer{572}{+206}{-350} & \aer{19.3}{+6.4}{-5.3} & \aer{15.5}{+7.0}{-7.0} & \aer{2.276}{+0.753}{-0.620} & \aer{1.824}{+0.825}{-0.819} & 1.5 & 1.7 & 3.02 & \aer{3.18}{+0.59}{-0.55} & 79.4 & \aer{80.8}{+14.9}{-14.1}\\ 
Cen~X$-$3 & 7511 & C & \aer{3.3}{a}{\quad} & \aer{3.5}{+0.2}{-3.6} & \aer{-153}{a}{\quad} & \aer{-166}{+168}{-7} & \aer{13.1}{a}{\quad} & \aer{14.9}{+2.6}{-2.4} & \aer{\ldots}{a}{\quad} & \aer{1.752}{+0.301}{-0.287} & 9.0 & 4.4 & 11.18 & \aer{17.88}{+1.24}{-1.21} & 59.3 & \aer{94.6}{+6.6}{-6.4}\\ 
%Cir~X$-$1 & 12235 & A & \aer{0.0}{f}{\quad} & \aer{0.0}{f}{\quad} & \aer{0}{f}{\quad} & \aer{0}{f}{\quad} & \aer{0.8}{f}{\quad} & \aer{0.8}{f}{\quad} & \aer{0.100}{f}{\quad} & \aer{0.100}{f}{\quad} & 2.9 & 4.5 & 0.20 & \aer{<0.43}{\quad}{\quad} & 5.9 & \aer{<4.9}{\quad}{\quad}\\ 
%Cir~X$-$1 & 1905 & B & \aer{0.0}{f}{\quad} & \aer{0.0}{f}{\quad} & \aer{0}{f}{\quad} & \aer{0}{f}{\quad} & \aer{0.8}{f}{\quad} & \aer{0.8}{f}{\quad} & \aer{0.100}{f}{\quad} & \aer{0.100}{f}{\quad} & 28.0 & 26.9 & 6.19 & \aer{3.95}{+3.86}{-3.80} & 3.3 & \aer{2.1}{+2.1}{-2.0}\\ 
%Cir~X$-$1 & 1906 & C & \aer{0.0}{f}{\quad} & \aer{0.0}{f}{\quad} & \aer{0}{f}{\quad} & \aer{0}{f}{\quad} & \aer{0.8}{f}{\quad} & \aer{0.8}{f}{\quad} & \aer{0.100}{f}{\quad} & \aer{0.100}{f}{\quad} & 38.3 & 36.4 & 2.55 & \aer{<9.07}{\quad}{\quad} & 1.2 & \aer{<4.2}{\quad}{\quad}\\ 
%Cir~X$-$1 & 1907 & D & \aer{0.0}{f}{\quad} & \aer{0.0}{f}{\quad} & \aer{0}{f}{\quad} & \aer{0}{f}{\quad} & \aer{0.8}{f}{\quad} & \aer{0.8}{f}{\quad} & \aer{0.100}{f}{\quad} & \aer{0.100}{f}{\quad} & 25.4 & 26.0 & 4.43 & \aer{<6.34}{\quad}{\quad} & 2.8 & \aer{<3.8}{\quad}{\quad}\\ 
%Cir~X$-$1 & 8993 & G & \aer{0.0}{f}{\quad} & \aer{0.0}{f}{\quad} & \aer{0}{f}{\quad} & \aer{0}{f}{\quad} & \aer{0.8}{f}{\quad} & \aer{0.8}{f}{\quad} & \aer{0.100}{f}{\quad} & \aer{0.100}{f}{\quad} & 0.6 & 0.6 & 0.15 & \aer{0.10}{+0.16}{-0.09} & 45.0 & \aer{23.7}{+38.9}{-20.6}\\ 
Cyg~X$-$1 & 11044 & A & \aer{7.2}{a}{\quad} & \aer{-9.7}{+1.5}{-3.1} & \aer{-338}{a}{\quad} & \aer{453}{+145}{-69} & \aer{0.8}{f}{\quad} & \aer{0.8}{f}{\quad} & \aer{\ldots}{a}{\quad} & \aer{0.100}{f}{\quad} & 13.9 & 13.4 & 1.72 & \aer{8.32}{+1.59}{-1.55} & 2.6 & \aer{12.5}{+2.4}{-2.3}\\ 
%Cyg~X$-$1 & 12313 & B & \aer{0.0}{f}{\quad} & \aer{0.0}{f}{\quad} & \aer{0}{f}{\quad} & \aer{0}{f}{\quad} & \aer{0.8}{f}{\quad} & \aer{0.8}{f}{\quad} & \aer{\ldots}{a}{\quad} & \aer{0.100}{f}{\quad} & 65.8 & 64.3 & 17.46 & \aer{27.52}{+28.95}{-17.49} & 5.8 & \aer{9.2}{+9.7}{-5.8}\\ 
%Cyg~X$-$1 & 12314 & C & \aer{0.0}{f}{\quad} & \aer{0.0}{f}{\quad} & \aer{0}{f}{\quad} & \aer{0}{f}{\quad} & \aer{0.8}{f}{\quad} & \aer{0.8}{f}{\quad} & \aer{\ldots}{a}{\quad} & \aer{0.100}{f}{\quad} & 50.7 & 49.8 & -1.15 & \aer{57.11}{+27.68}{-25.37} & -0.7 & \aer{35.9}{+17.4}{-15.9}\\ 
%Cyg~X$-$1 & 12472 & D & \aer{0.0}{f}{\quad} & \aer{0.0}{f}{\quad} & \aer{0}{f}{\quad} & \aer{0}{f}{\quad} & \aer{0.8}{f}{\quad} & \aer{0.8}{f}{\quad} & \aer{\ldots}{a}{\quad} & \aer{0.100}{f}{\quad} & 37.7 & 34.4 & 20.36 & \aer{2.65}{+100.79}{-2.35} & 15.4 & \aer{1.9}{+72.3}{-1.7}\\ 
%Cyg~X$-$1 & 13219 & E & \aer{0.0}{f}{\quad} & \aer{0.0}{f}{\quad} & \aer{0}{f}{\quad} & \aer{0}{f}{\quad} & \aer{0.8}{f}{\quad} & \aer{0.8}{f}{\quad} & \aer{\ldots}{a}{\quad} & \aer{0.100}{f}{\quad} & 29.8 & 29.0 & 10.25 & \aer{14.64}{+13.77}{-8.43} & 7.4 & \aer{10.6}{+10.0}{-6.1}\\ 
Cyg~X$-$1 & 1511 & F & \aer{0.0}{f}{\quad} & \aer{0.0}{f}{\quad} & \aer{0}{f}{\quad} & \aer{0}{f}{\quad} & \aer{0.8}{f}{\quad} & \aer{0.8}{f}{\quad} & \aer{0.100}{f}{\quad} & \aer{0.100}{f}{\quad} & 13.2 & 13.2 & 4.20 & \aer{6.71}{+2.51}{-2.44} & 6.8 & \aer{10.8}{+4.0}{-3.9}\\ 
Cyg~X$-$1 & 2415 & G & \aer{6.5}{a}{\quad} & \aer{6.0}{+4.6}{-2.0} & \aer{-303}{a}{\quad} & \aer{-282}{+92}{-213} & \aer{0.8}{f}{\quad} & \aer{0.8}{f}{\quad} & \aer{\ldots}{a}{\quad} & \aer{0.100}{f}{\quad} & 10.4 & 11.6 & 0.86 & \aer{12.99}{+2.31}{-2.27} & 0.9 & \aer{14.9}{+2.6}{-2.6}\\ 
%Cyg~X$-$1 & 2741 & H & \aer{0.0}{f}{\quad} & \aer{0.0}{f}{\quad} & \aer{0}{f}{\quad} & \aer{0}{f}{\quad} & \aer{0.8}{f}{\quad} & \aer{0.8}{f}{\quad} & \aer{\ldots}{a}{\quad} & \aer{0.100}{f}{\quad} & 42.0 & 45.0 & 16.42 & \aer{22.92}{+30.22}{-15.24} & 8.0 & \aer{11.1}{+14.6}{-7.4}\\ 
%Cyg~X$-$1 & 2742 & I & \aer{0.0}{f}{\quad} & \aer{0.0}{f}{\quad} & \aer{0}{f}{\quad} & \aer{0}{f}{\quad} & \aer{0.8}{f}{\quad} & \aer{0.8}{f}{\quad} & \aer{\ldots}{a}{\quad} & \aer{0.100}{f}{\quad} & 52.9 & 47.4 & 12.05 & \aer{<27.69}{\quad}{\quad} & 5.5 & \aer{<11.8}{\quad}{\quad}\\ 
%Cyg~X$-$1 & 2743 & J & \aer{0.0}{f}{\quad} & \aer{0.0}{f}{\quad} & \aer{0}{f}{\quad} & \aer{0}{f}{\quad} & \aer{0.8}{f}{\quad} & \aer{0.8}{f}{\quad} & \aer{\ldots}{a}{\quad} & \aer{0.100}{f}{\quad} & 36.4 & 36.4 & 8.49 & \aer{11.64}{+27.58}{-7.87} & 12.2 & \aer{17.1}{+40.5}{-11.6}\\ 
Cyg~X$-$1 & 3407 & K & \aer{9.6}{+6.9}{-9.1} & \aer{6.0}{+14.0}{-10.8} & \aer{-450}{+425}{-321} & \aer{-282}{+506}{-654} & \aer{0.8}{f}{\quad} & \aer{0.8}{f}{\quad} & \aer{0.100}{f}{\quad} & \aer{0.100}{f}{\quad} & 12.5 & 10.2 & 9.27 & \aer{6.95}{+7.23}{-4.47} & 6.9 & \aer{5.1}{+5.3}{-3.3}\\ 
Cyg~X$-$1 & 3814 & M & \aer{-3.0}{a}{\quad} & \aer{-4.5}{+8.0}{-6.9} & \aer{140}{a}{\quad} & \aer{211}{+325}{-377} & \aer{24.5}{a}{\quad} & \aer{21.3}{+9.0}{-7.4} & \aer{\ldots}{a}{\quad} & \aer{2.506}{+1.054}{-0.875} & 6.4 & 6.3 & 9.45 & \aer{11.26}{+2.02}{-1.92} & 14.6 & \aer{17.3}{+3.1}{-2.9}\\ 
Cyg~X$-$1 & 8525 & N & \aer{-7.2}{+5.3}{-6.4} & \aer{-12.2}{+7.4}{-0.7} & \aer{339}{+300}{-250} & \aer{575}{+31}{-345} & \aer{0.8}{f}{\quad} & \aer{0.8}{f}{\quad} & \aer{0.100}{f}{\quad} & \aer{0.100}{f}{\quad} & 12.9 & 12.0 & 5.88 & \aer{7.62}{+1.66}{-1.65} & 8.8 & \aer{11.4}{+2.5}{-2.5}\\ 
Cyg~X$-$1 & 9847 & O & \aer{-2.9}{+13.4}{-15.7} & \aer{-4.7}{+16.2}{-7.1} & \aer{137}{+740}{-627} & \aer{218}{+335}{-759} & \aer{41.0}{+17.9}{-21.4} & \aer{<40.5}{\quad}{\quad} & \aer{4.827}{+2.102}{-2.520} & \aer{<4.765}{\quad}{\quad} & 16.9 & 12.1 & 15.91 & \aer{12.01}{+5.39}{-3.11} & 19.5 & \aer{14.7}{+6.6}{-3.8}\\ 
Cyg~X$-$1 & 3815 & P & \aer{-1.9}{+7.1}{-5.4} & \aer{-3.1}{+6.8}{-1.7} & \aer{91}{+253}{-332} & \aer{144}{+82}{-317} & \aer{28.3}{+7.5}{-6.1} & \aer{24.9}{+7.2}{-6.0} & \aer{3.324}{+0.886}{-0.713} & \aer{2.930}{+0.849}{-0.706} & 7.1 & 6.7 & 15.01 & \aer{17.42}{+2.02}{-2.00} & 14.4 & \aer{16.7}{+1.9}{-1.9}\\ 
Cyg~X$-$3 & 101 & A & \aer{6.6}{a}{\quad} & \aer{6.0}{+14.3}{-10.4} & \aer{-309}{a}{\quad} & \aer{-281}{+486}{-665} & \aer{33.9}{a}{\quad} & \aer{32.3}{+13.0}{-8.9} & \aer{\ldots}{a}{\quad} & \aer{3.798}{+1.532}{-1.052} & 24.5 & 26.1 & 37.98 & \aer{48.74}{+15.41}{-12.07} & 81.2 & \aer{105.1}{+33.2}{-26.0}\\ 
$\gamma$~Cas & 1895 & A & \aer{-7.7}{a}{\quad} & \aer{-0.3}{+6.9}{-8.2} & \aer{360}{a}{\quad} & \aer{16}{+383}{-321} & \aer{0.8}{f}{\quad} & \aer{<24.2}{\quad}{\quad} & \aer{\ldots}{a}{\quad} & \aer{<2.846}{\quad}{\quad} & 1.4 & 1.0 & 0.34 & \aer{0.96}{+0.28}{-0.31} & 19.5 & \aer{54.9}{+16.0}{-17.9}\\ 
GX~301$-$2 & 2733 & B & \aer{-1.7}{+0.3}{-0.3} & \aer{-4.6}{+0.0}{-0.0} & \aer{81}{+13}{-13} & \aer{214}{+1}{-1} & \aer{9.4}{+0.6}{-0.7} & \aer{3.9}{+1.2}{-1.2} & \aer{1.107}{+0.074}{-0.079} & \aer{0.454}{+0.136}{-0.136} & 12.6 & 30.3 & 103.50 & \aer{87.74}{+3.86}{-1.93} & 515.0 & \aer{402.1}{+17.7}{-8.9}\\ 
GX~301$-$2 & 3433 & C & \aer{-1.6}{+0.8}{-0.4} & \aer{-3.1}{+2.9}{-0.1} & \aer{73}{+20}{-38} & \aer{145}{+3}{-134} & \aer{10.9}{+1.0}{-1.0} & \aer{5.3}{+1.6}{-1.9} & \aer{1.285}{+0.117}{-0.117} & \aer{0.626}{+0.192}{-0.220} & 6.4 & 5.4 & 40.89 & \aer{42.60}{+4.53}{-4.49} & 166.4 & \aer{166.6}{+17.7}{-17.6}\\ 
GX~1$+$4 & 2710 & A & \aer{3.0}{+3.3}{-4.1} & \aer{3.3}{+0.4}{-6.6} & \aer{-139}{+192}{-156} & \aer{-156}{+309}{-18} & \aer{<13.6}{\quad}{\quad} & \aer{<12.7}{\quad}{\quad} & \aer{<1.603}{\quad}{\quad} & \aer{<1.490}{\quad}{\quad} & 1.0 & 0.8 & 1.65 & \aer{1.88}{+0.37}{-0.28} & 67.1 & \aer{76.0}{+14.9}{-11.4}\\ 
GX~1$+$4 & 2744 & B & \aer{3.7}{a}{\quad} & \aer{3.5}{+0.1}{-3.4} & \aer{-173}{a}{\quad} & \aer{-165}{+159}{-5} & \aer{5.3}{a}{\quad} & \aer{<7.0}{\quad}{\quad} & \aer{\ldots}{a}{\quad} & \aer{<0.826}{\quad}{\quad} & 7.2 & 4.4 & 15.40 & \aer{20.51}{+3.97}{-3.33} & 162.7 & \aer{217.7}{+42.1}{-35.4}\\ 
Her~X$-$1 & 2703 & A & \aer{-1.4}{+5.2}{-3.3} & \aer{-4.6}{+8.1}{-3.4} & \aer{66}{+155}{-244} & \aer{214}{+159}{-378} & \aer{0.8}{f}{\quad} & \aer{0.8}{f}{\quad} & \aer{0.100}{f}{\quad} & \aer{0.100}{f}{\quad} & 4.2 & 4.6 & 3.23 & \aer{4.41}{+1.21}{-1.01} & 102.7 & \aer{157.6}{+43.2}{-36.0}\\ 
Her~X$-$1 & 2704 & B & \aer{5.3}{a}{\quad} & \aer{3.8}{+4.3}{-0.3} & \aer{-246}{a}{\quad} & \aer{-177}{+13}{-202} & \aer{15.4}{a}{\quad} & \aer{0.8}{f}{\quad} & \aer{\ldots}{a}{\quad} & \aer{0.100}{f}{\quad} & 9.3 & 8.0 & 4.00 & \aer{7.13}{+1.50}{-1.45} & 16.3 & \aer{28.9}{+6.1}{-5.9}\\ 
Her~X$-$1 & 2705 & C & \aer{18.0}{+12.0}{-11.1} & \aer{12.8}{+15.6}{-5.1} & \aer{-839}{+519}{-560} & \aer{-600}{+239}{-727} & \aer{<26.1}{\quad}{\quad} & \aer{<31.4}{\quad}{\quad} & \aer{<3.069}{\quad}{\quad} & \aer{<3.689}{\quad}{\quad} & 1.2 & 1.3 & 0.80 & \aer{1.12}{+0.41}{-0.35} & 125.0 & \aer{199.2}{+73.1}{-62.3}\\ 
Her~X$-$1 & 2749 & D & \aer{5.2}{a}{\quad} & \aer{3.6}{+0.1}{-3.5} & \aer{-245}{a}{\quad} & \aer{-167}{+163}{-7} & \aer{8.6}{a}{\quad} & \aer{<10.2}{\quad}{\quad} & \aer{\ldots}{a}{\quad} & \aer{<1.195}{\quad}{\quad} & 1.3 & 1.1 & 3.21 & \aer{3.62}{+0.63}{-0.53} & 456.1 & \aer{535.3}{+93.2}{-79.0}\\ 
Her~X$-$1 & 3821 & E & \aer{-1.4}{+4.0}{-3.1} & \aer{-4.5}{+5.1}{-0.3} & \aer{66}{+145}{-188} & \aer{210}{+12}{-240} & \aer{13.3}{+4.4}{-4.2} & \aer{13.2}{+4.8}{-4.7} & \aer{1.565}{+0.512}{-0.498} & \aer{1.551}{+0.564}{-0.551} & 2.0 & 1.9 & 3.63 & \aer{4.62}{+0.74}{-0.66} & 120.5 & \aer{161.5}{+25.7}{-23.0}\\ 
Her~X$-$1 & 3822 & F & \aer{0.5}{+4.0}{-3.0} & \aer{-3.1}{+6.4}{-1.5} & \aer{-25}{+139}{-185} & \aer{145}{+69}{-298} & \aer{15.8}{+4.2}{-3.8} & \aer{11.6}{+4.6}{-5.0} & \aer{1.862}{+0.494}{-0.446} & \aer{1.362}{+0.538}{-0.588} & 2.6 & 2.0 & 6.16 & \aer{5.55}{+6.61}{-5.15} & 187.8 & \aer{159.5}{+190.0}{-148.0}\\ 
Her~X$-$1 & 4375 & G & \aer{5.5}{+9.6}{-8.1} & \aer{6.0}{+14.7}{-9.3} & \aer{-258}{+377}{-447} & \aer{-283}{+436}{-687} & \aer{26.1}{+16.2}{-10.4} & \aer{35.5}{+17.5}{-15.8} & \aer{3.064}{+1.909}{-1.223} & \aer{4.179}{+2.063}{-1.861} & 3.8 & 4.3 & 6.36 & \aer{8.17}{+1.47}{-1.50} & 77.9 & \aer{104.0}{+18.7}{-19.0}\\ 
Her~X$-$1 & 4585 & H & \aer{5.3}{+12.7}{-9.8} & \aer{-4.8}{+8.4}{-7.5} & \aer{-248}{+457}{-595} & \aer{226}{+351}{-393} & \aer{28.8}{+13.0}{-11.8} & \aer{<22.7}{\quad}{\quad} & \aer{3.389}{+1.525}{-1.393} & \aer{<2.675}{\quad}{\quad} & 4.8 & 3.7 & 6.79 & \aer{4.25}{+9.32}{-1.39} & 88.1 & \aer{52.9}{+115.9}{-17.3}\\ 
Her~X$-$1 & 6149 & I & \aer{-2.1}{+6.0}{-5.5} & \aer{-8.1}{+11.2}{-4.5} & \aer{98}{+256}{-280} & \aer{378}{+209}{-527} & \aer{18.9}{+6.0}{-5.2} & \aer{22.0}{+5.8}{-5.3} & \aer{2.223}{+0.708}{-0.611} & \aer{2.592}{+0.677}{-0.621} & 4.5 & 4.2 & 6.62 & \aer{8.17}{+1.10}{-1.16} & 77.2 & \aer{102.0}{+13.7}{-14.5}\\ 
Her~X$-$1 & 6150 & J & \aer{-1.9}{+5.6}{-4.3} & \aer{-7.1}{+10.5}{-1.5} & \aer{91}{+202}{-263} & \aer{333}{+70}{-492} & \aer{13.1}{+6.1}{-5.5} & \aer{11.8}{+5.4}{-5.7} & \aer{1.544}{+0.714}{-0.646} & \aer{1.383}{+0.631}{-0.676} & 1.7 & 1.7 & 3.86 & \aer{3.76}{+0.73}{-0.71} & 133.4 & \aer{128.1}{+24.9}{-24.1}\\ 
LMC~X$-$4 & 9571 & A & \aer{4.3}{+7.4}{-7.7} & \aer{0.1}{+11.5}{-5.3} & \aer{-200}{+363}{-347} & \aer{-7}{+248}{-539} & \aer{<20.8}{\quad}{\quad} & \aer{<28.9}{\quad}{\quad} & \aer{<2.449}{\quad}{\quad} & \aer{<3.402}{\quad}{\quad} & 0.9 & 0.9 & 0.68 & \aer{0.83}{+0.27}{-0.24} & 54.2 & \aer{68.9}{+22.0}{-19.8}\\ 
LMC~X$-$4 & 9573 & B & \aer{-12.8}{a}{\quad} & \aer{-12.8}{+8.7}{-8.7} & \aer{600}{a}{\quad} & \aer{603}{+409}{-410} & \aer{24.0}{a}{\quad} & \aer{23.3}{+10.3}{-10.1} & \aer{\ldots}{a}{\quad} & \aer{2.739}{+1.211}{-1.193} & 0.4 & 0.4 & 0.43 & \aer{0.54}{+0.18}{-0.16} & 516.7 & \aer{702.1}{+235.8}{-207.5}\\ 
LMC~X$-$4 & 9574 & C & \aer{-1.4}{+6.0}{-4.7} & \aer{-6.5}{+4.5}{-5.9} & \aer{66}{+221}{-283} & \aer{303}{+278}{-212} & \aer{0.8}{f}{\quad} & \aer{0.8}{f}{\quad} & \aer{0.100}{f}{\quad} & \aer{0.100}{f}{\quad} & 0.6 & 0.6 & 0.36 & \aer{0.32}{+119.14}{-0.30} & 149.6 & \aer{129.1}{+47604.9}{-118.7}\\ 
OAO~1657$-$415 & 12460 & A & \aer{4.6}{a}{\quad} & \aer{3.4}{+7.9}{-8.0} & \aer{-217}{a}{\quad} & \aer{-159}{+376}{-369} & \aer{6.1}{a}{\quad} & \aer{<13.2}{\quad}{\quad} & \aer{\ldots}{a}{\quad} & \aer{<1.557}{\quad}{\quad} & 2.7 & 3.9 & 6.45 & \aer{6.87}{+1.77}{-0.12} & 1427.9 & \aer{1922.1}{+494.0}{-32.2}\\ 
OAO~1657$-$415 & 1947 & B & \aer{8.2}{+6.6}{-7.3} & \aer{6.5}{+5.7}{-7.0} & \aer{-385}{+340}{-308} & \aer{-303}{+325}{-266} & \aer{14.3}{+8.0}{-6.0} & \aer{<20.1}{\quad}{\quad} & \aer{1.680}{+0.941}{-0.705} & \aer{<2.368}{\quad}{\quad} & 3.5 & 3.8 & 7.02 & \aer{6.63}{+18.62}{-1.80} & 191.6 & \aer{170.1}{+478.1}{-46.2}\\ 
Vela~X$-$1 & 102 & A & \aer{-6.3}{a}{\quad} & \aer{-4.9}{+0.3}{-7.6} & \aer{297}{a}{\quad} & \aer{231}{+358}{-15} & \aer{4.3}{a}{\quad} & \aer{<13.0}{\quad}{\quad} & \aer{\ldots}{a}{\quad} & \aer{<1.525}{\quad}{\quad} & 1.0 & 1.0 & 1.39 & \aer{1.52}{+0.65}{-0.39} & 677.7 & \aer{529.8}{+226.9}{-136.2}\\ 
Vela~X$-$1 & 14654 & B & \aer{-1.7}{+2.1}{-1.2} & \aer{-0.3}{+0.1}{-4.2} & \aer{78}{+56}{-97} & \aer{12}{+198}{-5} & \aer{11.3}{+2.4}{-2.4} & \aer{<8.0}{\quad}{\quad} & \aer{1.335}{+0.283}{-0.279} & \aer{<0.940}{\quad}{\quad} & 3.5 & 4.1 & 11.69 & \aer{10.71}{+1.57}{-1.35} & 64.8 & \aer{57.4}{+8.4}{-7.2}\\ 
Vela~X$-$1 & 1926 & C & \aer{-1.8}{a}{\quad} & \aer{-0.4}{+0.3}{-2.9} & \aer{84}{a}{\quad} & \aer{20}{+134}{-13} & \aer{6.7}{a}{\quad} & \aer{0.8}{f}{\quad} & \aer{\ldots}{a}{\quad} & \aer{0.100}{f}{\quad} & 0.4 & 0.6 & 1.83 & \aer{2.12}{+0.27}{-0.27} & 740.8 & \aer{927.2}{+116.4}{-115.9}\\ 
Vela~X$-$1 & 1927 & D & \aer{-0.5}{+1.1}{-1.0} & \aer{-1.0}{+0.9}{-2.5} & \aer{23}{+48}{-49} & \aer{48}{+116}{-41} & \aer{8.1}{+1.5}{-1.5} & \aer{0.8}{f}{\quad} & \aer{0.955}{+0.177}{-0.176} & \aer{0.100}{f}{\quad} & 10.0 & 11.6 & 38.95 & \aer{39.82}{+7.59}{-4.78} & 121.6 & \aer{122.2}{+23.3}{-14.7}\\ 
Vela~X$-$1 & 1928 & E & \aer{1.7}{+1.5}{-1.5} & \aer{3.4}{+0.1}{-3.6} & \aer{-78}{+68}{-70} & \aer{-161}{+168}{-4} & \aer{<6.8}{\quad}{\quad} & \aer{<6.7}{\quad}{\quad} & \aer{<0.800}{\quad}{\quad} & \aer{<0.789}{\quad}{\quad} & 5.9 & 7.7 & 20.73 & \aer{25.03}{+1.92}{-1.76} & 52.6 & \aer{63.5}{+4.9}{-4.5}\\ 
\enddata
\tabletypesize{\small}
\tablecomments{\small
The table only includes significant detections.  Columns (1) and (2)
give the XRB system name, \chandra-HETG observation ID, and associated
alphabetical label used in this paper. Except for columns (1) and
(2), columns show pairs of results for the same parameter in the
spherical (\bnet) and \myt\ (\myty) model.  Thus, columns (3) and (4)
give the line's peak energy shift, and columns (5) and (6) the line's
velocity shift, all calculated from the models' best-fit redshift
parameters. Columns (7) and (8) give the energy width of the line's
Gaussian convolution kernel (\scr{sec-fekalinewidth}).  Columns (9) and
(10) give the line's velocity with, calculated from the values in
columns (7) and (8). Columns (11) and (12) give the maximum residuals
as a percentage of the model value in the vicinity of the \feka\ line
(energy range $6.3-6.5$ keV). Columns (13) and (14) give the flux of
the \feka\ line. Columns (15) and (16) give the \feka\ line equivalent
width.
}
\end{deluxetable*}
\end{longrotatetable}

%% file: taball_flux.tex
\startlongtable
\begin{deluxetable*}{C{1.4cm} c cc cc cc cc c}
  \tablecaption{Continuum fluxes\label{tab-flux}}
\tabletypesize{\scriptsize}
\tablehead{
\colhead{Source} & 
\multicolumn{2}{c}{obs.} &
\multicolumn{2}{c}{\fcobs}&
\multicolumn{2}{c}{\fcintr}&
\multicolumn{2}{c}{\lcobs}&
\multicolumn{2}{c}{\lcintr}
\\
\colhead{}&
\colhead{}&
\colhead{}& %for letter
\multicolumn{2}{c}{(log \funits)}&
\multicolumn{2}{c}{(log \funits)}&
\multicolumn{2}{c}{(log \lunits)}&
\multicolumn{2}{c}{(log \lunits)}
\\
\colhead{}&
\colhead{}&
\colhead{}& %for letter
\colhead{\bnet}&\colhead{\myty}&
\colhead{\bnet}&\colhead{\myty}&
\colhead{\bnet}&\colhead{\myty}&
\colhead{\bnet}&\colhead{\myty}
\\
\colhead{(1)}&
\multicolumn{2}{c}{(2)}&
\colhead{(3)}&
\colhead{(4)}&
\colhead{(5)}&
\colhead{(6)}&
\colhead{(7)}&
\colhead{(8)}&
\colhead{(9)}&
\colhead{(10)}
}
\startdata
4U1700$-$37 & 657 & A & $-9.09$ & $-9.09$ & $-8.88$ & $-8.87$ & $35.45$ & $35.45$ & $35.66$ & $35.67$\\ 
4U1822$-$371 & 671 & A & $-9.21$ & $-9.19$ & $-9.15$ & $-9.15$ & $31.87$ & $31.88$ & $31.93$ & $31.92$\\ 
4U1822$-$371 & 9076 & B & $-9.24$ & $-9.25$ & $-9.20$ & $-9.20$ & $31.83$ & $31.83$ & $31.87$ & $31.87$\\ 
4U1822$-$371 & 9858 & C & $-9.24$ & $-9.23$ & $-9.19$ & $-9.19$ & $31.84$ & $31.84$ & $31.88$ & $31.88$\\ 
4U1908$+$075 & 5476 & A & $-9.62$ & $-9.62$ & $-9.48$ & $-9.49$ & $34.62$ & $34.63$ & $34.76$ & $34.76$\\ 
4U1908$+$075 & 5477 & B & $-10.15$ & $-10.14$ & $-9.77$ & $-9.75$ & $34.09$ & $34.10$ & $34.48$ & $34.49$\\ 
4U1908$+$075 & 6336 & C & $-9.97$ & $-9.97$ & $-9.75$ & $-9.73$ & $34.27$ & $34.27$ & $34.49$ & $34.51$\\ 
Cen~X$-$3 & 1943 & A & $-8.19$ & $-8.20$ & $-8.16$ & $-8.17$ & $37.69$ & $37.69$ & $37.72$ & $37.71$\\ 
Cen~X$-$3 & 705 & B & $-9.60$ & $-9.60$ & $-9.58$ & $-9.43$ & $36.29$ & $36.29$ & $36.30$ & $36.46$\\ 
Cen~X$-$3 & 7511 & C & $-8.82$ & $-8.85$ & $-8.88$ & $-9.27$ & $37.06$ & $37.04$ & $37.00$ & $36.61$\\ 
Cir~X$-$1 & 12235 & A & $-9.71$ & $-9.70$ & $-9.59$ & $-9.60$ & $30.52$ & $30.54$ & $30.65$ & $30.63$\\ 
Cir~X$-$1 & 1905 & B & $-7.79$ & $-7.79$ & $-7.71$ & $-7.70$ & $32.45$ & $32.44$ & $32.53$ & $32.54$\\ 
Cir~X$-$1 & 1906 & C & $-7.77$ & $-7.77$ & $-7.69$ & $-7.69$ & $32.47$ & $32.46$ & $32.54$ & $32.55$\\ 
Cir~X$-$1 & 1907 & D & $-7.90$ & $-7.90$ & $-7.81$ & $-7.80$ & $32.34$ & $32.33$ & $32.42$ & $32.43$\\ 
Cir~X$-$1 & 8993 & G & $-10.65$ & $-10.65$ & $-10.53$ & $-10.51$ & $29.59$ & $29.59$ & $29.71$ & $29.72$\\ 
Cyg~X$-$1 & 11044 & A & $-8.22$ & $-8.22$ & $-8.22$ & $-8.22$ & $35.58$ & $35.58$ & $35.57$ & $35.57$\\ 
Cyg~X$-$1 & 12313 & B & $-7.37$ & $-7.38$ & $-7.69$ & $-7.72$ & $36.42$ & $36.42$ & $36.11$ & $36.08$\\ 
Cyg~X$-$1 & 12314 & C & $-7.61$ & $-7.61$ & $-7.61$ & $-7.61$ & $36.19$ & $36.19$ & $36.18$ & $36.18$\\ 
Cyg~X$-$1 & 12472 & D & $-7.52$ & $-7.51$ & $-8.11$ & $-8.11$ & $36.28$ & $36.29$ & $35.69$ & $35.69$\\ 
Cyg~X$-$1 & 13219 & E & $-7.66$ & $-7.66$ & $-8.03$ & $-8.03$ & $36.14$ & $36.14$ & $35.76$ & $35.77$\\ 
Cyg~X$-$1 & 1511 & F & $-8.23$ & $-8.23$ & $-8.22$ & $-8.22$ & $35.56$ & $35.56$ & $35.57$ & $35.58$\\ 
Cyg~X$-$1 & 2415 & G & $-8.06$ & $-8.08$ & $-8.50$ & $-8.08$ & $35.73$ & $35.71$ & $35.29$ & $35.71$\\ 
Cyg~X$-$1 & 2741 & H & $-7.57$ & $-7.57$ & $-7.83$ & $-7.87$ & $36.22$ & $36.22$ & $35.96$ & $35.93$\\ 
Cyg~X$-$1 & 2742 & I & $-7.54$ & $-7.56$ & $-7.82$ & $-7.66$ & $36.25$ & $36.24$ & $35.98$ & $36.14$\\ 
Cyg~X$-$1 & 2743 & J & $-7.89$ & $-7.88$ & $-8.35$ & $-8.41$ & $35.91$ & $35.91$ & $35.44$ & $35.39$\\ 
Cyg~X$-$1 & 3407 & K & $-7.79$ & $-7.79$ & $-7.93$ & $-8.12$ & $36.01$ & $36.00$ & $35.87$ & $35.67$\\ 
Cyg~X$-$1 & 3814 & M & $-8.25$ & $-8.25$ & $-8.24$ & $-8.24$ & $35.54$ & $35.54$ & $35.55$ & $35.55$\\ 
Cyg~X$-$1 & 8525 & N & $-8.26$ & $-8.26$ & $-8.23$ & $-8.23$ & $35.54$ & $35.53$ & $35.57$ & $35.56$\\ 
Cyg~X$-$1 & 9847 & O & $-8.16$ & $-8.16$ & $-8.15$ & $-8.15$ & $35.63$ & $35.63$ & $35.65$ & $35.65$\\ 
Cyg~X$-$1 & 3815 & P & $-8.01$ & $-8.01$ & $-7.96$ & $-7.96$ & $35.79$ & $35.79$ & $35.83$ & $35.83$\\ 
Cyg~X$-$3 & 101 & A & $-8.48$ & $-8.49$ & $-8.40$ & $-8.40$ & $35.73$ & $35.73$ & $35.82$ & $35.81$\\ 
$\gamma$~Cas & 1895 & A & $-9.80$ & $-9.80$ & $-9.76$ & $-9.78$ & $33.50$ & $33.51$ & $33.54$ & $33.53$\\ 
GX~301$-$2 & 2733 & B & $-8.99$ & $-8.93$ & $-8.16$ & $-7.83$ & $31.54$ & $31.60$ & $32.37$ & $32.70$\\ 
GX~301$-$2 & 3433 & C & $-8.84$ & $-8.84$ & $-8.53$ & $-8.51$ & $31.69$ & $31.70$ & $32.01$ & $32.03$\\ 
GX~1$+$4 & 2710 & A & $-9.75$ & $-9.75$ & $-9.67$ & $-9.67$ & $31.93$ & $31.93$ & $32.00$ & $32.00$\\ 
GX~1$+$4 & 2744 & B & $-9.19$ & $-9.19$ & $-9.08$ & $-9.10$ & $32.48$ & $32.49$ & $32.60$ & $32.58$\\ 
Her~X$-$1 & 2703 & A & $-9.70$ & $-9.72$ & $-9.17$ & $-9.75$ & $30.83$ & $30.82$ & $31.36$ & $30.79$\\ 
Her~X$-$1 & 2704 & B & $-8.71$ & $-8.70$ & $-8.69$ & $-8.70$ & $31.83$ & $31.83$ & $31.84$ & $31.83$\\ 
Her~X$-$1 & 2705 & C & $-10.28$ & $-10.36$ & $-9.82$ & $-10.39$ & $30.26$ & $30.17$ & $30.71$ & $30.14$\\ 
Her~X$-$1 & 2749 & D & $-10.34$ & $-10.33$ & $-10.09$ & $-10.43$ & $30.19$ & $30.20$ & $30.45$ & $30.11$\\ 
Her~X$-$1 & 3821 & E & $-9.73$ & $-9.71$ & $-9.53$ & $-9.74$ & $30.80$ & $30.83$ & $31.00$ & $30.80$\\ 
Her~X$-$1 & 3822 & F & $-9.68$ & $-9.68$ & $-9.45$ & $-9.19$ & $30.86$ & $30.86$ & $31.09$ & $31.34$\\ 
Her~X$-$1 & 4375 & G & $-9.20$ & $-9.20$ & $-9.55$ & $-9.22$ & $31.34$ & $31.33$ & $30.99$ & $31.32$\\ 
Her~X$-$1 & 4585 & H & $-9.33$ & $-9.32$ & $-8.60$ & $-8.44$ & $31.20$ & $31.21$ & $31.94$ & $32.10$\\ 
Her~X$-$1 & 6149 & I & $-9.24$ & $-9.22$ & $-8.72$ & $-9.24$ & $31.29$ & $31.31$ & $31.81$ & $31.30$\\ 
Her~X$-$1 & 6150 & J & $-9.75$ & $-9.72$ & $-9.10$ & $-9.34$ & $30.78$ & $30.81$ & $31.43$ & $31.19$\\ 
LMC~X$-$4 & 9571 & A & $-10.03$ & $-10.04$ & $-10.31$ & $-10.04$ & $31.91$ & $31.90$ & $31.63$ & $31.90$\\ 
LMC~X$-$4 & 9573 & B & $-11.32$ & $-11.33$ & $-10.89$ & $-11.47$ & $30.62$ & $30.61$ & $31.05$ & $30.47$\\ 
LMC~X$-$4 & 9574 & C & $-10.83$ & $-10.82$ & $-10.37$ & $-9.43$ & $31.11$ & $31.12$ & $31.57$ & $32.51$\\ 
OAO~1657$-$415 & 12460 & A & $-10.62$ & $-10.79$ & $-9.59$ & $-8.58$ & $30.06$ & $29.89$ & $31.09$ & $32.10$\\ 
OAO~1657$-$415 & 1947 & B & $-9.67$ & $-9.67$ & $-9.24$ & $-9.16$ & $31.01$ & $31.01$ & $31.44$ & $31.52$\\ 
Vela~X$-$1 & 102 & A & $-10.95$ & $-10.74$ & $-10.56$ & $-9.51$ & $29.57$ & $29.78$ & $29.96$ & $31.01$\\ 
Vela~X$-$1 & 14654 & B & $-8.96$ & $-8.95$ & $-8.67$ & $-8.65$ & $31.56$ & $31.56$ & $31.85$ & $31.87$\\ 
Vela~X$-$1 & 1926 & C & $-10.87$ & $-10.85$ & $-10.41$ & $-11.04$ & $29.65$ & $29.67$ & $30.10$ & $29.48$\\ 
Vela~X$-$1 & 1927 & D & $-8.70$ & $-8.68$ & $-8.53$ & $-8.52$ & $31.82$ & $31.84$ & $31.98$ & $32.00$\\ 
Vela~X$-$1 & 1928 & E & $-8.52$ & $-8.51$ & $-8.48$ & $-8.49$ & $32.00$ & $32.00$ & $32.04$ & $32.02$\\ 
\enddata
\tabletypesize{\small}
\tablecomments{\small
Columns (1) and (2) give the XRB system name, \chandra-HETG
observation ID, and associated alphabetical label used in this
paper. Except for columns (1) and (2), columns show pairs of results
for the same parameter in the spherical (\bnet) and \myt\ (\myty)
model. The energy band for all quantities is $2-10$ keV
(\scr{sec-contfluxes}).  Thus, columns (3) and (4) give the total
observed continuum flux, and columns (5) and (6) the total intrinsic
continuum flux. Columns (7) and (8) give the observed luminosities
corresponding to the fluxes in columns (3) and (4), and columns (9) and
(10) the intrinsic luminosities corresponding to the fluxes in columns
(5) and (6).
} 
\end{deluxetable*}

%% file: taball_rg.tex
\startlongtable
\begin{deluxetable*}{C{1.2cm} C{1cm} cc c cccccc}
  \tablecaption{FWHM and deduced radius of line emitting region for the
  \feka\ line\label{tab-rg}}
\tabletypesize{\scriptsize}
\tablehead{
\colhead{Source} & 
\multicolumn{2}{c}{obs.} &
\multicolumn{2}{c}{$r$}&
\colhead{\mcoa} &
\multicolumn{2}{c}{$r_1$}&
\colhead{\mcob} &
\multicolumn{2}{c}{$r_2$}
\\
\colhead{}&
\colhead{}&
\colhead{}& %for letter
\multicolumn{2}{c}{($10^5r_g$)}&
\colhead{(\msun)}&
\multicolumn{2}{c}{($10^5$~km)}&
\colhead{(\msun)}&
\multicolumn{2}{c}{($10^5$~km)}
\\
\colhead{}&
\colhead{}&
\colhead{}& %for letter
\colhead{\bnet}&\colhead{\myty}&
\colhead{}&
\colhead{\bnet}&\colhead{\myty}&
\colhead{}&
\colhead{\bnet}&\colhead{\myty}
\\
\colhead{(1)}&
\multicolumn{2}{c}{(2)}&
\colhead{(3)}&
\colhead{(4)}&
\colhead{(5)}&
\colhead{(6)}&
\colhead{(7)}&
\colhead{(8)}&
\colhead{(9)}&
\colhead{(10)}
}
\startdata
4U1700$-$37 & 657 & A & \aer{0.736}{+0.544}{-0.252} & \aer{>1.261}{\quad}{\quad} & $1.96\pm0.19^{1}$ & \aer{2.129}{+1.576}{-0.728} & \aer{>3.651}{\quad}{\quad} & $2.44\pm0.27^{2}$ & \aer{2.651}{+1.962}{-0.907} & \aer{>4.545}{\quad}{\quad}\\ 
4U1822$-$371 & 671 & A & \aer{0.203}{+0.268}{-0.088} & \aer{0.416}{+4.005}{-0.240} & $0.97\pm0.24^{3}$ & \aer{0.291}{+0.384}{-0.127} & \aer{0.596}{+5.738}{-0.343} & $2.32^{4}$ & \aer{0.695}{+0.919}{-0.303} & \aer{1.426}{+13.724}{-0.821}\\ 
4U1822$-$371 & 9076 & B & \aer{0.268}{+0.220}{-0.114} & \aer{0.451}{+0.951}{-0.233} & $0.97\pm0.24^{3}$ & \aer{0.384}{+0.315}{-0.163} & \aer{0.646}{+1.363}{-0.334} & $2.32^{4}$ & \aer{0.918}{+0.753}{-0.389} & \aer{1.545}{+3.260}{-0.798}\\ 
4U1822$-$371 & 9858 & C & \aer{0.377}{+0.269}{-0.143} & \aer{0.584}{+1.020}{-0.284} & $0.97\pm0.24^{3}$ & \aer{0.541}{+0.385}{-0.205} & \aer{0.837}{+1.461}{-0.406} & $2.32^{4}$ & \aer{1.293}{+0.921}{-0.490} & \aer{2.002}{+3.494}{-0.972}\\ 
4U1908$+$075 & 5476 & A & \aer{0.153}{+0.177}{-0.077} & \aer{0.198}{+0.367}{-0.104} & $1.4^{c}$ & \aer{0.315}{+0.366}{-0.159} & \aer{0.410}{+0.758}{-0.214} & \quad & \aer{\quad}{\quad}{\quad} & \aer{\quad}{\quad}{\quad}\\ 
4U1908$+$075 & 5477 & B & \aer{>0.533}{\quad}{\quad} & \aer{>1.064}{\quad}{\quad} & $1.4^{c}$ & \aer{>1.102}{\quad}{\quad} & \aer{>2.200}{\quad}{\quad} & \quad & \aer{\quad}{\quad}{\quad} & \aer{\quad}{\quad}{\quad}\\ 
4U1908$+$075 & 6336 & C & \aer{0.271}{+0.595}{-0.160} & \aer{0.384}{+2.183}{-0.243} & $1.4^{c}$ & \aer{0.560}{+1.230}{-0.331} & \aer{0.795}{+4.514}{-0.503} & \quad & \aer{\quad}{\quad}{\quad} & \aer{\quad}{\quad}{\quad}\\ 
Cen~X$-$3 & 1943 & A & \aer{\ldots}{a}{\quad} & \aer{1.227}{+4.317}{-0.620} & $1.21\pm0.21^{5}$ & \aer{\ldots}{a}{\quad} & \aer{2.192}{+7.715}{-1.108} & $1.57\pm0.16^{1}$ & \aer{\ldots}{a}{\quad} & \aer{2.845}{+10.010}{-1.438}\\ 
Cen~X$-$3 & 705 & B & \aer{0.231}{+0.205}{-0.101} & \aer{0.360}{+0.825}{-0.189} & $1.21\pm0.21^{5}$ & \aer{0.413}{+0.367}{-0.180} & \aer{0.643}{+1.475}{-0.338} & $1.57\pm0.16^{1}$ & \aer{0.536}{+0.476}{-0.234} & \aer{0.835}{+1.914}{-0.439}\\ 
Cen~X$-$3 & 7511 & C & \aer{\ldots}{a}{\quad} & \aer{0.391}{+0.168}{-0.106} & $1.21\pm0.21^{5}$ & \aer{\ldots}{a}{\quad} & \aer{0.698}{+0.300}{-0.190} & $1.57\pm0.16^{1}$ & \aer{\ldots}{a}{\quad} & \aer{0.906}{+0.389}{-0.246}\\ 
Cyg~X$-$1 & 3814 & M & \aer{\ldots}{a}{\quad} & \aer{0.191}{+0.260}{-0.096} & $10.1^{6}$ & \aer{\ldots}{a}{\quad} & \aer{2.846}{+3.876}{-1.436} & $14.8\pm1.0^{8}$ & \aer{\ldots}{a}{\quad} & \aer{4.171}{+5.680}{-2.104}\\ 
%Cyg~X$-$1 & 8525 & N & \aer{119.892}{f}{\quad} & \aer{119.892}{f}{\quad} & $10.1^{7}$ & \aer{1788.600}{f}{\quad} & \aer{1788.600}{f}{\quad} & $14.8\pm1.0^{8}$ & \aer{2620.919}{f}{\quad} & \aer{2620.919}{f}{\quad}\\ 
Cyg~X$-$1 & 9847 & O & \aer{0.051}{+0.174}{-0.026} & \aer{>0.053}{\quad}{\quad} & $10.1^{6}$ & \aer{0.767}{+2.593}{-0.395} & \aer{>0.787}{\quad}{\quad} & $14.8\pm1.0^{7}$ & \aer{1.124}{+3.800}{-0.579} & \aer{>1.154}{\quad}{\quad}\\ 
Cyg~X$-$1 & 3815 & P & \aer{0.108}{+0.067}{-0.041} & \aer{0.140}{+0.103}{-0.056} & $10.1^{6}$ & \aer{1.618}{+1.003}{-0.609} & \aer{2.083}{+1.531}{-0.830} & $14.8\pm1.0^{7}$ & \aer{2.371}{+1.470}{-0.893} & \aer{3.052}{+2.244}{-1.217}\\ 
Cyg~X$-$3 & 101 & A & \aer{\ldots}{a}{\quad} & \aer{0.083}{+0.076}{-0.041} & $1.0^{8}$ & \aer{\ldots}{a}{\quad} & \aer{0.123}{+0.112}{-0.060} & $8.0^{8}$ & \aer{\ldots}{a}{\quad} & \aer{0.982}{+0.896}{-0.483}\\ 
$\gamma$~Cas & 1895 & A & \aer{\ldots}{a}{\quad} & \aer{>0.148}{\quad}{\quad} & $0.4^{9}$ & \aer{\ldots}{a}{\quad} & \aer{>0.087}{\quad}{\quad} & $1.9^{10}$ & \aer{\ldots}{a}{\quad} & \aer{>0.415}{\quad}{\quad}\\ 
GX~301$-$2 & 2733 & B & \aer{0.978}{+0.155}{-0.118} & \aer{5.806}{+6.015}{-2.364} & $1.35^{12}$ & \aer{1.949}{+0.309}{-0.236} & \aer{11.578}{+11.995}{-4.714} & $2.4\pm0.7^{11}$ & \aer{3.465}{+0.549}{-0.420} & \aer{20.584}{+21.324}{-8.381}\\ 
GX~301$-$2 & 3433 & C & \aer{0.725}{+0.153}{-0.116} & \aer{3.059}{+4.225}{-1.266} & $1.85\pm0.6^{11}$ & \aer{1.982}{+0.418}{-0.317} & \aer{8.360}{+11.546}{-3.459} & $2.4\pm0.7^{11}$ & \aer{2.571}{+0.542}{-0.411} & \aer{10.845}{+14.978}{-4.487}\\ 
GX~1$+$4 & 2710 & A & \aer{>0.467}{\quad}{\quad} & \aer{>0.540}{\quad}{\quad} & $1.35^{12}$ & \aer{>0.930}{\quad}{\quad} & \aer{>1.077}{\quad}{\quad} & \quad & \aer{\quad}{\quad}{\quad} & \aer{\quad}{\quad}{\quad}\\ 
GX~1$+$4 & 2744 & B & \aer{\ldots}{a}{\quad} & \aer{>1.755}{\quad}{\quad} & $1.35^{12}$ & \aer{\ldots}{a}{\quad} & \aer{>3.500}{\quad}{\quad} & \quad & \aer{\ldots}{a}{\quad} & \aer{\quad}{\quad}{\quad}\\ 
%Her~X$-$1 & 2703 & A & \aer{119.892}{f}{\quad} & \aer{119.892}{f}{\quad} & $1.3^{14}$ & \aer{230.216}{f}{\quad} & \aer{230.216}{f}{\quad} & $1.7^{14}$ & \aer{301.052}{f}{\quad} & \aer{301.052}{f}{\quad}\\ 
%Her~X$-$1 & 2704 & B & \aer{\ldots}{a}{\quad} & \aer{119.892}{f}{\quad} & $1.3^{14}$ & \aer{\ldots}{a}{\quad} & \aer{230.216}{f}{\quad} & $1.7^{14}$ & \aer{\ldots}{a}{\quad} & \aer{301.052}{f}{\quad}\\ 
Her~X$-$1 & 2705 & C & \aer{>0.127}{\quad}{\quad} & \aer{>0.088}{\quad}{\quad} & $1.3^{13}$ & \aer{>0.244}{\quad}{\quad} & \aer{>0.169}{\quad}{\quad} & $1.7^{13}$ & \aer{>0.319}{\quad}{\quad} & \aer{>0.221}{\quad}{\quad}\\ 
Her~X$-$1 & 2749 & D & \aer{\ldots}{a}{\quad} & \aer{>0.840}{\quad}{\quad} & $1.3^{13}$ & \aer{\ldots}{a}{\quad} & \aer{>1.613}{\quad}{\quad} & $1.7^{13}$ & \aer{\ldots}{a}{\quad} & \aer{>2.109}{\quad}{\quad}\\ 
Her~X$-$1 & 3821 & E & \aer{0.490}{+0.565}{-0.212} & \aer{0.498}{+0.701}{-0.230} & $1.3^{13}$ & \aer{0.940}{+1.084}{-0.406} & \aer{0.957}{+1.345}{-0.442} & $1.7^{13}$ & \aer{1.229}{+1.418}{-0.531} & \aer{1.251}{+1.759}{-0.578}\\ 
Her~X$-$1 & 3822 & F & \aer{0.346}{+0.252}{-0.130} & \aer{0.646}{+1.355}{-0.314} & $1.3^{13}$ & \aer{0.664}{+0.485}{-0.249} & \aer{1.241}{+2.602}{-0.603} & $1.7^{13}$ & \aer{0.868}{+0.634}{-0.326} & \aer{1.622}{+3.402}{-0.788}\\ 
Her~X$-$1 & 4375 & G & \aer{0.128}{+0.226}{-0.079} & \aer{0.069}{+0.154}{-0.038} & $1.3^{13}$ & \aer{0.245}{+0.433}{-0.152} & \aer{0.132}{+0.296}{-0.073} & $1.7^{13}$ & \aer{0.320}{+0.567}{-0.199} & \aer{0.172}{+0.388}{-0.095}\\ 
Her~X$-$1 & 4585 & H & \aer{0.104}{+0.196}{-0.055} & \aer{>0.167}{\quad}{\quad} & $1.3^{13}$ & \aer{0.200}{+0.377}{-0.105} & \aer{>0.322}{\quad}{\quad} & $1.7^{13}$ & \aer{0.262}{+0.493}{-0.137} & \aer{>0.421}{\quad}{\quad}\\ 
Her~X$-$1 & 6149 & I & \aer{0.243}{+0.218}{-0.103} & \aer{0.178}{+0.130}{-0.066} & $1.3^{13}$ & \aer{0.466}{+0.419}{-0.198} & \aer{0.343}{+0.250}{-0.127} & $1.7^{13}$ & \aer{0.609}{+0.549}{-0.259} & \aer{0.448}{+0.327}{-0.166}\\ 
Her~X$-$1 & 6150 & J & \aer{0.503}{+0.982}{-0.268} & \aer{0.627}{+1.768}{-0.331} & $1.3^{13}$ & \aer{0.965}{+1.885}{-0.514} & \aer{1.203}{+3.395}{-0.636} & $1.7^{13}$ & \aer{1.262}{+2.465}{-0.672} & \aer{1.573}{+4.439}{-0.831}\\ 
LMC~X$-$4 & 9571 & A & \aer{>0.200}{\quad}{\quad} & \aer{>0.104}{\quad}{\quad} & $1.25\pm0.11^{14}$ & \aer{>0.369}{\quad}{\quad} & \aer{>0.191}{\quad}{\quad} & $1.57\pm0.11^{1}$ & \aer{>0.463}{\quad}{\quad} & \aer{>0.240}{\quad}{\quad}\\ 
LMC~X$-$4 & 9573 & B & \aer{\ldots}{a}{\quad} & \aer{0.160}{+0.342}{-0.083} & $1.25\pm0.11^{14}$ & \aer{\ldots}{a}{\quad} & \aer{0.295}{+0.631}{-0.153} & $1.57\pm0.11^{1}$ & \aer{\ldots}{a}{\quad} & \aer{0.370}{+0.793}{-0.192}\\ 
%LMC~X$-$4 & 9574 & C & \aer{119.892}{f}{\quad} & \aer{119.892}{f}{\quad} & $1.25\pm0.11^{15}$ & \aer{221.361}{f}{\quad} & \aer{221.361}{f}{\quad} & $1.57\pm0.11^{1}$ & \aer{278.030}{f}{\quad} & \aer{278.030}{f}{\quad}\\ 
OAO~1657$-$415 & 12460 & A & \aer{\ldots}{a}{\quad} & \aer{>0.494}{\quad}{\quad} & $1.74\pm0.30^{1}$ & \aer{\ldots}{a}{\quad} & \aer{>1.270}{\quad}{\quad} & $1.42\pm0.26^{15}$ & \aer{\ldots}{a}{\quad} & \aer{>1.037}{\quad}{\quad}\\ 
OAO~1657$-$415 & 1947 & B & \aer{0.424}{+0.835}{-0.250} & \aer{>0.214}{\quad}{\quad} & $1.74\pm0.30^{1}$ & \aer{1.091}{+2.146}{-0.643} & \aer{>0.549}{\quad}{\quad} & $1.42\pm0.26^{15}$ & \aer{0.890}{+1.751}{-0.524} & \aer{>0.448}{\quad}{\quad}\\ 
Vela~X$-$1 & 102 & A & \aer{\ldots}{a}{\quad} & \aer{>0.515}{\quad}{\quad} & $1.77\pm0.08^{16}$ & \aer{\ldots}{a}{\quad} & \aer{>1.347}{\quad}{\quad} & $2.12\pm0.16^{2}$ & \aer{\ldots}{a}{\quad} & \aer{>1.613}{\quad}{\quad}\\ 
Vela~X$-$1 & 14654 & B & \aer{0.673}{+0.403}{-0.215} & \aer{>1.357}{\quad}{\quad} & $1.77\pm0.08^{16}$ & \aer{1.759}{+1.055}{-0.562} & \aer{>3.547}{\quad}{\quad} & $2.12\pm0.16^{2}$ & \aer{2.107}{+1.263}{-0.673} & \aer{>4.248}{\quad}{\quad}\\ 
%Vela~X$-$1 & 1926 & C & \aer{\ldots}{a}{\quad} & \aer{119.892}{f}{\quad} & $1.77\pm0.08^{17}$ & \aer{\ldots}{a}{\quad} & \aer{313.448}{f}{\quad} & $2.12\pm0.16^{2}$ & \aer{\ldots}{a}{\quad} & \aer{375.429}{f}{\quad}\\ 
Vela~X$-$1 & 1927 & D & \aer{1.313}{+0.659}{-0.379} & \aer{119.892}{f}{\quad} & $1.77\pm0.08^{16}$ & \aer{3.433}{+1.724}{-0.991} & \aer{313.448}{f}{\quad} & $2.12\pm0.16^{2}$ & \aer{4.112}{+2.065}{-1.187} & \aer{375.429}{f}{\quad}\\ 
Vela~X$-$1 & 1928 & E & \aer{>1.873}{\quad}{\quad} & \aer{>1.925}{\quad}{\quad} & $1.77\pm0.08^{16}$ & \aer{>4.897}{\quad}{\quad} & \aer{>5.034}{\quad}{\quad} & $2.12\pm0.16^{2}$ & \aer{>5.865}{\quad}{\quad} & \aer{>6.029}{\quad}{\quad}\\ 
\enddata
\tablecomments{
Cases for which the FWHM was frozen in the fit are not
included, unless FWHM errors were obtained for one of the
models.
Columns (1) and (2) give the XRB system name, \chandra-HETG
observation ID, and associated alphabetical label used in this
paper. Columns (3), (4), (6), (7), (9) and (10) show pairs of results
for the same parameter in the spherical (\bnet) and \myt\ (\myty)
model.
Columns (3) and (4) give the radius $r$ in units of $10^5 r_g$, where
the gravitational radius $r_g = GM_{\rm CO} / c^2$.
Column (5) gives the lowest compact object mass estimate, \mcoa,
available in the literature.  Columns (6) and (7) give the radius in
km, assuming the \mcoa\ value from column (5).  Column (8) gives the
highest compact object mass estimate, \mcob, if available in the
literature.  Columns (9) and (10) give the radius in km, assuming
the \mcob\ value from column (8), where available.
Letter notes are as in
\tr{tab-trans}, except for $^c$, indicating that a ``canonical''
compact object mass was assumed. \mcoa\ and \mcob\ references- $^1$:
\citet{falanga2015}, $^2$: \citet{clark2002}, $^3$:
\citet{jonker2003}, $^4$: \citet{munoz-darias2005}, $^5$:
\citet{ash1999}, $^6$:
\citet{herrero1995}, $^7$: \citet{orosz2011}, $^8$:
\citet{vilhu2009}; $^{9}$: \citet{smith2012}; $^{10}$:
\citet{harmanec2000}; $^{11}$: \citet{kaper2006}; $^{12}$:
\citet{hinkle2006}; $^{13}$: \citet{leahy2014}; $^{14}$:
\citet{vandermeer2007}; $^{15}$: \citet{mason2012}; $^{16}$:
\citet{rawls2011}.
}
\end{deluxetable*}